\documentclass[%
 reprint,
 superscriptaddress,
 amsmath,amssymb,
 prl,
]{revtex4-2}

\usepackage{graphicx}
\usepackage{dcolumn}
\usepackage{bm}
\usepackage{gensymb}
\usepackage{svg}
\usepackage{afterpage}
\usepackage{physics}
\usepackage{tabularx}
\usepackage{xr}
\usepackage{natbib}
\usepackage{amsmath}
\usepackage{amsfonts}
\usepackage[T1]{fontenc}
\usepackage[utf8]{inputenc}

\DeclareUnicodeCharacter{2061}{}

\newcommand{\nyuphysics}{Center for Soft Matter Research, Department of Physics, New York University, New York 10003, USA}
\newcommand{\nyusimons}{Simons Center for Computational Physical Chemistry, Department of Chemistry, New York University, New York 10003, USA}
\newcommand{\nyucourant}{Courant Institute of Mathematical Sciences, New York University, New York 10003, USA}
\newcommand{\nyucns}{Center for Neural Science, New York University, New York 10003, USA}

\begin{document}

\preprint{APS/123-QED}

\author{Mathias Casiulis}
\thanks{Equal contribution}
\email{mc9287@nyu.edu}
\affiliation{\nyuphysics}
\affiliation{\nyusimons}
\author{Satyam Anand}
\thanks{Equal contribution}
\email{sa7483@nyu.edu}
\affiliation{\nyucourant}
\affiliation{\nyuphysics}
\author{Stefano Martiniani}
\email{sm7683@nyu.edu}
\affiliation{\nyuphysics}
\affiliation{\nyusimons}
\affiliation{\nyucourant}
\affiliation{\nyucns}

\title{Hyperuniform systems are maximally irreversible}

\begin{abstract}

Hyperuniform systems, defined by the anomalous suppression of large-scale density fluctuations, are a paradigm of non-equilibrium self-assembly.
While mechanisms underlying the self-assembly of hyperuniform states have been widely studied, the energetics of this process remain unexplored.
This raises a fundamental question: what is the energetic cost of self-assembling a hyperuniform system?
Here, we address this question across several noisy particle systems drawn from soft matter and machine learning, in which hyperuniformity can be induced by tuning noise correlations.
Despite their distinct microscopic dynamics, we uncover a universal behavior across all systems: hyperuniform states are maximally irreversible, as quantified by the entropy production rate.
Further, we develop a path integral formulation of the entropy production rate directly from the microscopic dynamics, which explains our observations.
Our work establishes a direct link between emergent long-range structure and time irreversibility and opens a new avenue of probing the energetic cost of hyperuniform self-assembly, ubiquitous across physics, biology, and materials science.
\end{abstract}

\maketitle

In physics, ``order'' spans a spectrum ranging from perfectly ordered crystals at zero temperature to completely disordered ideal gases.
Between these extremes lie systems with correlated disorder, such as disordered hyperuniformity~\cite{Torquato2018}.
Hyperuniform systems are isotropic and, while locally disordered (like ideal gases), exhibit suppression of density fluctuations at large length scales (like crystals)~\cite{Torquato2018}. 
Hyperuniform systems have attracted significant attention across physics~\cite{Torquato2018}, chemistry~\cite{Martelli2017}, biology~\cite{Liu2024b}, and mathematics~\cite{Lachieze-Rey2025}, notably because of their applications in transport~\cite{Leseur2016} and sampling~\cite{Pilleboue2015, Shih2023}.

Hyperuniformity arises across a wide range of physical systems, including equilibrium systems with long-range interactions~\cite{Torquato2015} and at critical points of certain phase transitions~\cite{Hexner2015}.
It also emerges in non-equilibrium steady states (NESS) without requiring either long-range interactions or criticality, forming a rapidly growing class of hyperuniform systems~\cite{Lei2024}.
Representative examples span active matter systems~\cite{Lei2019}, diffusive processes~\cite{Jack2015,Ikeda2023}, and shaken granulars~\cite{Maire2025a}.

In systems at NESS, the entropy production rate (EPR) quantifies the energetic cost of maintaining the steady state and the ``distance'' from equilibrium~\cite{OByrne2022}.
Despite substantial progress along two largely independent directions, viz., identifying non-equilibrium routes to hyperuniformity~\cite{Lei2024} and quantifying EPR in out-of-equilibrium systems~\cite{OByrne2022}, the connection between hyperuniformity and time irreversibility remains unexplored.
This motivates a deceptively simple yet fundamental question: how irreversible are hyperuniform systems?

Here, we answer this question combining simulations and theory.
We study several non-equilibrium particle systems in which hyperuniformity is induced by tuning noise correlations, either through inter-particle noisy interactions or via an active bath~\cite{Anand2025,Anand2026}. 
Remarkably, despite having microscopically distinct dynamics, we uncover a universal behavior: all systems exhibit maximal time-irreversibility precisely when they become hyperuniform.
Further, we construct an analytical theory for the EPR using a path-integral formulation, which explains our observations.
Altogether, our results demonstrate that, across a broad class of systems, hyperuniform states are maximally irreversible and therefore energetically expensive to sustain.
More generally, our work opens up an entirely new avenue of studying the energetics of hyperuniform systems, with potential applications for biological~\cite{Liu2024b}, ecological~\cite{Ge2023}, and social~\cite{Dong2023} systems, where hyperuniformity spontaneously emerges despite the energetic cost.

\begin{figure*}[htpb!]
    \centering
    \includegraphics[width=\linewidth]{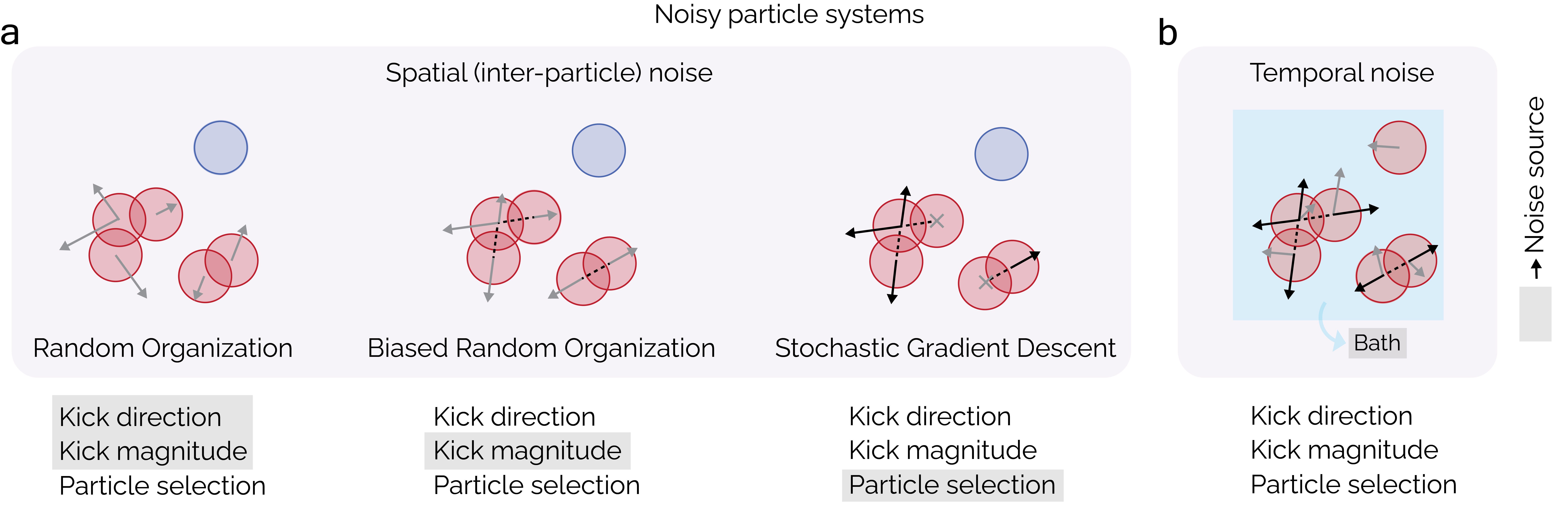}
    \caption{\small \textbf{Noisy particle systems.}
    $(a)$ Schematics of systems having spatial (inter-particle) noise---RO, BRO, and SGD (left to right).
    Particles shown in blue are free of overlaps, while those in red overlap with at least one particle.
    Gray crosses in SGD denote particles that are not selected for update.
    In the RO dynamics, both the direction and magnitude of particle displacements are stochastic; in BRO, randomness is only in the kick magnitude; and in SGD, noise enters solely through the selection of particles that are updated.
    ($b$) Schematic of a system having temporal noise.
    In this system, interacting passive particles are immersed in a non-equilibrium bath (light blue), modeled as a colored noise acting on particles.
    In both ($a$) and ($b$), gray represents noise, whereas black indicates deterministic interactions.
    Dashed black lines connect the centers of overlapping particle pairs.
}
    \label{fig:figure1}
\end{figure*}

\section{Setup}

We study two classes of systems: (i) systems with noisy interactions between particles~\cite{Anand2025}, and (ii) systems in which noise arises from an active bath~\cite{Anand2026}.
Within the first class, called random-organizing systems, we consider three specific systems: random organization (RO)~\cite{Corte2008}, biased random organization (BRO)~\cite{Milz2013,Wilken2020}, and stochastic gradient descent (SGD)~\cite{Zhang2024a,Anand2025}.
In these systems, the noise is correlated across particles but uncorrelated in time~\cite{Anand2025}.
In contrast, the second class consists of passive particles immersed in an active bath, where the noise is uncorrelated between particles but exhibits temporal correlations~\cite{Anand2026}.
These contrasting correlation structures make the two classes fundamentally distinct.
All systems are discrete-time systems composed of $N$ spherical particles of radius $R$ in $d$-dimensional space.
We describe each system in detail below.

\subsection{Spatial (inter-particle) noise}

RO, BRO, and SGD are random-organizing systems in which noise enters through inter-particle interactions.
At each time-step, particles that do not overlap with any neighbors remain stationary (blue in Fig.~\ref{fig:figure1}a), while overlapping particles (red in Fig.~\ref{fig:figure1}a) are displaced according to system-specific rules designed to resolve overlaps.
When initialized from random configurations, all three systems exhibit a transition from an absorbing to an active state as the packing fraction $\phi = N V_s/V_c$ is increased, where $V_s$ is the volume of a particle and $V_c$ is the total available volume~\cite{Hinrichsen2000}.
For $\phi < \phi_c$, the dynamics eventually resolve all overlaps, and the motion stops completely: this is an absorbing state.
For $\phi > \phi_c$, however, overlaps cannot be fully resolved, and the system eventually reaches a NESS characterized by a non-zero, stationary fraction of overlapping particles.
We now describe each of these three systems in detail.

\textit{\textbf{Random organization.}} RO was originally introduced to model experiments on sheared colloidal suspensions, where interactions between colloids during a shear cycle are represented as effective ``random kicks''~\cite{Corte2008}.
Here, we consider a simpler, isotropic version of RO without external shear, which retains all the essential features of the original model (Fig.~\ref{fig:figure1}a)~\cite{Milz2013,Wilken2020,Anand2025}.

In RO, overlapping (``active'') particles impart pairwise kicks to one another at every time-step. Both the magnitude and direction of the kicks are random, while there is no stochasticity in the selection of active  particles, i.e., all active particles are moved at every time step (Fig.~\ref{fig:figure1}a, Methods).
The total noisy kick exerted by particle $i$ on particle $j$ at time-step $m$ is denoted by $\omega_{ij,\alpha}^m$ (Methods). 
We introduce a Pearson correlation coefficient $c^s \in [-1,1]$ between corresponding components of the pairwise noise vectors $\omega_{ij,\alpha}^m$ and $\omega_{ji,\alpha}^m$.
The limit $c^s = 0$ corresponds to uncorrelated pairwise kicks, whereas $c^s = -1$ enforces anti-correlated kicks, implying equal magnitudes and opposite directions. Here, we focus on $c^s \in [-1,0]$, corresponding to a regime in which particles effectively repel one another (for positive values, particles effectively move together, much like in Ref.~\cite{DalCengio2025}).

\textit{\textbf{Biased random organization.}}
BRO was introduced as a dynamical model to study disordered packings of spheres~\cite{Wilken2020}.
In addition, its active phase has been shown to exhibit crystalline order in $2d$, forbidden in equilibrium systems by the Mermin-Wagner argument~\cite{Galliano2023,Guo2026}. 

In BRO, overlapping particles impart pairwise kicks to one another at every time-step, similar to RO. However, the pairwise inter-particle kicks are anisotropic: they act along the line connecting the centers of overlapping particles (Fig.~\ref{fig:figure1}a, Methods).
Hence, stochasticity arises solely from the magnitude of the kicks, while both the direction of the kicks and the selection of active particles are deterministic (Fig.~\ref{fig:figure1}a).
As defnied for RO, we introduce a Pearson correlation coefficient $c^s$ between components of the pairwise noise vectors $\omega_{ij,\alpha}^m$ and $\omega_{ji,\alpha}^m$. The limit $c = 0$ corresponds to uncorrelated kick magnitudes, whereas $c = -1$ implies equal kick magnitudes.

\textit{\textbf{Stochastic gradient descent.}} SGD is a canonical optimization method widely used in machine learning to train artificial neural networks by minimizing objective functions expressed as sums of individual contributions~\cite{Bishop2023}.
The stochasticity in SGD arises from evaluating only a randomly selected subset of these contributions at each iteration, rather than the full sum.

For particle systems, the objective function can be taken as the total energy $E = \sum_i \sum_{j \geq i} V(\mathbf{x}_i, \mathbf{x}_j)$, where $V(\mathbf{x}_i, \mathbf{x}_j)$ is an arbitrary pair potential~\cite{Zhang2024a, Anand2025}.
An SGD step then consists of selecting a random subset of interaction terms and updating the positions of the associated particles (either individually or jointly) to reduce the corresponding partial energy (Fig.~\ref{fig:figure1}a)~\cite{Zhang2024a, Anand2025}. 
This procedure contrasts with deterministic gradient descent, in which all particles are updated simultaneously according to the gradient of the full energy.

In SGD, particles interact via a short-range pairwise potential (Methods). For any overlapping pair $(i,j)$ at a given time step, there are four possible update outcomes: (i) only particle $i$ moves, (ii) only particle $j$ moves, (iii) both particles move, and (iv) neither particle moves (Fig.~\ref{fig:figure1}a, Methods). Thus, stochasticity enters through the selection of active particles, whereas both the direction and magnitude of the resulting kicks are deterministic (Fig.~\ref{fig:figure1}a).
Analogous to RO and BRO, we introduce a Pearson correlation coefficient $c^s$ between components of the pairwise noise vectors $\omega_{ij,\alpha}^m$ and $\omega_{ji,\alpha}^m$. The limit $c^s = 0$ corresponds to uncorrelated selection noise for a particle pair (all four outcomes are equally likely), whereas $c = -1$ implies perfectly correlated selection (only outcomes (iii) and (iv) occur).

\begin{figure*}[htpb!]
    \centering
    \includegraphics[width=\linewidth]{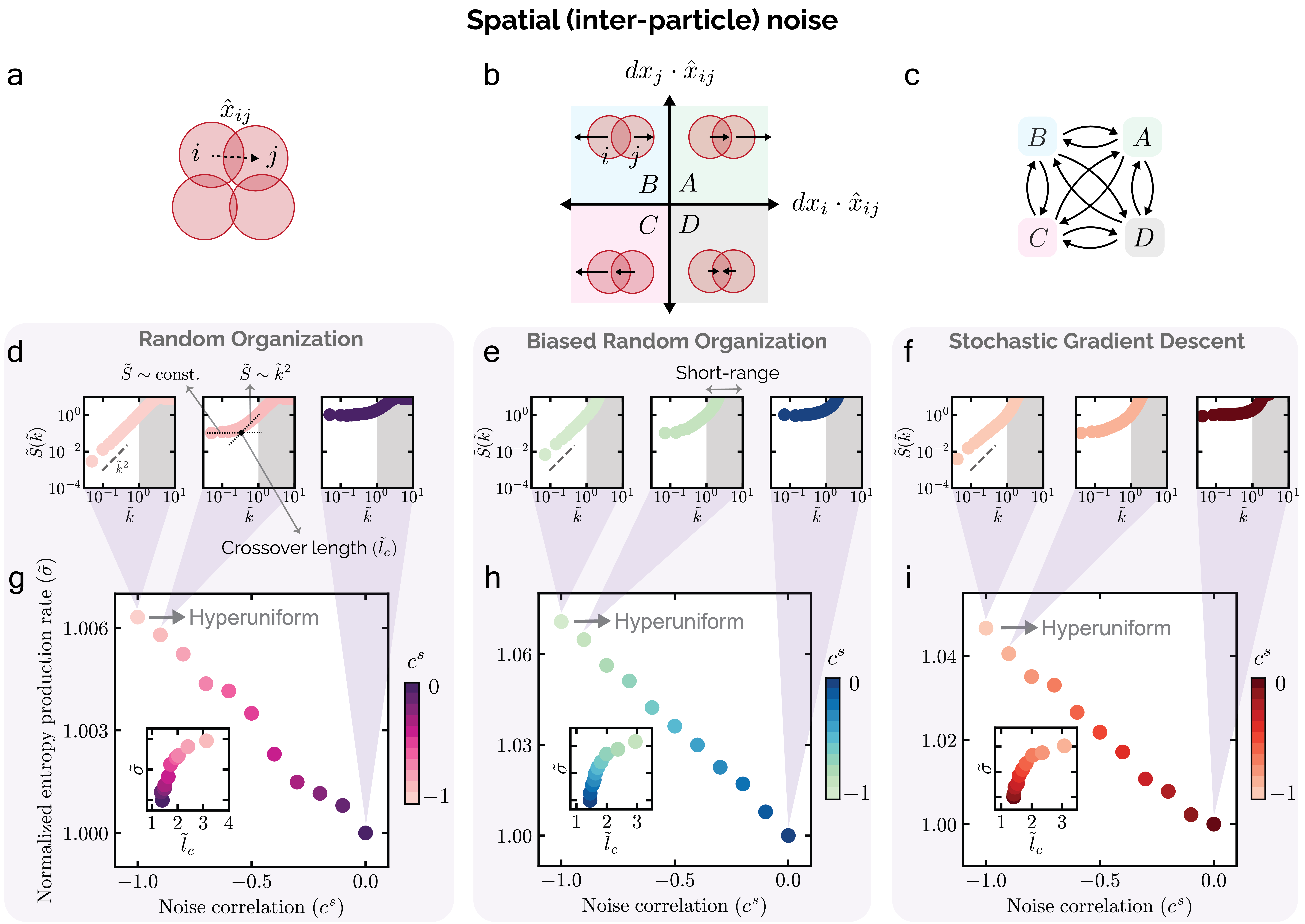}
    \caption{\small \textbf{Time-irreversibility and hyperuniformity in systems with spatial noise.} To measure the EPR, we track overlapping particle pairs $(a)$.
    We coarse-grain the dynamics into four states ($A, B, C, D$) based on the relative direction of motion within each particle pair $(b)$.
    We then compute the EPR of the resulting four-state Markov process $(c)$.
    The normalized, radially averaged structure factor $\widetilde{S}(\widetilde{k})$ plotted as a function of the normalized wavenumber $\widetilde{k}$ for RO $(d)$, BRO $(e)$, and SGD $(f)$.
    Here, $\widetilde{S} = S(k)/S_{0}(2\pi/L)$, where $S_{0}(2\pi/L)$ denotes the structure factor evaluated at $k = 2\pi/L$ in the case of uncorrelated noise ($c^s=0$), and $L$ is the system size.
    The dimensionless wavenumber is defined as $\widetilde{k} = k/k_0$, where $k_0$ satisfies $\widetilde{S}(k_0) = 1$ for the same system with anti-correlated noise ($c^s = -1$).
    In simulations, the crossover wavenumber $\widetilde{k}_c = 1/\widetilde{l}_c$ is determined as the intersection point (in log-log plot) between a low-$\widetilde{k}$ regime characterized by a zero slope and an intermediate regime near $\widetilde{k} \approx 1$ exhibiting a slope of $2$.
    Dashed black lines denote $\widetilde{k}^2$ scaling.
    Shaded gray regions highlight the short-range regime defined by $\widetilde{k} > 1$.
    The normalized EPR $\widetilde{\sigma}$ plotted as a function of the inter-particle noise correlation coefficient $c^s$ for RO $(g)$, BRO $(h)$, and SGD $(i)$.
    The EPR $\sigma$ is normalized as: $\widetilde{\sigma} (c^s) = \sigma (c^s) / \sigma (c^s = 0)$.
    Insets in $(g)$, $(h)$, and $(i)$ display $\widetilde{\sigma}$ as a function of the normalized crossover length $\widetilde{l}_c$.
}
    \label{fig:figure2}
\end{figure*}

\subsection{Temporal noise}

The second class of systems we consider consists of interacting passive particles immersed in a non-equilibrium bath (Fig.~\ref{fig:figure1}b)~\cite{Anand2026}. The bath is active and exhibits temporal correlations. A canonical example is that of colloids suspended in an active (e.g., bacterial) bath, where fluctuations display exponentially decaying temporal correlations~\cite{Wu2000}.

Here, we consider a minimal non-equilibrium setting in which particles interact via a short-range pairwise potential and are driven by bath noise having only lag-$1$ temporal correlations (Methods)~\cite{Anand2026}.
The noise acting on particle $i$ at time-step $m$ is denoted by $\boldsymbol{\zeta}_i^{m}$ (Methods). 
We introduce a Pearson correlation coefficient $c^t \in [-1/2, 1/2]$ between $\zeta_{i,\alpha}^m$ and $\zeta_{j,\alpha}^{m+1}$.
The limit $c^t = 0$ corresponds to temporally uncorrelated noise (i.e., equilibrium dynamics), whereas $c^t = -1/2$ corresponds to maximally anti-correlated kicks across consecutive time steps. 
Note that in this system, stochasticity arises solely from the active bath, while interparticle interactions remain deterministic, as the particles themselves are passive (unlike random-organizing systems where particles are ``active'') (compare Figs.~\ref{fig:figure1}a and b).

\begin{figure*}[htpb!]
    \centering
    \includegraphics[width=\linewidth]{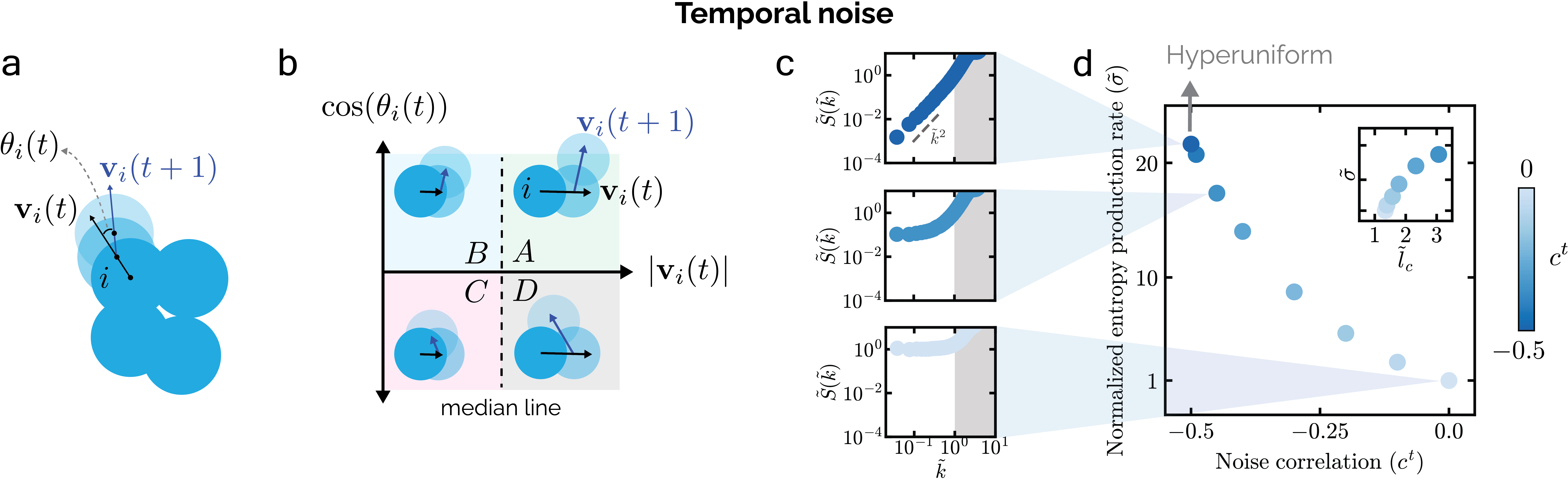}
    \caption{\small \textbf{Time-irreversibility and hyperuniformity in systems with temporal noise.} To measure the EPR, we track the dynamics of particles across two consecutive time steps, $t$ and $t+1$.
    We coarse-grain the dynamics into four states ($A, B, C, D$) based on particle speed (fast or slow) and the angle between the velocities at $t$ and $t+1$ (persistent or reversing), and compute the corresponding EPR $(b)$.
    $(c)$ The normalized, radially averaged structure factor $\widetilde{S}(\widetilde{k})$ plotted as a function of the normalized wavenumber $\widetilde{k}$.
    Here, $\widetilde{S} = S(k)/S_{0}(2\pi/L)$, where $S_{0}(2\pi/L)$ denotes the structure factor evaluated at $k = 2\pi/L$ in the case of uncorrelated noise ($c^t=0$), and $L$ is the system size.
    The dimensionless wavenumber is defined as $\widetilde{k} = k/k_0$, where $k_0$ satisfies $\widetilde{S}(k_0) = 1$ for the same system with $c^t = -1/2$.
    In simulations, the crossover wavenumber $\widetilde{k}_c = 1/\widetilde{l}_c$ is determined as the intersection point (in log-log plot) between a low-$\widetilde{k}$ regime characterized by a zero slope and an intermediate regime near $\widetilde{k} \approx 1$ exhibiting a slope of $2$.
    Dashed black line denotes $\widetilde{k}^2$ scaling.
    Shaded gray regions highlight the short-range regime defined by $\widetilde{k} > 1$.
    $(d)$ The normalized EPR $\widetilde{\sigma}$ plotted as a function of the lag-$1$ temporal noise correlation coefficient $c^t$. The EPR $\sigma$ is normalized as: $\widetilde{\sigma} (c^t) = \sigma (c^t) / \sigma (c^t = 0)$. Inset displays $\widetilde{\sigma}$ as a function of the normalized crossover length $\widetilde{l}_c$.
}
    \label{fig:figure3}
\end{figure*}

\section{Time irreversibility}

We perform particle simulations for all systems in the dense phase and evolve them until they reach a NESS.
Convergence to a NESS is identified by the emergence of time-independent spatial structure, which we quantify using the radially averaged static structure factor $S(k)$, where $k = 2\pi / l$ is the wavenumber and $l$ is the diameter of the hypersphere (Methods).

For random-organizing systems (RO, BRO, and SGD), it is known that the interparticle noise correlation coefficient $c^s$ governs the emergent long-range structure~\cite{Anand2025}.
For systems with temporally correlated noise, the temporal correlation coefficient $c^t$ plays an analogous role~\cite{Anand2026}.
Specifically, there exists a characteristic length scale $l_c = 2\pi / k_c$ such that, for $l < l_c$ (i.e., $k > k_c$), the structure factor scales as $S(k) \sim k^2$, while for $l > l_c$ (i.e., $k < k_c$), it saturates to a constant, $S(k) \sim \mathrm{const.}$ (Figs.~\ref{fig:figure2}d, e, f, and \ref{fig:figure3}c).
The crossover length $l_c$ increases monotonically with both $c^s$ and $c^t$, and diverges in the limiting cases, leading to strongly hyperuniform behavior (class III~\cite{Torquato2018}, $S(k) \sim k^2$) (Figs.~\ref{fig:figure2}d, e, f, and \ref{fig:figure3}c)~\cite{Anand2025, Anand2026}.
Random-organizing systems become hyperuniform at $c^s = -1$, while systems with temporally correlated noise become hyperuniform at $c^t = -1/2$ (Figs.~\ref{fig:figure2}d, e, f, and \ref{fig:figure3}c)~\cite{Anand2025, Anand2026}.

We now investigate the connection between long-range structure and time irreversibility, quantified by the EPR.
For random-organizing systems, where the noise is pairwise, we track the dynamics of all overlapping particle pairs at a given time $t$.
We coarse-grain the dynamics into four states based on the relative direction of motion within each particle pair and compute the EPR of the resulting four-state Markov process (Fig.~\ref{fig:figure2}a, b, c, Methods).
For systems with temporally correlated noise, where the noise acts on individual particles, we instead track the dynamics of individual particles across two consecutive time steps, $t$ and $t+1$ (Fig.~\ref{fig:figure3}a).
We coarse-grain these trajectories into four states based on particle speed (fast or slow) and the angle between the velocities at $t$ and $t+1$ (persistent or reversing), and compute the corresponding EPR (Fig.~\ref{fig:figure3}b, Methods).

Note that we track the dynamics of overlapping particle pairs in random-organizing systems and of individual particles in systems with temporally correlated noise (Figs.~\ref{fig:figure2}b, and \ref{fig:figure3}b).
This distinction arises because, in random-organizing systems, the dynamics are inherently pairwise (non-overlapping particles do not evolve) (Eqs.~\ref{eq:discrete_ro}, \ref{eq:discrete_bro}, and \ref{eq:discrete_sgd}), whereas in temporal-noise systems, the dynamics are particle-wise, with the bath acting independently on each particle (Eq.~\ref{eq:discrete_pgd}).

Despite their distinct microscopic dynamics, the EPR increases monotonically across all systems as the noise becomes more anti-correlated, either in inter-particle or temporal correlations (Figs.~\ref{fig:figure2}g, h, i, and \ref{fig:figure3}d).
Consequently, the EPR also increases monotonically with the crossover length scale $l_c$, indicating that the systems become more irreversible as the length scale over which they self-organize grows (Figs.~\ref{fig:figure2}g, h, i, and \ref{fig:figure3}d insets). 
We thus uncover a universal behavior across all systems: the EPR attains its maximum precisely when the system becomes hyperuniform, corresponding to $c^s = -1$ for random-organizing systems and $c^t = -1/2$ for systems with temporally correlated noise.

\section{Theory}

Why does the system become most irreversible precisely at the onset of hyperuniformity?
To address this question, we now develop a theory for the EPR. 
Starting from the continuous-time approximation of the discrete-time dynamics for all systems, we construct the Martin--Siggia--Rose--Janssen--De Dominicis path-integral representation of the dynamics, and subsequently derive the corresponding Onsager--Machlup action $\mathcal{A}_{\text{OM}}$~\cite{Martin1973,Janssen1976,DeDominicis1976,deDominicis1978,Onsager1953}.
This action is a functional of the system’s trajectory and determines the probability of observing the system in state $\bm{X}_T$ at time $t = T$ given the initial condition $\bm{X}_0$ at time $t = 0$, via
\begin{align}
    \mathbb{P}[\bm{X}_T | \bm{X}_0] \propto e^{\mathcal{A}_{OM}[\bm{X}_T| \bm{X}_0]}.
\end{align}
The ratio of the probability of a trajectory and its time-reversed version is then given by,
\begin{align}
    \frac{\mathbb{P}[\bm{X}_0 | \bm{X}_T]}{\mathbb{P}[\bm{X}_T | \bm{X}_0]} = e^{ \mathcal{T}\mathcal{A}_{OM}[\bm{X}_T|\bm{X}_0] - \mathcal{A}_{OM}[\bm{X}_T|\bm{X}_0]} \equiv e^{-\Delta S[\bm{X}_T|\bm{X}_0]},
\end{align}
where $\mathcal{T}$ is the time-reversal operator and $\Delta S$ denotes the entropy production over a time interval $T$~\cite{Seifert2005}.
The EPR, $\sigma(t)$, is then defined as $\Delta S[\bm{X}_T|\bm{X}_0] = \int_0^T dt\, \sigma(t)$.

\begin{figure}
    \centering
    \includegraphics[width=\columnwidth]{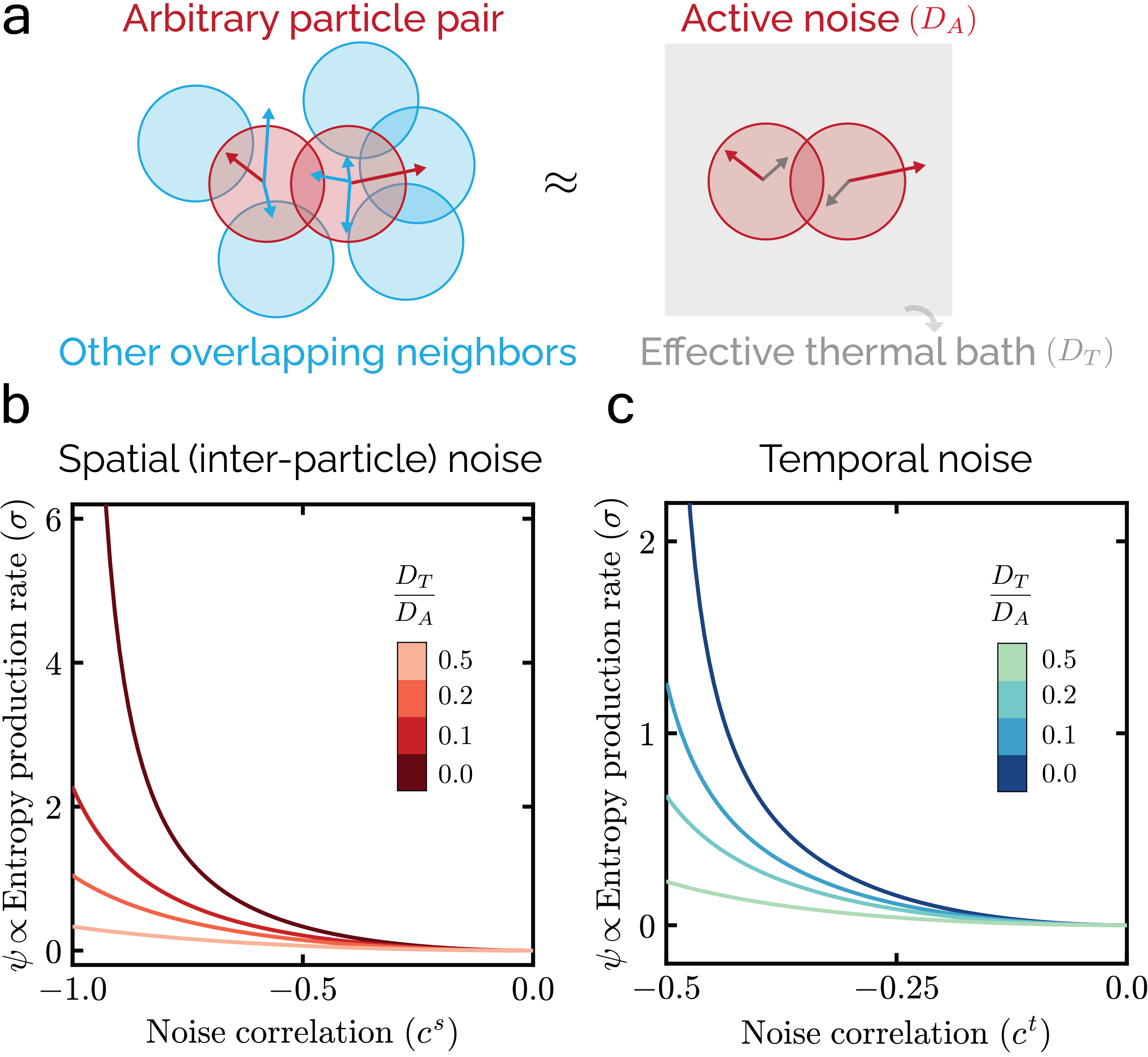}
    \caption{\textbf{Analytical entropy production rate.}
    $(a)$ Schematic depicting an arbitrarily chosen pair of particles (red) surrounded by several other overlapping particles (blue). The effect of the neighboring blue particles can be approximated by a thermal bath.
    $(b)$ The prefactor $\psi \sim D^{-1}_{ij}$ of the EPR plotted as a function of the inter-particle noise correlation coefficient $c^s$ (Eq.~\ref{eq:Doffdiagminus1}). 
    $(c)$ The prefactor $\psi = \sigma / \beta \dot{W}_t$ of the EPR plotted as a function of the temporal noise correlation coefficient $c^t$ (Eq.~\ref{eq:M=1_sametime_EPR}).
    }
    \label{fig:figure4}
\end{figure}

Before proceeding further, it is worth noting that the analytical EPR does not carry the same status as the numerical EPR measurements presented thus far.
The numerical estimates are \textit{local}, in the sense that they focus on coupled subsystems analyzed independently.
In contrast, the analytical EPR is a \textit{global} quantity, as it incorporates information from the entire system simultaneously. 
To better approximate the local nature of the numerical estimates, we follow an approach inspired by the Zwanzig--Mori formalism and model the effect of couplings to unobserved degrees of freedom by introducing an effective thermal bath, which enters dynamics through a standard thermal diffusion constant $D_T \propto k_BT$ (Fig.~\ref{fig:figure4}$a$) ~\cite{Zwanzig1961,Mori1965,Mori1965a}.

The derivations of the dynamical actions are cumbersome but follow standard strategies~\cite{ArnoulxdePirey2022} for multiplicative~\cite{GonzalezArenas2012} and colored~\cite{Hänggi1995,ArnoulxdePirey2022} noise terms (see SI Sec.I).
We find that all systems are truly irreversible as soon as the noise correlation is non-zero, and in particular that they don't admit a generalized time-reversal symmetry, which happens in some systems~\cite{Casiulis2026} (see SI Secs. I and VI).

In both classes of systems, the EPR assumes the generic form, 
\begin{align}
    \sigma(t) \sim \psi(c,D_T/D_A) \ \mathcal{P}_t[\bm{X}_t], \label{eq:EPR_Sketch}
\end{align}
where $\mathcal{P}_t$ is the instantaneous dissipated power that depends on the full state of the system, and $\psi$ is a prefactor that depends on two factors: the noise correlation coefficient $c$ of the relevant noise (either spatial or temporal), and a ratio between the thermal diffusion constant $D_T$ and the diffusion constant that encodes the amplitude of the spatial or temporal noise, $D_A$.
$\mathcal{P}_t$ is, in general, intractable in many-body dynamics, as it depends on the full state of the system.
We thus focus only on the prefactor $\psi$---this is also justified by the fact that $\psi$ is independent of the system state, and hence would survive any averaging over trajectories.

In the case of inter-particle (spatial) noise, the EPR is generically given by (see SI Sec.I.A.5),
\begin{align}
    \sigma(t) &= -\frac{c^s}{2}\,\,{}^t\dot{\bm{X}}_t\overline{\overline{D}}{}^{-1}(c^s,D_T,D_A,\bm{X}_t)  \bm{\Phi}(\bm{X}_t), \label{eq:EPR_Spatial}
\end{align}
where $\bm{\Phi}$ is an effective force generated by the stochastic part of the interaction, $^t$ indicates a transpose, and $\overline{\overline{D}}$ is a diffusion tensor with shape $dN\times dN$ that couples it to the velocity of the system. This result is generic across a large family of models that encompass RO, BRO, and SGD. 
Further, $\overline{\overline{D}}$ is a doubly stochastic tensor for $c^s = -1$ with each row and column summing to $D_T$, and as a result, is non-invertible for $D_T \to 0$ (SI Sec.I.A.5). 
In other words, without the presence of a thermal noise, the EPR diverges for $c^s \to -1$.
Furthermore, the EPR vanishes in the limit $c^s \to 0$.

We now illustrate this result with a minimal example in which the diffusion tensor is tractable: one-dimensional RO-like dynamics with $N$ particles and diverging radii (see SI Sec. II). 
In this simple limit, 
\begin{align}
 D^{-1}_{ij} 
 &= \frac{2}{2D_T + (N-1-c^s)D_A}\Bigg[\delta_{ij}  \nonumber \\  
 &\quad - \frac{c}{2D_T/D_A + (c^s+1)(N-1)}\Bigg].
   \label{eq:Doffdiagminus1}
\end{align}
In particular, treating the off-diagonal elements of $D^{-1}_{ij}$ like $\psi$ of Eq.~\eqref{eq:EPR_Sketch} (the diagonal part is essentially identical), we find that it is maximal for $c^s = -1$, and this maximum diverges as $D_T \to 0$ (Fig.~\ref{fig:figure4}$b$).

We now turn to systems with temporal noise.
In this setting, the entropy production decomposes into $\Delta S = \int dt \sigma(t) +\int dt dt' \varsigma(t,t')$ where $\sigma$ is an EPR and $\varsigma$ is a delayed contribution with no $t=t'$ contribution (SI Sec.I.B).
The EPR is then given as (SI Sec.I.B),
\begin{align}
    \sigma(t) &= \frac{1}{1+D_T/D_A} \left(\frac{1}{\sqrt{1 - 4 {c^t}^2/(1 + D_T/D_A)^2}} - 1 \right)\beta\dot{W}_t,
    \label{eq:M=1_sametime_EPR}
\end{align}
where $\beta \propto 1/(D_A+D_T)$ is the effective inverse temperature when $c^t = 0$, and the instantaneous power is simply given by the time derivative of the work of external conservative forces, $\dot{W}_t = -\dot{\bm{X}}_t \cdot \nabla V(\bm{X}_{t})$.
The EPR vanishes when $c^t = 0$ irrespective of $D_T/D_A$ as expected, since the system reduces to a system of interacting Brownian particles at equilibrium.
Furthermore, $\psi =\sigma/\beta\dot{W}_t$ is maximal for $c^t = - 1/2$, with a maximum that diverges as $D_T/D_A \to 0$ (Fig.~\ref{fig:figure4}$c$).
We further show that the EPR dependence on the noise correlation remains qualitatively unchanged for noise with arbitrarily long temporal correlations, rather than being restricted to lag-$1$ correlation (SI Sec.II.B). 

Thus, both classes of systems exhibit maximal EPR precisely when they become hyperuniform, in agreement with our numerical observations.

\section{Discussion}

Combining simulations and theory, we investigate the energetics of hyperuniform self-assembly across several non-equilibrium systems.
We uncover a universal phenomenon: the onset of hyperuniformity coincides with maximal irreversibility.
This suggests that hyperuniformity is associated with a high thermodynamic cost, indicating that energy must be continuously supplied to assemble and maintain such structures.

Where does the energy required to maintain a hyperuniform state come from in the noisy systems considered here?
In systems with inter-particle noise, hyperuniformity emerges precisely when the pairwise noise conserves the local center of mass~\cite{Anand2025}.
This result on stochastic pair interactions goes against popular wisdom, as deterministic non-reciprocal interactions are usually more irreversible~\cite{Loos2020}.
In systems with temporal noise, hyperuniformity emerges when the mean-square displacement of the whole system saturates at long times, i.e., when diffusion becomes effectively caged (SI Sec.III) ~\cite{Anand2026}.
This link between hyperuniformity and conserved quantities ~\cite{Hexner2017a,Maire2025b}, implies that the noise that generates hyperuniformity is not fully \textit{random}, but instead has a constrained structure.
Producing and sustaining such structured noise requires energy, which in our spatial-noise models may be supplied by external shearing of colloids, as in RO~\cite{ Corte2008,Hexner2015}, and in temporal-noise models by the energetic cost of maintaining an active bath~\cite{Anand2026}.

Beyond the two classes of noisy particle systems studied here, hyperuniformity also emerges in a variety of other systems at NESS, including chiral active matter~\cite{Lei2019},hyperuniform fluids~\cite{Lei2019a}, pulsating cell tissues~\cite{Li2025a}, harmonic chains~\cite{Ikeda2024}.
Investigating the energetics of these systems will shed further light on the generality of the correlation between self-organized hyperuniformity and time irreversibility.
This is crucial to design self-assembly strategies for hyperuniform structures, a necessary step to bring the exciting (\textit{e.g.} optical~\cite{Leseur2016}) properties of hyperuniform structures to large-scale applications.

\bibliography{ref.bib,PostDoc-StefanoMartiniani}

\section{Acknowledgements}
We thank Guanming Zhang for numerous enlightening discussions on these models, and we acknowledge insightful comments from Julien Tailleur.
M.C. and S.M. acknowledge the Simons Center for Computational Physical Chemistry for financial support.
This work was supported in part through the NYU IT High Performance Computing resources, services, and staff expertise.

\section{Author contributions}

M.C and S.A. conceptualized the project. M.C. developed the theory and S.A. performed numerical simulations. M.C. and S.A. analyzed data. M.C., S.A. and S.M. wrote the manuscript. S.M. acquired funding and supervised the research. M.C. and S.A. contributed equally.

\section{Methods}

All systems consist of $N$ hyperspherical particles of radius $R$ in a $d$-dimensional hypercubic domain of side length $L$, with periodic boundary conditions imposed along all directions. The fundamental units of length, time, and energy were chosen to be $2R$, $\tau$, and $\mathcal{E}$, respectively.
Here, $\mathcal{E}=1$ defines the characteristic interaction energy scale associated with the pair potential in Eq.~\ref{eq:v_ji_family}, and $\tau=1$ corresponds to the time scale corresponding to one discrete time-step.
Throughout, the particle number and radius are fixed to $N = 10^5$ and $R=1$, respectively.
The packing fraction, defined as $\phi = N V_s / V_c$, is controlled by adjusting the box size $L$, where $V_s$ and $V_c$ denote the volumes of a $d$-dimensional sphere of radius $R$ and a hypercube of side length $L$, respectively.
Initial particle positions at $t=0$ are drawn from a uniform distribution within the simulation box.
All systems are evolved until a stationary state was reached, identified by the total energy $E = \sum_i \sum_{j \geq i} V_{ij}$ fluctuating around a steady mean value, and a steady static structure factor $S(k)$.

For SGD and systems with temporal noise, the pairwise interaction potential $V_{ij}$ is described by a class of finite-range, purely repulsive potentials of the form
\begin{equation}
    V_{ij}(r) =     
    \begin{cases}
      \frac{\mathcal{E}}{p}\left(1 - \frac{r_{ij}}{2R}\right)^p, & \text{if } 0 < r_{ij} < 2R, \\
      0, & \text{otherwise},
    \end{cases}
    \label{eq:v_ji_family}
\end{equation}
where $r_{ij} = |\mathbf{x}_j - \mathbf{x}_i|$ is the distance between particles $i$ and $j$, $\mathcal{E}$ sets the energy scale, and the exponent $p$ determines the stiffness of the interaction. 

We now outline the numerical methods and parameter choices for each system individually.

\textit{\textbf{Spatial (inter-particle) noise.}} Discrete-time simulations were performed for the RO, BRO, and SGD by updating particle positions according to Eqs.~\ref{eq:discrete_ro}, \ref{eq:discrete_bro}, and \ref{eq:discrete_sgd}, respectively. 

\textit{Random Organization.} For RO, the position of particle $i$ at time-step $m+1$ evolves according to \cite{Anand2025},
\begin{equation}
    \mathbf{x}_{i}^{m+1} = \mathbf{x}_{i}^{m} + \epsilon \sum_{j \in \Gamma_i^m} u_{ji}^m \boldsymbol{\zeta}^m_{ji}, 
    \label{eq:discrete_ro}
\end{equation}
where $\epsilon$ sets the magnitude of the pairwise kick between particles.
$u_{ji}^m$ is a random number that introduces noise in the kick magnitude at time-step $m$, here drawn from a uniform distribution ($U[0,1]$).
$\boldsymbol{\zeta}^m_{ji}$ is a random unit vector which introduces noise in the kick direction at time-step $m$, and is drawn uniformly from the surface of a unit $d$-dimensional hypersphere.
The interaction set $\Gamma_i^m = \{j \mid |\mathbf{x}_j^m - \mathbf{x}_i^m| < 2R,\ j \neq i \}$ contains all particles that overlap with particle particle $i$ at time-step $m$.
To ensure that overlapping particles separate from each other, we set $\boldsymbol{\zeta}^m_{ji} = -\boldsymbol{\zeta}^m_{ij}$.

The total noisy kick exerted by particle $i$ on particle $j$ can be written component-wise as $\omega_{ij,\alpha}^m = \epsilon \, u_{ij}^m \, \zeta_{ij,\alpha}^m$. 
We introduce a Pearson correlation coefficient $c^s \in [-1,0]$ between corresponding components of the pairwise noise vectors $\omega_{ij,\alpha}^m$ and $\omega_{ji,\alpha}^m$. The limit $c = 0$ corresponds to uncorrelated pairwise kicks, whereas $c = -1$ enforces anti-correlated kicks, implying equal magnitudes ($u_{ij}^m = u_{ji}^m$) and opposite directions.

For the RO dynamics, the behavior is governed by four parameters: the kick amplitude $\epsilon$, the packing fraction $\phi$, the spatial dimensionality $d$, and the pairwise noise correlation coefficient $c^s$.
In the results presented in the main text, we fixed $\epsilon = 0.1$, $\phi = 8.0$, and $d = 2$, while varying $c^s$ within the range $[-1, 0]$. 

\textit{Biased Random Organization.} The position of particle $i$ at time-step $m+1$ evolves according to \cite{Anand2025},
\begin{equation}
    \mathbf{x}_{i}^{m+1} = \mathbf{x}_{i}^{m} + \epsilon \sum_{j \in \Gamma_i^m} u_{ji}^m \mathbf{\hat{x}}^m_{ji}, 
    \label{eq:discrete_bro}
\end{equation}
where $\epsilon$ and $u_{ji}^m$ are defined as in RO.
The unit vector $\mathbf{\hat{x}}^m_{ji} = -(\mathbf{x}^m_{j} - \mathbf{x}^m_{i})/|\mathbf{x}^m_{j} - \mathbf{x}^m_{i}|$ points from particle $j$ to particle $i$ at time step $m$. 

The total noisy kick exerted by particle $i$ on particle $j$ can be written component-wise as $\omega_{ij,\alpha}^m = \epsilon \, u_{ij}^m \, \hat{x}_{ij,\alpha}^m$.
As in RO, we introduce a Pearson correlation coefficient $c^s \in [-1,0]$ between corresponding components of the pairwise noise vectors $\omega_{ij,\alpha}^m$ and $\omega_{ji,\alpha}^m$. The limit $c = 0$ corresponds to uncorrelated magnitudes $u_{ij}^m$ and $u_{ji}^m$, whereas $c = -1$ implies $u_{ij}^m = u_{ji}^m$ (since $\hat{\mathbf{x}}_{ij}^m = - \hat{\mathbf{x}}_{ji}^m$). 

The BRO dynamics depend on the same set of control parameters as RO, namely $\epsilon$, $\phi$, $d$, and $c$.
In the results presented in the main text, we set $\epsilon = 0.1$, $\phi = 8.0$, and $d = 2$, with $c^s$ spanning $[-1, 0]$.

\textit{Stochastic Gradient Descent.} For SGD, the position of particle $i$ at time-step $m+1$ evolves according to \cite{Anand2025},
\begin{equation}
    \begin{aligned}
        \mathbf{x}_{i}^{m+1} = \mathbf{x}_{i}^{m} - \alpha \sum_{j \in \Gamma_i^m} \theta_{ji}^m \nabla_{\mathbf{x}_{i}} V_{ji}^{m},
    \end{aligned}
\label{eq:discrete_sgd}
\end{equation}
where $V_{ji}^m = V\bigl(| \mathbf{x}_j^m - \mathbf{x}_i^m | \bigr)$ is the pairwise interaction potential, and $\alpha$ is the learning rate, which sets the displacement scale (with units of $\text{length}/\text{force}$). 
The variable $\theta_{ji}^m$ is a Bernoulli random variable with parameter $b_f$ at time step $m$.
The batch fraction $b_f$ controls the average fraction of overlapping particle pairs $(i,j)$ that are updated at a given time.
Consequently, for any overlapping pair $(i,j)$ at a given time step, there are four possible update outcomes: (i) only particle $i$ moves ($\theta_{ji}^m = 1,\, \theta_{ij}^m = 0$), (ii) only particle $j$ moves ($\theta_{ji}^m = 0,\, \theta_{ij}^m = 1$), (iii) both particles move ($\theta_{ji}^m = 1,\, \theta_{ij}^m = 1$), and (iv) neither particle moves ($\theta_{ji}^m = 0,\, \theta_{ij}^m = 0$). 

The total noisy kick exerted by particle $i$ on particle $j$ can be written component-wise as $\omega_{ij,\alpha}^m = - \alpha \, \theta_{ij}^m \, \partial_{i,\alpha} V_{ij}^m$.
Analogous to RO and BRO, we introduce a Pearson correlation coefficient $c \in [-1,0]$ between corresponding components of the pairwise noise vectors $\omega_{ij,\alpha}^m$ and $\omega_{ji,\alpha}^m$. The limit $c = 0$ corresponds to uncorrelated $\theta_{ij}^m$ and $\theta_{ji}^m$, whereas $c = -1$ implies $\theta_{ij}^m = \theta_{ji}^m$ (since $\nabla_{\mathbf{x}_{i}} V_{ji}^m = - \nabla_{\mathbf{x}_{j}} V_{ji}^m$).
While $V_{ij}$ can, in principle, be any pairwise potential, we focus here on a class of purely repulsive, short-range potentials with a finite cutoff (Eq.~\ref{eq:v_ji_family}).

For SGD dynamics, six parameters determine the evolution: the learning rate $\alpha$, the packing fraction $\phi$, the batch fraction $b_f$, the exponent $p$ characterizing the interaction potential, the spatial dimension $d$, and the noise correlation coefficient $c^s$.  In the results presented in the main text, we set $\alpha = 0.1$, $\phi = 8.0$, $b_f = 0.5$, $p = 1$, and $d = 2$, while varying $c^s$ over the interval $[-1, 0]$.

\textit{\textbf{Temporal noise.}} The position of particle $i$ at time-step $m+1$ evolves according to \cite{Anand2026},
\begin{equation}
  \mathbf{x}_{i}^{m+1} = \mathbf{x}_{i}^{m} - \alpha \sum_{j \in \Gamma_i^{m}} \nabla_{\mathbf{x}_{i}} V_{ji}^{m} + \boldsymbol{\zeta}_i^{m+1},
\label{eq:discrete_pgd}
\end{equation}
where $\alpha$ and $V_{ji}^m$ are defined as in SGD.
The term $\boldsymbol{\zeta}_i^{m+1}$ represents the Gaussian noise exerted by the active bath on particle $i$ at time step $m+1$.
If $\boldsymbol{\zeta}_i^{m+1}$ exhibits temporal correlations, the system is driven out of equilibrium.

We consider the most minimal non-equilibrium setting in which $\boldsymbol{\zeta}_i^{m+1}$ exhibits only lag-$1$ temporal correlations~\cite{Anand2026}.
We introduce a Pearson correlation coefficient $c^t \in [-1/2, 0]$ between corresponding components of the noise vectors $\zeta_{i,\alpha}^m$ and $\zeta_{j,\alpha}^{m+1}$.
The limit $c^t = 0$ corresponds to temporally uncorrelated noise (i.e., equilibrium dynamics), whereas $c^t = -1/2$ corresponds to maximally anti-correlated kicks across consecutive time steps. 

For systems with temporal noise having lag-$1$ correlations, the evolution is characterized by six parameters: the learning rate $\alpha$, the interaction exponent $p$ controlling potential stiffness, the noise amplitude $\sigma$, the packing fraction $\phi$, the spatial dimensionality $d$, and the lag-$1$ noise correlation coefficient $c^t$. 
In the results presented in the main text, we used $\alpha = 0.1$, $p = 1.5$, $\sigma = 0.2$, $\phi = 8.0$, and $d = 2$, while varying $c^t$ over the interval $[-0.5, 0]$.

\subsection{Structure factor}

The structure factor was evaluated as $S(\mathbf{k}) = |\hat{\rho}(\mathbf{k})|^2 / N$, where the microscopic density field is defined as $\rho(\mathbf{x}) = \sum_{i=1}^{N} \delta(\mathbf{x} - \mathbf{x}_i)$. For an arbitrary function $f(\mathbf{x})$, its spatial Fourier transform is given by $\hat{f}(\mathbf{k}) = \int d\mathbf{x}\, f(\mathbf{x}) e^{-i \mathbf{k} \cdot \mathbf{x}}$. $S(\mathbf{k})$ was obtained using a nonuniform fast Fourier transform algorithm, and radially averaged to yield the radial structure factor $S(k)$~\cite{Barnett2019, Barnett2020fixed}. Results in the main text were obtained by averaging over $100$ configurations sampled in steady state.

\subsection{Entropy production rate}

For all systems, to quantify the irreversibility, we coarse-grain the dynamics into four states $i \in \{1,2,3,4\}$ and compute the EPR of the associated discrete-time Markov process. The system evolves according to a transition matrix $P_{ij}$, where $P_{ij}$ denotes the probability of transitioning from state $i$ to state $j$ in one time step.
The transition probabilities satisfy $\sum_{j=1}^4 P_{ij} = 1$ for all $i$.
Furthermore, $\pi_i$ denotes the stationary probability of state $i$, defined as the normalized solution of
\begin{equation}
\pi_j = \sum_{i=1}^4 \pi_i P_{ij}, \qquad \sum_{i=1}^4 \pi_i = 1.
\end{equation}

The total EPR was computed using the standard expression
\begin{equation}
\sigma =
\frac{k_B}{\tau} \sum_{i<j}
\left(\pi_i P_{ij} - \pi_j P_{ji}\right)
\ln\!\left(\frac{\pi_i P_{ij}}{\pi_j P_{ji}}\right),
\end{equation}
where $k_B$ is Boltzmann constant, and $\tau$ is the time scale corresponding to a discrete time-step.
Results in the main text were obtained by averaging over $1000$ configurations sampled in steady state.

\textit{\textbf{Spatial (inter-particle) noise.}} For RO, BRO, and SGD, where the noise is pairwise, the dynamics of all overlapping particle pairs at a given time $t$ was tracked.
The four coarse-grained states were then constructed based on the relative direction of motion within each particle pair.

\textit{\textbf{Temporal noise.}} For systems with temporally correlated noise, where the noise acts on individual particles, the dynamics of individual particles across two consecutive time steps, $t$ and $t+1$, was tracked. The four coarse-grained states were then constructed based on based on the particle speed (fast or slow) and the angle between the velocities at $t$ and $t+1$ (persistent or reversing).

\end{document}


\preprint{APS/123-QED}

\author{Mathias Casiulis}
\thanks{Equal contribution}
\email{mc9287@nyu.edu}
\affiliation{\nyuphysics}
\affiliation{\nyusimons}
\author{Satyam Anand}
\thanks{Equal contribution}
\email{sa7483@nyu.edu}
\affiliation{\nyucourant}
\affiliation{\nyuphysics}
\author{Stefano Martiniani}
\email{sm7683@nyu.edu}
\affiliation{\nyuphysics}
\affiliation{\nyusimons}
\affiliation{\nyucourant}
\affiliation{\nyucns}

\title{Supplementary Material for ``Hyperuniform systems are maximally irreversible''}

\renewcommand{\figurename}{FIG.}
\renewcommand{\thefigure}{S\arabic{figure}}
\renewcommand{\thetable}{S\arabic{figure}}
\newtagform{S}{(S})
\usetagform{S}

\date{\today}

\maketitle

\begin{spacing}{0.95}
\tableofcontents
\end{spacing}

\section{Path-Integral derivation of entropy production}

In this section, we reproduce the complete derivation of path-integral actions and entropy productions for the models studied in this paper.
This section explicitly splits the derivation for the case of spatial noise, Sec.~\ref{sec:Spatial_EP_Analytical},
and that of temporal noise, Sec.~\ref{sec:Temporal_EP_Analytical}.

\subsection{Spatial noise \label{sec:Spatial_EP_Analytical}}

In this subsection, we study the dynamics with spatial noise, as introduced in Refs~\cite{Zhang2024a,Anand2025}.
First, we recall the stochastic differential equations (SDEs) and derive the relevant Fokker-Planck equation in Sec.~\ref{sec:Spatial_Dynamics_FPE}.
Then, we construct the Martin-Siggia-Rose~\cite{Martin1973}-Janssen~\cite{Janssen1976}-de Dominicis~\cite{DeDominicis1976,DeDominicis1978} (MSRJD) path integral and review the effect of correlations on time-reversal symmetry in Sec.~\ref{sec:Spatial_MSRJD}.
Finally, we establish the Onsager-Machlup~\cite{Onsager1953} (OM) action and the corresponding Entropy Production (EP) and Entropy Production Rate (EPR) in Sec.~\ref{sec:Spatial_OM_EPR}.

\subsubsection{Dynamics and Fokker-Planck Equation \label{sec:Spatial_Dynamics_FPE}}

Random-organizing systems (RO, BRO, and SGD) are formulated as discrete-time dynamics (see Methods section in the main text). It was recently shown that their discrete-time dynamics can be approximated by a continuous-time stochastic differential equation (SDE) \cite{Zhang2024a,Anand2025}. Our starting point is the continuous-time SDE for random-organizing systems along with an additional thermal bath.

We consider dynamics that model a system of $N$ particles in a $d$-dimensional space, in which they interact through a central pair potential $V(r_{ij})$,
\begin{align}
    \dot{{r}}_{i,a}(t) &= -\mu_i \sum\limits_{j \neq i} \partial_{i,a} V(\bm{r}_{ij}) + \sum\limits_{j \neq i} \sqrt{\Lambda_{ij,ab}(\bm{r}_{ij})} \xi_{ij}^b(t) + \sqrt{2 D_0} \eta_{i}^a(t) \label{eq:BROSGD_dynamics_proj}
\end{align}
where $\bm{r}_i$ is the position of particle $i$ ($1 \leq i \leq N$), ${r}_{i,a}$ is its Cartesian projection along direction $a$,$\mu_i$ is its mobility, $\Lambda_{ijab}$ is a function of positions in each pair that stems from stochasticity in the instantaneous interactions, $D_0$ is a diffusion constant associated to a thermal bath, and the two (mutually independent) sources of noise verify
\begin{align}
    \left\langle \xi_{ij}^a (t)  \right\rangle_\xi &= 0  \\
    \left\langle \xi_{ij}^a (t) \xi_{kl}^b (t') \right\rangle_\xi &= \delta_{ab} \delta(t - t') \left( \delta_{ik} \delta_{jl} + c \delta_{il}\delta_{jk}\right) \\
    \left\langle \eta_{i}^a (t)  \right\rangle_\eta &= 0  \\
    \left\langle \eta_{i}^a (t) \eta_{j}^b (t') \right\rangle_\eta &= \delta_{ab} \delta_{ij} \delta(t - t'),
\end{align}
with $c \in [-1;1]$ encoding the degree of correlations between the pairwise-random part of the updates -- $c = 0$ encodes uncorrelated updates between particles in the pair, $c = -1$ center-of-mass conserving updates, and $c = 1$ identical updates.

In practice, following Refs.~\cite{Zhang2024a,Anand2025}, we assume that $\Lambda_{ijab}$ acts on Cartesian components of the noise as a projector, so that it does not affect the total noise variance, so that
\begin{align}
    \sqrt{\Lambda_{ij,ab}} = P_{ab} \sqrt{\Lambda_{ij}}
\end{align}
with a projection matrix $\overline{\overline{P}}$, that verifies
\begin{align}
    \overline{\overline{P}}{}^2 = \overline{\overline{P}}.
\end{align}
Two particular examples are the identity tensor $\delta_{ab}$, which was proposed~\cite{Zhang2024a,Anand2025} as a model of Random Organization~\cite{Corte2008,Wilken2020} and yields the simpler dynamics
\begin{align}
    \dot{\bm{r}}_i(t) &= -\mu_i \sum\limits_{j \neq i} \bm{\nabla}_i V(\bm{r}_{ij}) + \sum\limits_{j \neq i} \sqrt{\Lambda_{ij}(\bm{r}_{ij})} \bm{\xi}_{ij}(t) + \sqrt{2 D_0} \bm{\eta}_i(t), \label{eq:BROSGD_dynamics}
\end{align}
and a projection onto the axis that links particles $i$ and $j$ as in Biased Random Organization~\cite{Wilken2020,Wilken2021},
\begin{align}
    \overline{\overline{P}}(\bm{r}_{i}, \bm{r}_j) = \hat{\bm{r}}_{ij} \otimes \hat{\bm{r}}_{ij}.
\end{align}

Since the dynamics at hand contain multiplicative noise, heed must be taken not just of the discretization convention for the SDE, but also of correcting possible spurious terms in the dynamics.
These difficulties are illustrated in a simpler $1d$ case in Sec.~\ref{sec:Multiplicative_Aside} for readers who may not be familiar with these issues.
The first step is establishing the Fokker-Planck equation for the (here, Markovian) dynamics.
The Fokker-Planck equation is traditionally obtained~\cite{Lau2007,ArnoulxdePirey2022} by writing that for a Markov process, the probability density function for the set of $N$ positions at time $t + dt$ can be written as a master equation on time $t$,
\begin{align}
    P(\bm{Y},t+dt) = \int d\bm{X}\, P(\bm{Y}, t+dt | \bm{X}, t) P(\bm{X}, t) \label{eq:Markov}
\end{align}
where the joint probability density function at time $t_2$ conditional on the state at time $t_1$, $P(\bm{Y},t_2 | \bm{X}, t_1)$, verifies, for an infinitesimal time difference,
\begin{align}
    P(\bm{Y}, t+dt | \bm{X}, t) = \left\langle \delta\left(\bm{Y} - \bm{X}(t + dt)\vphantom{\sum}\right)\right\rangle
\end{align}
where the short-hand notation indicates that $\bm{Y}$ has to follow from the dynamics applied to $\bm{X}$ over a time $dt$.
Writing a Taylor expansion of the dynamical term around $t$ then yields
\begin{align}
     P(\bm{Y}, t+dt | \bm{X}, t) =
     &\delta\left(\bm{Y} - \bm{X}(t )\vphantom{\sum}\right) + \sum\limits_{i=1}^N \sum\limits_{a=1}^d \partial_{x_{i,a}}\left[\delta\left(\bm{Y} - \bm{X}(t)\right)\vphantom{\sum}\right] \left\langle x_{i,a}(t+dt) - x_{i,a} (t)\vphantom{\sum}\right\rangle \nonumber \\
     &+ \frac{1}{2} \sum\limits_{i,j = 1}^N\sum\limits_{a,b = 1}^d \partial_{x_{i,a}}\partial_{x_{j,b}}\left[\delta\left(\bm{Y} - \bm{X}(t)\right)\vphantom{\sum}\right] \left\langle \left( x_{i,a}(t+dt) - x_{i,a} (t)\vphantom{\sum}\right)\left( x_{j,b}(t+dt) - x_{j,b} (t)\vphantom{\sum}\right)\right\rangle + \ldots \label{eq:Taylor_expanded_conditional}
\end{align}
Then, injecting Eq.~\ref{eq:Taylor_expanded_conditional} into Eq.~\ref{eq:Markov} yields the Fokker-Planck equation in its general form,
\begin{align}
    P(\bm{Y},t+dt) 
    &= P(\bm{Y},t) \nonumber \\
    &+  \int d\bm{X} \sum\limits_{i=1}^N \sum\limits_{a=1}^d\partial_{x_{i,a}}\left[\delta\left(y_{i,a} - x_{i,a}(t)\right)\vphantom{\sum}\right]\cdot \left\langle \delta x_{i,a}\vphantom{\sum}\right\rangle P(\bm{X}, t) \nonumber \\
    &+\frac{1}{2}\sum\limits_{i,j = 1}^N \sum\limits_{a,b = 1}^d\int d\bm{X} \partial_{x_i,a}\partial_{x_j,b}\left[\delta\left(\bm{Y} - \bm{X}(t)\right)\vphantom{\sum}\right]\cdot \left\langle \delta x_{i,a} \delta x_{j,b}\right\rangle P(\bm{X}, t), \\
    &= P(\bm{Y},t) - \sum\limits_{i = 1}^N \sum\limits_{a = 1}^d \partial_{x_{i,a}} \left[ \left\langle \delta y_{i,a}\vphantom{\sum}\right\rangle P(\bm{Y}, t) \right] + \frac{1}{2}\sum\limits_{i,j = 1}^N \sum\limits_{a,b = 1}^d \partial_{x_i,a}\partial_{x_j,b} \left[ \left\langle \delta y_{i,a} \delta y_{j,b}\right\rangle  P(\bm{Y}, t) \right]. \label{eq:Proto_FPE}
\end{align}

One then needs to evaluate the averages in this Taylor expansion for the process of interest.
To do so, due to the presence of multiplicative noise, one has to heed the discretization convention.
Using short-hand notations for the dynamics, one may write
\begin{align}
    \bm{\delta X} = \bm{X}(t+dt) - \bm{X}(t) 
    &= \left( \bm{F}(\bm{X} + \alpha \bm{\delta X}) + \overline{\overline{\Pi}}(\bm{X}(t) + \alpha \bm{\delta X})\bm{\xi}(t + \alpha dt) + \sqrt{2 D_0}\, \bm{\eta}(t + \alpha dt)\right) dt, \label{eq:ShortHand_Update}
\end{align}
with $\bm{F}$ the conservative force part stemming from $V$, $\overline{\overline{\Pi}}$ a tensor that encodes the effect of $\Lambda_{ijab}$, and $\bm{\xi}$, $\bm{\eta}$ vectors containing all elements of the corresponding noise sources (all pairs and all particles, respectively). 
Focusing on a single component of this vector and Taylor-expanding the functions of $\bm{X} + \alpha \bm{\delta X}$ then yields
\begin{align}
    \delta x_{i,a} &=  \left( F_{i,a}(\bm{X}) + \sum\limits_{j \neq i} \sum\limits_{b=1}^d\sqrt{\Lambda_{ijab} (\bm{X})}\xi^b_{ij}(t + \alpha dt) + \sqrt{2 D_0}\, \eta_i^a(t + \alpha dt)\right) dt  \nonumber \\
    &\hphantom{aaa}+ \alpha \sum\limits_{m=1}^N \sum\limits_{p = 1}^d \partial_{x_{m,p}}\left[ F_{i,a}(\bm{X}) + \sum\limits_{j \neq i} \sum\limits_{b=1}^d\sqrt{\Lambda_{ijab} (\bm{X})}\xi^b_{ij}(t + \alpha dt) \right] \delta x_{m,p} dt \label{eq:deltax_FPE_proj}
\end{align}
This implicit equation over the components of $\bm{\delta X}$ can be iterated once to yield all terms of order up to $dt$, since the covariance of noise sources yields factors of order $1/dt$,
\begin{align}
\delta x_{i,a} &=  \left( F_{i,a}(\bm{X}) + \sum\limits_{j \neq i}\sum\limits_{b=1}^d \sqrt{\Lambda_{ijab} (\bm{X})}\xi^b_{ij}(t + \alpha dt) + \sqrt{2 D_0}\, \eta_i^a(t + \alpha dt)\right) dt  \nonumber \\
    &\hphantom{aaa}+ \alpha \sum\limits_{m=1}^N \sum\limits_{p = 1}^d \partial_{x_{m,p}}\left[ F_{i,a}(\bm{X}) + \sum\limits_{j \neq i}\sum\limits_{b=1}^d \sqrt{\Lambda_{ijab} (\bm{X})}\xi^b_{ij}(t + \alpha dt) \right] \nonumber \\
    &\hphantom{aaaaaaaaaaaaaaaa}\times\left( F_{m,p}(\bm{X}) + \sum\limits_{n \neq m} \sum\limits_{q=1}^d \sqrt{\Lambda_{mnpq} (\bm{X})}\xi^q_{mn}(t + \alpha dt) + \sqrt{2 D_0}\, \eta_m^p(t + \alpha dt)\right) dt^2 + o(dt) \\
    &= \left( F_{i,a}(\bm{X}) + \sum\limits_{j \neq i}\sum\limits_{b=1}^d \sqrt{\Lambda_{ijab} (\bm{X})}\xi^b_{ij}(t + \alpha dt) + \sqrt{2 D_0}\, \eta_i^a(t + \alpha dt)\right) dt  \nonumber \\
    &\hphantom{aaa}+ \alpha \sum\limits_{m=1}^N \sum\limits_{p = 1}^d \partial_{x_{m,p}}\left[ \sum\limits_{j \neq i}\sum\limits_{b=1}^d \sqrt{\Lambda_{ijab} (\bm{X})}\xi^b_{ij}(t + \alpha dt) \right]\times\left( \sum\limits_{n \neq m}\sum\limits_{q=1}^d \sqrt{\Lambda_{mnpq} (\bm{X})}\xi^q_{mn}(t + \alpha dt) \right) dt^2 + o(dt)\label{eq:deltaX_multnoise_expansion_proj}
\end{align}

From this last expression, one may evaluate the moments of the infinitesimal displacement due to dynamics.
The mean infinitesimal displacement reads, for each component
\begin{align}
    \left\langle \delta x_{i,a} \right\rangle &= F_{i,a}(\bm{X}) dt + \alpha \left\langle \sum\limits_{m=1}^N \sum\limits_{p = 1}^d \partial_{x_{m,p}}\left[ \sum\limits_{j \neq i}\sum\limits_{b=1}^d \sqrt{\Lambda_{ijab} (\bm{X})}\xi^b_{ij}(t + \alpha dt) \right]\times\left( \sum\limits_{n \neq m}\sum\limits_{q=1}^d \sqrt{\Lambda_{mnpq} (\bm{X})}\xi^q_{mn}(t + \alpha dt) \right)  \right\rangle dt^2 \\
    &= F_{i,a}(\bm{X}) dt + \sum\limits_{m=1}^N \sum\limits_{b=1}^d \sum\limits_{p = 1}^d \sum\limits_{q=1}^d\sum\limits_{n \neq m}\sum\limits_{j \neq i} \alpha  \left[ \partial_{x_{m,p}} \sqrt{\Lambda_{ijab} (\bm{X})} \right] \sqrt{\Lambda_{mnpq} (\bm{X})} \left\langle  \xi^b_{ij}(t + \alpha dt)\xi^q_{mn}(t + \alpha dt)  \right\rangle dt^2\\
    &= F_{i,a}(\bm{X}) dt + \sum\limits_{m=1}^N \sum\limits_{b=1}^d \sum\limits_{p = 1}^d \sum\limits_{q=1}^d \sum\limits_{n \neq m}\sum\limits_{j \neq i} \alpha  \left[ \partial_{x_{m,p}} \sqrt{\Lambda_{ijab} (\bm{X})} \right] \sqrt{\Lambda_{mnpq} (\bm{X})} \delta_{bq} (\delta_{im}\delta_{jn} + c \delta_{in}\delta_{jm}) dt \\
    &= F_{i,a}(\bm{X}) dt + \alpha \sum\limits_{j \neq i}\sum\limits_{b=1}^d \sum\limits_{p=1}^d \left(   \left[ \partial_{x_{i,p}} \sqrt{\Lambda_{ijab} (\bm{X})} \right] \sqrt{\Lambda_{ijpb} (\bm{X})} + c   \left[ \partial_{x_{j,p}} \sqrt{\Lambda_{ijab} (\bm{X})} \right] \sqrt{\Lambda_{jipb} (\bm{X})}\right) dt, \label{eq:avg_deltaX_proj}
\end{align}
while the mean quadratic infinitesimal displacements read, up to order $dt$
\begin{align}
    \left\langle \left( \vphantom{\sum}x_{i,a}(t+dt) - x_{i,a}(t)\right) \left( \vphantom{\sum}x_{j,b}(t+dt) - x_{j,b}(t)\right)\right\rangle &= 2 D_0\left\langle  \eta_{i,a}(t +\alpha dt) \eta_{j,b}(t +\alpha dt)\vphantom{\sum}\right\rangle dt^2 \nonumber \\
    &+ \left\langle \left( \sum\limits_{k (\neq i) = 1}^N\sum\limits_{l (\neq j) = 1}^N \sum\limits_{p=1}^d \sum\limits_{q = 1}^d  \sqrt{\Lambda_{ikap} \Lambda_{jlbq}} \xi_{ik,p}(t + \alpha dt)\xi_{jl,q}(t + \alpha dt)\right)\right\rangle dt^2  \\
    &= 2D_0 \delta_{ij}\delta_{ab}dt + \left(\sum\limits_{k \neq i}\sum\limits_{p=1}^d\sqrt{\Lambda_{ikap} \Lambda_{ikbp}}\delta_{ij} +c\sum\limits_{p=1}^d\sqrt{\Lambda_{ijap}\Lambda_{jibp}}(1-\delta_{ij})\right) dt. \label{eq:avg_deltaX_squared_proj}
\end{align}

Injecting Eqs.~\ref{eq:avg_deltaX_proj} and~\ref{eq:avg_deltaX_squared_proj} into Eq.~\ref{eq:Proto_FPE} yields the Fokker-Planck equation
\begin{align}
    \partial_t P(\bm{X},t) &= - \sum\limits_{i=1}^N \sum\limits_{a = 1}^d \partial_{x_{i,a}}\left[\left( F_{i,a} + \alpha \sum\limits_{j \neq i}\sum\limits_{b,p=1}^d \left(   \left[ \partial_{x_{i,p}} \sqrt{\Lambda_{ijab}} \right] \sqrt{\Lambda_{ijpb}} + c   \left[ \partial_{x_{j,p}} \sqrt{\Lambda_{ijab} } \right] \sqrt{\Lambda_{jipb}}\right)  \right) P(\bm{X},t)\right] \nonumber \\
    &\hphantom{aaaa}+ \frac{1}{2} \sum\limits_{i,j=1}^N \sum\limits_{a,b=1}^d \partial_{x_{i,a}}\partial_{x_{j,b}}\left[ \left(  \left[ 2D_0 \delta_{ab} + \sum\limits_{k\neq i} \sum\limits_{p=1}^d \sqrt{\Lambda_{ikap}\Lambda_{ikbp}}\right] \delta_{ij}+c\sum\limits_{p=1}^d\sqrt{\Lambda_{ijap}\Lambda_{jibp}}(1-\delta_{ij})\right) P(\bm{X}, t) \right], \\
    &= - \sum\limits_{i=1}^N \sum\limits_{a = 1}^d \partial_{x_{i,a}}\left[\left( F_{i,a} + \alpha \sum\limits_{j \neq i}\sum\limits_{b,p=1}^d \left(   \left[ \partial_{x_{i,p}} \sqrt{\Lambda_{ijab} } \right] \sqrt{\Lambda_{ijpb}} + c   \left[ \partial_{x_{j,p}} \sqrt{\Lambda_{ijab}} \right] \sqrt{\Lambda_{jipb}}\right)  \right) P(\bm{X},t)\right] \nonumber \\
    &\hphantom{aaaa}+ \sum\limits_{i=1}^N \sum\limits_{a,b=1}^d \partial_{x_{i,a}}\partial_{x_{i,b}}\left[ \left( D_0 \delta_{ab} + \frac{1}{2}\sum\limits_{k\neq i} \sum\limits_{p=1}^d \sqrt{\Lambda_{ikap}\Lambda_{ikbp}}\right) P(\bm{X},t)\right] \nonumber \\
    &\hphantom{aaaa}+ \frac{c}{2} \sum\limits_{i=1}^N\sum\limits_{j \neq i} \sum\limits_{a,b,p=1}^d \partial_{x_{i,a}}\partial_{x_{j,b}}\left[\left(\sqrt{\Lambda_{ijap}\Lambda_{jibp}}\right) P(\bm{X}, t) \right].\label{eq:FokkerPlanck_proj}
\end{align}

\subsubsection{Diffusion tensor and spurious force}

The Fokker-Planck equation, Eq.~\ref{eq:FokkerPlanck_proj}, is a conservation equation for probability mass.
As such, it may be generically written by introducing a probability current $\bm{J}$ such that 
\begin{align}
    \partial_t P(\bm{X},t) + \bm{\nabla}\cdot\bm{J}(\bm{X},t) = 0.
\end{align}
Extracting the probability current $\bm{J}$ by factoring the $\partial_{x_{i,a}}$ in Eq.~\ref{eq:FokkerPlanck_proj}, one may then write the equilibrium conditions $J_{i,a} = 0$ on the equilibrium distribution $P_{eq}$,
\begin{align}
    0 &= \left( F_{i,a} + \alpha \sum\limits_{j \neq i}\sum\limits_{b,p=1}^d \left(   \left[ \partial_{x_{i,p}} \sqrt{\Lambda_{ijab} } \right] \sqrt{\Lambda_{ijpb}} + c   \left[ \partial_{x_{j,p}} \sqrt{\Lambda_{ijab}} \right] \sqrt{\Lambda_{jipb}}\right)  \right) P_{eq}(\bm{X}) \nonumber \\
    &\hphantom{aaaa}- \sum\limits_{b=1}^d \partial_{x_{i,b}}\left[ \left( D_0 \delta_{ab} + \frac{1}{2}\sum\limits_{k\neq i} \sum\limits_{p=1}^d \sqrt{\Lambda_{ikap}\Lambda_{ikbp}}\right) P_{eq}(\bm{X})\right] - \frac{c}{2} \sum\limits_{j\neq i} \sum\limits_{b,p=1}^d \partial_{x_{j,b}}\left[\left(\sqrt{\Lambda_{ijap}\Lambda_{jibp}}\right) P_{eq}(\bm{X}) \right].
\end{align}
Writing $P_{eq} = e^{-S_{eq}(\bm{X})}/\mathcal{N}$, this yields a simpler equation on $S$,
\begin{align}
    0 &=  F_{i,a} + \alpha \sum\limits_{j \neq i}\sum\limits_{b,p=1}^d \left(   \left[ \partial_{x_{i,p}} \sqrt{\Lambda_{ijab} } \right] \sqrt{\Lambda_{ijpb}} + c   \left[ \partial_{x_{j,p}} \sqrt{\Lambda_{ijab}} \right] \sqrt{\Lambda_{jipb}}\right) \nonumber \\
    &\hphantom{aaaa}-  \frac{1}{2}\sum\limits_{k\neq i} \sum\limits_{b,p=1}^d\partial_{x_{i,b}}\left[ \sqrt{\Lambda_{ikap}\Lambda_{ikbp}}\right]  -\frac{c}{2} \sum\limits_{j\neq i} \sum\limits_{b,p=1}^d \partial_{x_{j,b}} \left[\left(\sqrt{\Lambda_{ijap}\Lambda_{jibp}}\right)  \right]  \nonumber \\
    &\hphantom{aaaa}+ \sum\limits_{b=1}^d \left[ \left( D_0 \delta_{ab} + \frac{1}{2}\sum\limits_{k\neq i} \sum\limits_{p=1}^d \sqrt{\Lambda_{ikap}\Lambda_{ikbp}}\right) \partial_{x_{i,b}}S_{eq}(\bm{X})\right] + \frac{c}{2} \sum\limits_{j\neq i} \sum\limits_{b,p=1}^d \left[\left(\sqrt{\Lambda_{ijap}\Lambda_{jibp}}\right) \partial_{x_{j,b}} S_{eq}(\bm{X}) \right].
\end{align}

This last expression may be rewritten by introducing a tensor acting on the gradient of $S$,
\begin{align}
    \left.\overline{\overline{D}}\cdot \nabla S_{eq}\right|_{i,a} &= - F_{i,a} - \alpha \sum\limits_{j \neq i}\sum\limits_{b,p=1}^d \left(   \left[ \partial_{x_{i,p}} \sqrt{\Lambda_{ijab} } \right] \sqrt{\Lambda_{ijpb}} + c   \left[ \partial_{x_{j,p}} \sqrt{\Lambda_{ijab}} \right] \sqrt{\Lambda_{jipb}}\right) \nonumber \\
    &\hphantom{aaaa}+ \frac{1}{2}\sum\limits_{k\neq i} \sum\limits_{b,p=1}^d\partial_{x_{i,b}}\left[ \sqrt{\Lambda_{ikap}\Lambda_{ikbp}}\right]  +\frac{c}{2} \sum\limits_{j\neq i} \sum\limits_{b,p=1}^d \partial_{x_{j,b}} \left[\left(\sqrt{\Lambda_{ijap}\Lambda_{jibp}}\right)  \right], \label{eq:Equilibrium_Condition_Spatial}
\end{align}
with $\overline{\overline{D}}$ formally a rank-4 tensor with elements
\begin{align}
    D_{iiab} &= D_0 \delta_{ab} + \frac{1}{2} \sum\limits_{k\neq i} \sum\limits_{p=1}^d \sqrt{\Lambda_{ikap}\Lambda_{ikbp}} \\
    D_{ijab} &=\frac{c}{2} \sum\limits_{p=1}^d \sqrt{\Lambda_{ijap}\Lambda_{jibp}} \hphantom{aa} \text{(if $j\neq i$)}. \label{eq:D_general_definition_proj}
\end{align}

This tensor, which we shall henceforth call a diffusion tensor, is a central object in the study of the entropy production of the system.
It is thus useful to note a few of its properties.
First, it is symmetric under $i\leftrightarrow j$ and $a \leftrightarrow b$ permutations by construction.
Furthermore, we consider the special case of pairwise-symmetric multiplicative noise $\Lambda_{ijab} = \Lambda_{jiab}$, as in all previously considered examples of such dynamics~\cite{Zhang2024a, Anand2025}.
With that symmetry, in the special case $c=-1$, the operator becomes akin to a doubly stochastic matrix, as
\begin{align}
    \sum_{i,a} D_{ijab} =     \sum_{j,b} D_{ijab} = D_0.
\end{align}
As a result, for $D_0 = 0$ and $c = -1$,
\begin{align}
    \sum_{i,a} D_{ijab} =     \sum_{j,b} D_{ijab} = 0.
\end{align}
This implies that the operator, seen as a $dN \times dN$ matrix, stops being invertible in this situation.
Indeed, consider the generic case where the sums over rows and columns is $D_0$, this can be written as
\begin{align}
    \overline{\overline{D}} \bm{J} = D_0 \bm{J}
\end{align}
with $\bm{J} = (1, 1,\ldots ,1)$ a vector of ones.
Assuming that the inverse of $\overline{\overline{D}}$ exists, this implies
\begin{align}
    \bm{J} = D_0 \overline{\overline{D}}{}^{-1}\bm{J},
\end{align}
and likewise for the left-multiplication.
As a result, the inverse $\overline{\overline{D}}{}^{-1}$ also has rows and columns summing to a constant, given by $1/D_0$.
As a result, in the limit $D_0 \to 0$, the sum of rows and columns of $\overline{\overline{D}}{}^{-1}$ diverges, and the matrix becomes non-invertible.
This is a crucial aspect that will affect entropy production in Sec.~\ref{sec:Spatial_OM_EPR}.
Note that the case of a generic $c$ with $D_0 = 0$ is not guaranteed to be invertible, as will be illustrated in a simple example in Sec.~\ref{sec:Examples}.

Furthermore, note that writing explicitly the projector component of $\Lambda$ via
\begin{align}
    \sqrt{\Lambda_{ijab}} = \sqrt{\Lambda_{ij}(\bm{r}_{ij})}P_{ab}(\bm{r}_{ij})
\end{align}
and that, since projection matrices are symmetric,
\begin{align}
    \sum\limits_{p=1}^d P_{ap}P_{bp} = (P^2)_{ab},
\end{align}
 as $P^2 = P$, simplifies the expression of the elements of the diffusion tensor to
\begin{align}
    D_{iiab} &= D_0 \delta_{ab} + \frac{1}{2} \sum\limits_{k\neq i}  \Lambda_{ik} P_{ab} \nonumber \\
    D_{ijab} &=\frac{c}{2} \sqrt{\Lambda_{ij}\Lambda_{ji}} P_{ab} \hphantom{aa} \text{(if $j\neq i$)}. \label{eq:D_general_definition_proj_simpler}
\end{align}

Having elucidated the nature of $\overline{\overline{D}}$, we now turn our attention back to the equilibrium condition, Eq.~\ref{eq:Equilibrium_Condition_Spatial}.
When $\overline{\overline{D}}$ admits an inverse, this equation yields an expression for the equilibrium distribution through,
\begin{align}
    S_{eq}(\bm{X}) = - \int  d\bm{X}\cdot \left[\overline{\overline{D}}{}^{-1}\widetilde{\bm{F}}(\bm{X}) \right]
\end{align}
where $\widetilde{\bm{F}}$ is a corrected force field that removes spurious effects due to multiplicativeness of the noise~\cite{Aron2016} and allows the system to relax to an equilibrium (see also the simple example in Sec.~\ref{sec:Multiplicative_Aside}).
In the general case of projective $\Lambda_{ijab}$, its expression is cumbersome but can be deduced from Eq.~\ref{eq:Equilibrium_Condition_Spatial}.

\subsubsection{Special cases\label{sec:Special_Spatial_Cases}}

The diffusion tensor and corrected force are however easier to express when making assumptions about the precise shape of $\Lambda_{ijab}$.
In particular, suppose that the projector part is identity, $P_{ab} = \delta_{ab}$, which we will use in later sections as an illustration to make notations lighter.
Then, the diffusion tensor becomes
\begin{align}
    D_{iiab} &= \delta_{ab} \left( D_0 + \frac{1}{2}\sum\limits_{j \neq i} \Lambda_{ij}(\bm{X})\right), \\
    D_{ijab} &\underset{i\neq j}{=} \delta_{ab} \frac{c}{2} \sqrt{\Lambda_{ij}(\bm{X})\Lambda_{ji}(\bm{X})}. \label{eq:D_general_definition_NoProj}
\end{align}
The equilibrium distribution thus verifies, when the diffusion tensor is invertible,
\begin{align}
    S_{eq}(\bm{X}) = \int \overline{\overline{D}}{}^{-1}\left( - \bm{F} (\bm{X}) + \frac{1}{2}(1 - \alpha) \nabla\cdot \overline{\overline{\Pi}}{}^2 + c \left( \alpha - \frac{1}{2}\right) \nabla\cdot (\overline{\overline{\Pi}}{}^t \overline{\overline{\Pi}}) - \alpha c\overline{\overline{\Pi}}\cdot\nabla ^t\overline{\overline{\Pi}}\right) \cdot d\bm{X},
\end{align}
where the short-hand notation $\overline{\overline{\Pi}}$, that was introduced in Eq.~\ref{eq:ShortHand_Update}, is used for brevity.
For the sake of clarity, note that
\begin{align}
    \left.\nabla\cdot \overline{\overline{\Pi}}{}^2\right|_{ia} &=\sum\limits_{j \neq i} \partial _{x_{i,a}} \Lambda_{ij} (\bm{X}). \\
    \left. \nabla\cdot (\overline{\overline{\Pi}}{}^t \overline{\overline{\Pi}})\right|_{i,a} &= \sum\limits_{j \neq i} \left[        \partial_{x_{i,a}}\sqrt{\Lambda_{ij}(\bm{X}) \Lambda_{ji}(\bm{X})} \right] \\
    \left.\overline{\overline{\Pi}}\cdot\nabla ^t\overline{\overline{\Pi}}\right|_{i,a} &= \sum\limits_{j \neq i} \left[ \partial _{x_{i,a}} \sqrt{\Lambda_{ji} (\bm{X})} \right] \left(    \sqrt{\Lambda_{ij} (\bm{X})}\right)
\end{align}
One may then introduce an effective force that ensures a well-behaved equilibrium distribution,
\begin{align}
    \widetilde{\bm{F}}(\bm{X}) = \overline{\overline{D}}{}^{-1}\left(  \bm{F} (\bm{X}) + \frac{1}{2}(\alpha - 1) \nabla\cdot \overline{\overline{\Pi}}{}^2 - c \left( \alpha - \frac{1}{2}\right) \nabla\cdot (\overline{\overline{\Pi}}{}^t \overline{\overline{\Pi}}) + \alpha c\overline{\overline{\Pi}}\cdot\nabla ^t\overline{\overline{\Pi}}\right). \label{eq:EffectiveForce_GenericLambda}
\end{align}

Suppose additionally that $\Lambda_{ij} = \Lambda_{ji}$, like in Random Organization~\cite{Corte2008,Wilken2020}.
In that case, 
\begin{align}
    D_{iiab} &= \delta_{ab}(D_0 + \frac{1}{2}\sum\limits_{j \neq i} \Lambda_{ij}(\bm{X})), \\
    D_{ijab} &\underset{i\neq j}{=} \delta_{ab}\frac{c}{2} \Lambda_{ij} (\bm{X}).   \label{eq:D_general_definition_RO}
\end{align}
The equilibrium distribution then verifies
\begin{align}
    S_{eq}(\bm{X}) = \int \overline{\overline{D}}{}^{-1}\left( - \bm{F} (\bm{X}) + \frac{1}{2}(1 - \alpha) (1-c)\nabla\cdot \overline{\overline{\Pi}}{}^2\right) \cdot d\bm{X}.
\end{align}
The effective force $\widetilde{\bm{F}}$ then becomes
\begin{align}
    \widetilde{\bm{F}}(\bm{X}) = \overline{\overline{D}}{}^{-1}\left(  \bm{F} (\bm{X}) + \frac{1}{2}(\alpha - 1 ) (1-c)\nabla\cdot \overline{\overline{\Pi}}{}^2\right). \label{eq:EffectiveForce_SymmetricLambda}
\end{align}

Notice that the special case $c = 0$ leaves only elements $D_{iiab}$ non-zero, regardless of the level of assumptions (\textit{i.e.} even in Eq.\ref{eq:D_general_definition_proj_simpler}).
If, in addition, $P_{ab} = \delta_{ab}$, $\overline{\overline{D}}$ is diagonal with non-negative elements: it is invertible as long as its diagonal entries are not zero (which is guaranteed if $D_0>0$).
Additionally, $\overline{\overline{D}}{}^{-1}$ itself is then diagonal.
The effect of $\Lambda_{ij}(\bm {X})$ is then essentially that of a generalized temperature: this can be seen by choosing the ``thermal'' convention $\alpha = 1$, and writing the corresponding expression of $S_{eq}$,
\begin{align}
    S_{eq}(\bm{X}) = \frac{(\mu/D_0)V(\bm{X})}{\left(1 + \frac{1}{2D_0} \sum\limits_{j \neq i}\Lambda_{ij}(\bm{X})\right)},
\end{align}
where we recalled that forces are conservative and assumed a single mobility value $\mu$.
This expression is written like the usual Boltzmann distribution but with a rescaled, coordinate-dependent generalized temperature $T_{\text{gen}}(\bm{X})$, such that
\begin{align}
    k_B T_{\text{gen}}(\bm{X}) = \frac{D_0}{\mu} \left(1 + \frac{1}{2D_0} \sum\limits_{j \neq i}\Lambda_{ij}(\bm{X})\right).
\end{align}
One may recognize the standard temperature defined through Einstein's relation in the prefactor, $k_B T = D_0 / \mu$.

A number of simple examples of expressions of $\overline{\overline{D}}$ and its inverse are given in Sec.~\ref{sec:Examples} to further discuss the inversibility of the diffusion tensor, as we now turn our attention to path-integral representation.

\subsubsection{MSRJD Action \label{sec:Spatial_MSRJD}}

Following the standard construction of the dynamical MSRJD action~\cite{Aron2016,ArnoulxdePirey2022}, the first step is to write the probability to observe state $\bm{X}_T = (\bm{r}_1, \ldots, \bm{r}_N)$ at time $T$ starting from $\bm{X}_0$ at time $0$ in a path integral representation,
\begin{align}
    \mathbb{P}(\bm{X}_T,T|\bm{X}_0, 0) &= \int \prod_{n = 1}^{N} \left[ \mathcal{D}\bm{r}_n \mathcal{D}\bm{\eta}_n p_\eta(\bm{\eta}_n) \prod_{m<n} \mathcal{D} \bm{\xi}_{mn} p_\xi(\bm{\xi}_{mn}) \right] \delta(\bm{X}(0) - \bm{X}_0) \delta(\bm{X}(T) - \bm{X}_T)  \prod_{0 < t < T} \delta\left(\bm{X}(t) - \bm{X}_{\text{dyn}}(t)  \right) 
\end{align}
where $\mathcal{D} x = \prod\limits_{0<t<T} dx(t)$ represents a path-integral differential element, $p_\eta$ and $p_\xi$ are the probability distribution functions (pdf) of noise amplitudes, and $\bm{X}_{\text{dyn}}(t)$ is the value imposed at time $t$ by the dynamics (which is a deterministic value once the noise values have been drawn within the integral).
This conditional probability may be integrated over all possible values of $X_T$, and over an arbitrary distribution $p_0$ of initial conditions, yielding the more common expression
\begin{align}
    1 = \int \prod_{n = 1}^{N} \left[ d^d\bm{r}_n(\bm{0}) p_0\left(\bm{r}_n (\bm{0})\right) \mathcal{D}\bm{r}_n \mathcal{D}\bm{\eta}_n p_\eta(\bm{\eta}_n) \prod_{m<n} \mathcal{D} \bm{\xi}_{mn} p_\xi(\bm{\xi}_{mn}) \right]  \prod_{0 \leq t \leq T} \delta\left(\bm{X}(t) - \bm{X}_{\text{dyn}}(t)  \right).
\end{align}
This expression is used to compute dynamical averages of observables -- for an arbitrary function $\mathcal{O}(\bm{X})$, the dynamical average reads
\begin{align}
    \left\langle \mathcal{O} \right\rangle_{\text{dyn}} = \int \prod_{n = 1}^{N} \left[ d^d\bm{r}_n(\bm{0}) p_0\left(\bm{r}_n (\bm{0})\right) \mathcal{D}\bm{r}_n \mathcal{D}\bm{\eta}_n p_\eta(\bm{\eta}_n) \prod_{m<n} \mathcal{D} \bm{\xi}_{mn} p_\xi(\bm{\xi}_{mn}) \right]  \prod_{0 \leq t \leq T} \mathcal{O}(\bm{X})\,\delta\left(\bm{X}(t) - \bm{X}_{\text{dyn}}(t)  \right).
\end{align}

Then, introducing the notation
\begin{align}
    \dot{\bm{r}}_{n}(t) &= \bm{U}_{n} \left[\left\{ \bm{r}_{mn}\right\},\left\{ \bm{\xi}_{mn}\right\}, \bm{\eta}_n\right]
\end{align}
for Eq.~\ref{eq:BROSGD_dynamics_proj}, it is common to change variables in the Dirac delta explicitly, yielding
\begin{align}
    1 = \int \prod_{n = 1}^{N} \left[ d^d\bm{r}_n(\bm{0}) p_0\left(\bm{r}_n (\bm{0})\right) \mathcal{D}\bm{r}_n \mathcal{D}\bm{\eta}_n p_\eta(\bm{\eta}_n) \prod_{m<n} \mathcal{D} \bm{\xi}_{mn} p_\xi(\bm{\xi}_{mn}) \right]  \prod_{0 \leq t \leq T}  \det \mathcal{J}(t, t')  \prod_{n=1}^N \delta\left(\dot{\bm{r}}_n - U_n \left[\left\{ \bm{r}_{mn}\right\},\left\{ \bm{\xi}_{mn}\right\}, \bm{\eta}_n\right] \right) 
\end{align}
where the Jacobian of the change of variables is given by
\begin{align}
    \mathcal{J}(t,t') &= \left[\frac{\delta  }{\delta \bm{X}(t')}\left(\dot{\bm{X}}(t) - \bm{U}(t)  \right) \right] \\
    &= \frac{d}{dt} \delta(t-t') \mathcal{I} - \frac{\delta \bm{U}(t) }{\delta \bm{X}(t)}  \delta(t - t'),
\end{align}
with $\mathcal{I}$ is the identity rank-$4$ tensor such that
\begin{align}
    \mathcal{I}_{ijab} = \delta_{ij} \delta_{ab}
\end{align}
Introducing the short-hand notation for the components of $\bm{U}$,
\begin{align}
    \bm{U}(t) = \overline{\overline{M}}\bm{\nabla} V(\bm{X}) + \overline{\overline{\Pi}}(\bm{X})\bm{\xi}(t) + \sqrt{2 D_0}\, \overline{\overline{I}}\bm{\eta}(t), \label{eq:U_def} 
\end{align}
where the tensors $\overline{\overline{M}}$ and $\overline{\overline{\Pi}}$ respectively encode the neg-mobilities and the amplitudes of the pairwise multiplicative noise, one may write
\begin{align}
    \frac{\delta \bm{U}(t) }{\delta \bm{X}(t)} = \overline{\overline{M}}\bm{\nabla} \frac{\delta V(\bm{X})}{\delta \bm{X} } + \frac{\delta \overline{\overline{\Pi}}(\bm{X})}{\delta \bm{X}}\bm{\xi}(t). \label{eq:U_grad} 
\end{align}
One element of this derivative can be written as
\begin{align}
    \mathcal{G}_{ijab}(\bm{X},t) \equiv\frac{\delta U_{i,a}}{\delta x_{j,b}}(\bm{X},t) &= - \mu_i \frac{\partial^2 V}{\partial x_{i,a}\partial x_{j,b}}(\bm{X}) + \sum\limits_{c=1}^d\sum\limits_{k\neq i} \left[ \partial_{x_{j,b}}\sqrt{\Lambda_{ikac}(\bm{X})} \right] \xi_{ik}^c(t) \\
    &= - \mu_i \mathcal{H}_{ijab}(\bm{X}) + \sum\limits_{c=1}^d\sum\limits_{k\neq i} \left[ \partial_{x_{j,b}}\sqrt{\Lambda_{ikac}(\bm{X})} \right] \xi_{ik}^c(t)
\end{align}
where $\mathcal{H}$ is the Hessian of interactions and $\mathcal{G}$ is the gradient tensor defined by the derivative.
One may rewrite the determinant by factorizing the Jacobian,
\begin{align}
    \mathcal{J}(t,t') = \int dt'' \delta(t-t'')   \frac{d}{dt''} \left[\delta(t''-t') \mathcal{I} - \Theta(t'' - t')\frac{\delta \bm{U}(t) }{\delta \bm{X}(t)}  \right],
\end{align}
with $\Theta$ a Heaviside distribution, so that
\begin{align}
    \det \mathcal{J} = \det \left[ \delta(t-t'')   \frac{d}{dt''}\right] \det \left[ \delta(t''-t')\mathcal{I} - \Theta(t'' - t')\frac{\delta \bm{U}(t) }{\delta \bm{X}(t)} \right].
\end{align}
The first factor is just a constant that can be absorbed into a normalization constant, but the second one will play a role due to the multiplicative nature of the noise.
One may use the identity
\begin{align}
    \det(\mathbb{1} + A) = \exp \text{Tr} \ln (\mathbb{1} + A).
\end{align}
Finally, one may expand the logarithm up to quadratic order (so as to catch the variance of the noise dependence of $A$).
All in all,
\begin{align}
    \det \mathcal{J} \propto \det \left[ \delta(t''-t')\mathcal{I} - \Theta(t'' - t')\frac{\delta \bm{U}(t) }{\delta \bm{X}(t)} \right] &= \exp \underset{t't'',i j, a b}{\text{Tr}} \ln\left[ \delta(t''-t')\mathcal{I} - \Theta(t'' - t')\frac{\delta \bm{U}(t) }{\delta \bm{X}(t)}\right] \\
    &\approx \exp \underset{t't'',i j, a b}{\text{Tr}}\left[ - \Theta(t'' - t')\frac{\delta \bm{U}(t) }{\delta \bm{X}(t)} - \frac{1}{2} A^2(t',t'') \right] \\
    &= \exp \left[ - \int dt \left[ \Theta(0) \underset{i j, a b}{\text{Tr}}\frac{\delta \bm{U}(t) }{\delta \bm{X}(t)} + \frac{1}{2} A^2(t,t) \right] \right]
\end{align}
where
\begin{align}
    A^2(t,t') = \underset{i j, a b}{\text{Tr}}\int dt'' \Theta(t - t'')\frac{\delta \bm{U}(t) }{\delta \bm{X}(t)} \Theta(t'' - t')\frac{\delta \bm{U}(t'') }{\delta \bm{X}(t'')}
\end{align}
so that
\begin{align}
    A^2(t,t) &=  \underset{i j, a b}{\text{Tr}}\int dt'' \Theta(t - t'')\frac{\delta \bm{U}(t) }{\delta \bm{X}(t)} \Theta(t'' - t)\frac{\delta \bm{U}(t'') }{\delta \bm{X}(t'')} \label{eq:A2_def} \\
    &= \underset{i j, a b}{\text{Tr}}\int dt'' \Theta(t - t'') \Theta(t'' - t) \sum\limits_{k=1}^N \sum\limits_{c=1}^d \mathcal{G}_{ikac}(\bm{X}(t),t) \mathcal{G}_{kjcb} (\bm{X}(t''),t'') \\
    &= \sum\limits_{i,k=1}^N \sum\limits_{a,c = 1}^d\int dt'' \Theta(t - t'') \Theta(t'' - t) \mathcal{G}_{ikac}(\bm{X}(t),t) \mathcal{G}_{kica}(\bm{X}(t''),t'')
\end{align}
where the value of the Heaviside at the step is ambiguous and thus left as is for now.

To progress further with the expression of the Jacobian, it is necessary to go back to the full expression of the path integral.
Introducing $\mathcal{N}$ a normalization that will be used to absorb constant factors, and introducing shorter notations for convenience we get
\begin{align}
     \left\langle \mathcal{O} \right\rangle_{\text{dyn}} = \frac{1}{\mathcal{N}} \int &d^{dN}\bm{X}_0 p_0\left(\bm{X}_0\right) \mathcal{D}\bm{X}  \prod\limits_{n=1}^N\left[\mathcal{D}\bm{\eta}_n p_\eta(\bm{\eta}_n) \prod_{m<n} \mathcal{D} \bm{\xi}_{mn} p_\xi(\bm{\xi}_{mn})\right] \nonumber \\
     &\times  \exp \left[ - \int\limits_0^T d\tau \left[ \Theta(0) \underset{i j, a b}{\text{Tr}}\frac{\delta \bm{U}(\tau) }{\delta \bm{X}(\tau)} + \frac{1}{2} A^2(\tau,\tau) \right] \right] \prod_{0 \leq t \leq T}  \mathcal{O}(\bm{X}(t))\delta\left(\dot{X} - \bm{U}(t) \right) 
\end{align}
The Dirac-deltas that enforce the dynamics may be represented as their Fourier decomposition, leading (after absorbing a multiple of $2\pi$ into $\mathcal{N}$) to
\begin{align}
    \left\langle \mathcal{O} \right\rangle_{\text{dyn}} = \frac{1}{\mathcal{N}} \int &d^{dN}\bm{X}_0 p_0\left(\bm{X}_0\right) \mathcal{D}\bm{X} \mathcal{D}\bm{Q} \prod\limits_{n=1}^N\left[\mathcal{D}\bm{\eta}_n p_\eta(\bm{\eta}_n) \prod_{m<n} \mathcal{D} \bm{\xi}_{mn} p_\xi(\bm{\xi}_{mn})\right] \nonumber \\
     &\times \exp \left[ - \int\limits_0^T d\tau \left[ \Theta(0) \underset{i j, a b}{\text{Tr}}\frac{\delta \bm{U}(\tau) }{\delta \bm{X}(\tau)} + \frac{1}{2} A^2(\tau,\tau) \right] \right] \prod_{0 \leq t \leq T}  \mathcal{O}(\bm{X}(t))\exp\left[ - i \bm{Q}(t) \cdot\left(\dot{X} - \bm{U}(t) \right) \right]
\end{align}
where $\bm{Q}(t) = (\bm{q}_1(t), \ldots, \bm{q}_N (t))$ is the Fourier variable associated to $\bm{X}$ at time $t$.
The product over time elements may then be exponentiated, and the size of time steps be taken to be very small, leading to 
\begin{align}
    \left\langle \mathcal{O} \right\rangle_{\text{dyn}} = \frac{1}{\mathcal{N}} \int &d^{dN}\bm{X}_0 p_0\left(\bm{X}_0\right) \mathcal{D}\bm{X} \mathcal{D}\bm{Q} \prod\limits_{n=1}^N\left[\mathcal{D}\bm{\eta}_n p_\eta(\bm{\eta}_n) \prod_{m<n} \mathcal{D} \bm{\xi}_{mn} p_\xi(\bm{\xi}_{mn})\right] \nonumber \\
     &\times \exp \left[ - \int\limits_0^T dt \left[ \Theta(0) \underset{i j, a b}{\text{Tr}}\frac{\delta \bm{U}(t) }{\delta \bm{X}(t)} + \frac{1}{2} A^2(t,t) + i\bm{Q}(t) \cdot\left(\dot{X} - \bm{U}(t) \right) - \ln \mathcal{O}(\bm{X}(t)) \right] \right] \label{eq:Dynamical_Average}
\end{align}
where for convenience we here assume that $\mathcal{O}$ has a well-defined logarithm (that will be of no consequence eventually).
This expression can be used to introduce the dynamical action, $\mathcal{A}$, such that
\begin{align}
    \left\langle 1 \right\rangle_{\text{dyn}} = 1=  \frac{1}{\mathcal{N}} \int &d^{dN}\bm{X}_0 \mathcal{D}\bm{X} \mathcal{D}\bm{Q} e^{\mathcal{A}[\bm{X}, \bm{Q} | \bm{X}_0]}.
\end{align}

We now focus on the expression of $\mathcal{A}$, which must be obtained by integrating the path-integral representation over realizations of noise histories.
We follow the path outlined in Refs.~\cite{Aron2016, ArnoulxdePirey2022}.
We thus focus on
\begin{align}
    e^\mathcal{A} = \int & \prod\limits_{n=1}^N\left[\mathcal{D}\bm{\eta}_n p_\eta(\bm{\eta}_n) \prod_{m<n} \mathcal{D} \bm{\xi}_{mn} p_\xi(\bm{\xi}_{mn})\right] \exp \left[ - \int\limits_0^T dt \left[ \Theta(0)\text{Tr}\frac{\delta \bm{U}(t) }{\delta \bm{X}(t)} + \frac{1}{2} A^2(t,t) + i\bm{Q}(t) \cdot\left(\dot{X} - \bm{U}(t) \right)  \right] + \ln p_0(\bm{X}_0)\right]. \label{eq:Noise_Integral_def}
\end{align}
For a choice of convention $\alpha$ for the discretized SDE, the consistent value for the center of the Heaviside is $\Theta(0) = \alpha$, as it results from the mixing of the two sides of the steps.
Recalling the expressions of $\bm{U}$, Eq.~\ref{eq:U_def}, its derivative, Eq.~\ref{eq:U_grad}, and the value of $A^2$ found in Eq.~\ref{eq:A2_def}, the expression in Eq.~\ref{eq:Noise_Integral_def} can be split between a deterministic part, a part due to additive noise, and a part due to multiplicative noise,
\begin{align}
    \mathcal{A}\left[\bm{X}, \dot{\bm{X}}, \bm{Q}; \bm{X}_0 \right] &= \mathcal{A}_{det}\left[\bm{X}, \dot{\bm{X}}, \bm{Q}; \bm{X}_0 \right] + \mathcal{A}_{\text{add}}\left[\bm{X}, \dot{\bm{X}}, \bm{Q} \right] + \mathcal{A}_{\text{mul}}\left[\bm{X}, \dot{\bm{X}}, \bm{Q} \right]\label{eq:Action_decomp}
\end{align}
with
\begin{align}
    e^{\mathcal{A}_{\text{det}}} &= \exp\left[\ln p_0 (\bm{X}_0)- i  \int\limits_{0}^T dt \bm{Q}(t)\cdot \left[\dot{\bm{X}}(t) - \overline{\overline{M}} \bm{\nabla} V(\bm{X}) \right)\right], \label{eq:Adet_def} \\
    e^{\mathcal{A}_{\text{add}}} &= \int  \prod\limits_{n=1}^N\left[\mathcal{D}\bm{\eta}_n p_\eta(\bm{\eta}_n) \right] \exp \left[  i \sqrt{2 D_0}\int\limits_0^T dt   \bm{Q}(t) \cdot \overline{\overline{I}} \bm{\eta}(t) \right], \label{eq:Aadd_def}\\
    e^{\mathcal{A}_{\text{mul}}} &= \int  \left[\prod\limits_{n=1}^N \prod_{m<n}  \mathcal{D} \bm{\xi}_{mn} p_\xi(\bm{\xi}_{mn})\right] \exp \left[ - \int\limits_0^T dt \left[ \alpha\text{Tr}\frac{\delta \bm{U}(t) }{\delta \bm{X}(t)} + \frac{1}{2} A^2(t,t) - i\bm{Q}(t) \cdot \overline{\overline{\Pi}}(t)\bm{\xi}(t)  \right]\right], \label{eq:Amul_def}
\end{align}
where the last equation contains all terms that usually do not appear within the additive noise case, in spite of the fact that some terms seemingly do not depend explicitly on $\bm{\xi}$ but only on $\alpha$.

To compute the averages over noise trajectories, it is useful to note a property of averages over Gaussian variables: if an average is performed over a Gaussian random variable $x$ with zero mean and variance $\sigma^2$, one can write that
\begin{align}
    \left\langle e^{- a x} \right\rangle &= \frac{1}{\sqrt{2\pi\sigma^2}} \int dx \ e^{- a x - x^2/2\sigma^2} = e^{a^2 \sigma^2/2}
    \; .
\end{align}
In particular, a simple example is that of additive noise,
\begin{align}
    e^{\mathcal{A}_{\text{add}}} &= \prod\limits_{n=1}^N \prod\limits_{0\leq t \leq T}\left\langle \exp \left[  i \sqrt{2 D_0}   \bm{q}_n(t) \cdot  \bm{\eta}_n(t) \right]\right\rangle_{\eta_n} , \\
    &= \prod\limits_{n=1}^N \prod\limits_{0\leq t \leq T} e^{-D_0q_n(t)^2} \\
    &=\exp\left[ - D_0 \int\limits_{0}^T dt Q(t)^2 \right].
\end{align}
As for the multiplicative term, one has
\begin{align}
    e^{\mathcal{A}_{\text{mul}}} &= \int  \left[\prod\limits_{n=1}^N \prod_{m<n}  \mathcal{D} \bm{\xi}_{mn} p_\xi(\bm{\xi}_{mn})\right] \exp \left[ - \int\limits_0^T dt \left[ \alpha \left(\overline{\overline{M}}\bm{\nabla}^2 V(\bm{X}) + \bm{\nabla}\overline{\overline{\Pi}}(\bm{X})\bm{\xi}(t)\right)  - i\bm{Q}(t) \cdot \overline{\overline{\Pi}}(t)\bm{\xi}(t) \vphantom{\int\limits_0} + \frac{1}{2} A^2(t,t) \right]\right], \\
    &= \left\langle\prod\limits_{0 \leq t \leq T}  \exp \left[ -\alpha \left[\overline{\overline{M}}\bm{\nabla}^2 V(\bm{X},t)\right]\right]  \exp\left[ \left( i \bm{Q}(t) \cdot \overline{\overline{\Pi}} - \alpha \nabla \overline{\overline{\Pi}} \right) \cdot \bm{\xi}(t) - \frac{1}{2} A^2(t,t) \right]\right\rangle_{\xi}. \label{eq:Amul_average}
\end{align}
In that last expression, the components of $A^2$ can be made more explicit using Eqs.~\ref{eq:A2_def} and~\ref{eq:U_grad},
\begin{align}
    A^2(t,t) &= \sum\limits_{i,k=1}^N \sum\limits_{a,c = 1}^d\int dt'' \Theta(t - t'') \Theta(t'' - t) \mathcal{G}_{ikac}(\bm{X}(t),t) \mathcal{G}_{kica}(\bm{X}(t''),t'') \\
    &= \sum\limits_{i,k=1}^N \sum\limits_{a,c = 1}^d\int dt'' \Theta(t - t'') \Theta(t'' - t) \left[ \mu_i \mu_k \mathcal{H}_{ikac}(\bm{X}(t))  \mathcal{H}_{kica}(\bm{X}(t''))  \right] \nonumber \\
    &\hphantom{a} -\sum\limits_{i,k=1}^N \sum\limits_{a,c,p = 1}^d\int dt'' \Theta(t - t'') \Theta(t'' - t) \mu_i  \mathcal{H}_{ikac}(\bm{X}(t))   \sum\limits_{l \neq k} \partial_{x_{i,a}}\sqrt{\Lambda_{klcp}(\bm{X}(t''))}\xi_{kl}^p(t'')\nonumber  - \left\{i\leftrightarrow k, a\leftrightarrow c, t\leftrightarrow t'' \right\}\\
    &\hphantom{a}+ \sum\limits_{i,k=1}^N \sum\limits_{a,c,p,q = 1}^d\int dt'' \Theta(t - t'') \Theta(t'' - t)\left[\sum\limits_{l\neq i}\sum\limits_{m\neq j}\left(\partial_{x_{kc}}\sqrt{\Lambda_{ilap} (\bm{X}(t))}\right) \left(\partial_{x_{ia}}\sqrt{\Lambda_{kmcq} (\bm{X}(t''))}\right) \xi_{il}^p(t) \xi_{km}^q (t'')\right]
\end{align}
where the middle line contains two symmetric terms.
The last equation can be written in short-hand form for convenience, as
\begin{align}
     A^2(t,t)&=\int dt'' \Theta(t - t'')\Theta(t'' - t) \left( \overline{\overline{M}}\bm{\nabla}^2 V(\bm{X}(t)) + \bm{\nabla}\overline{\overline{\Pi}}(\bm{X}(t))\bm{\xi}(t)\right)\left( \overline{\overline{M}}\bm{\nabla}^2 V(\bm{X}(t'')) + \bm{\nabla}\overline{\overline{\Pi}}(\bm{X}(t''))\bm{\xi}(t'')\right).
\end{align}

From this point on, it is convenient to assume that $P_{ab} = \delta_{ab}$ to make the calculation lighter -- in the case of a general projector, the same calculation can otherwise be reproduced but becomes more cumbersome without yielding a particularly more enlightening result.
Then, writing the distribution of the pairwise noise as a Gaussian such that
\begin{align}
    \int \mathcal{D}\bm{\xi} \left\langle f(t)\right\rangle_{\bm{\xi}} \equiv \prod\limits_{0\leq t \leq T}\frac{1}{Z_\xi}\int d\bm{\xi}(t) f(t) \exp\left[ -\frac{\xi(t)^2}{2} \right],
\end{align}
one may simplify the expression the average in Eq.~\ref{eq:Amul_average} by performing a change of variables to ``complete'' the square with the imaginary part of the integral,
\begin{align}
    \bm{\zeta}(t) \equiv \bm{\xi}(t) -\left( i \bm{Q}(t) \cdot \overline{\overline{\Pi}} - \alpha \nabla \overline{\overline{\Pi}} \right),
\end{align}
or equivalently
\begin{align}
    \bm{\zeta}(t)^2 \equiv \bm{\xi}(t)^2 -2\left( i \bm{Q}(t) \cdot \overline{\overline{\Pi}} - \alpha \nabla \overline{\overline{\Pi}} \right)\cdot\bm{\xi(t)} + \left( i \bm{Q}(t) \cdot \overline{\overline{\Pi}} - \alpha \nabla \overline{\overline{\Pi}} \right)^2,
\end{align}
which should be understood as,
\begin{align}
    \zeta_{mn}^a(t)^2 \equiv \xi_{mn}^a(t)^2 -2\left( i q_{m}^a(t) \sqrt{\Lambda_{mn}}- \alpha \partial_{m,a} \sqrt{\Lambda_{mn}} \right)\xi_{mn}^a(t) + \left( i q_{m}^a(t) \sqrt{\Lambda_{mn}} - \alpha \partial_{m,a} \sqrt{\Lambda_{mn}} \right)^2.
\end{align}
Injecting this change of variable into the integral yields
\begin{align}
    e^{\mathcal{A}_{\text{mul}}}
    &= \exp \left[\int\limits_{0}^T dt \left(-\alpha \left[\overline{\overline{M}}\bm{\nabla}^2 V(\bm{X},t)\right] +\frac{1}{2}\left( i\bm{Q}(t) \cdot \overline{\overline{\Pi}} - \alpha \nabla \overline{\overline{\Pi}} \right)^2\right)\right] \left\langle \exp\left[- \int\limits_{0}^T dt \frac{1}{2}  A^2_\zeta(t,t) \right]\right\rangle_{\zeta}.
\end{align}
The remaining noise integral may be studied on its own,
\begin{align}
    \left\langle \exp\left[- \int\limits_{0}^T dt \frac{1}{2}  A^2_\zeta(t,t) \right]\right\rangle_{\zeta} &= \nonumber \\
    & \hspace{-4cm}\left\langle \exp\left[- \frac{1}{2}\int\limits_{0}^T \int\limits_{0}^T dt dt''  \Theta(t - t'')\Theta(t'' - t) \left( \overline{\overline{M}}\bm{\nabla}^2 V(\bm{X}(t)) + \bm{\nabla}\overline{\overline{\Pi}}(\bm{X})\bm{\xi}(t)\right)\left( \overline{\overline{M}}\bm{\nabla}^2 V(\bm{X}(t'')) + \bm{\nabla}\overline{\overline{\Pi}}(\bm{X})\bm{\xi}(t'')\right)\right]\right\rangle_{\zeta} \\
    &\hspace{-3.5cm} = \prod\limits_{0\leq t, t'' \leq T} \left\langle \exp\left[- \frac{1}{2}  \Theta(t - t'')\Theta(t'' - t) \left( \overline{\overline{M}}\bm{\nabla}^2 V(\bm{X}(t)) + \bm{\nabla}\overline{\overline{\Pi}}(\bm{X})\bm{\xi}(t)\right)\left( \overline{\overline{M}}\bm{\nabla}^2 V(\bm{X}(t'')) + \bm{\nabla}\overline{\overline{\Pi}}(\bm{X})\bm{\xi}(t'')\right)\right]\right\rangle_{\zeta} 
\end{align}
The product of Heavisides may only have weight if it is measured by a $\delta(t - t'')$.
As a result, the only contribution in the average comes from the covariance of $\xi(t)\xi(t'')$ and
\begin{align}
    \left\langle \exp\left[- \int\limits_{0}^T dt \frac{1}{2}  A^2_\zeta(t,t) \right]\right\rangle_{\zeta} = \prod\limits_{0\leq t, t'' \leq T}  \exp\left[- \frac{1}{2}  \Theta(t - t'')\Theta(t'' - t) \left\langle\bm{\nabla}\overline{\overline{\Pi}}(\bm{X})\bm{\xi}(t)\bm{\nabla}\overline{\overline{\Pi}}(\bm{X})   \bm{\xi}(t'')\right\rangle_{\zeta} \right].
\end{align}
To progress further, it is necessary to re-establish the precise meaning of short-hand notations in the average.
\begin{align}
    \left\langle \exp\left[- \int\limits_{0}^T dt \frac{1}{2}  A^2_\zeta(t,t) \right]\right\rangle_{\zeta} &= \nonumber \\
    &\hspace{-3cm}\prod\limits_{0\leq t, t'' \leq T} \prod\limits_{n=1}^N\prod\limits_{m<n} \prod\limits_{n'=1}^N\prod\limits_{m'<n'} \prod\limits_{a,b=1}^d \exp\left[- \frac{1}{2}  \Theta(t - t'')\Theta(t'' - t) \left\langle\partial_{n,a} \sqrt{\Lambda_{mn}}\xi_{mn}^a(t)\partial_{n',b} \sqrt{\Lambda_{m'n'}}\xi_{m'n'}^b(t'')\right\rangle_{\zeta} \right], \\
    &\hspace{-4cm}= \prod\limits_{0\leq t, t'' \leq T} \prod\limits_{n=1}^N\prod\limits_{m<n} \prod\limits_{n'=1}^N\prod\limits_{m'<n'} \prod\limits_{a,b=1}^d \exp\left[- \frac{1}{2}  \Theta(t - t'')\Theta(t'' - t) \partial_{n,a} \sqrt{\Lambda_{mn}(t)} \partial_{n',b} \sqrt{\Lambda_{m'n'}(t'') }\left\langle\xi_{mn}^a(t)\xi_{m'n'}^b(t'')\right\rangle_{\zeta} \right].
\end{align}
Using the definition of the correlations in the pairwise noise, as well as the value $\Theta(0) = \alpha$, one finds
\begin{align}
   \left\langle \exp\left[- \int\limits_{0}^T dt \frac{1}{2}  A^2_\zeta(t,t) \right]\right\rangle_{\zeta} = \prod\limits_{0\leq t \leq T} \prod\limits_{n=1}^N\prod\limits_{m<n}  \prod\limits_{a=1}^d \exp\left[- \frac{1}{2}  \alpha^2 \left[ \left(\partial_{n,a} \sqrt{\Lambda_{mn}(t)}\right)^2 +  c \left(\partial_{n,a} \sqrt{\Lambda_{mn}(t)} \partial_{m,a} \sqrt{\Lambda_{nm }(t)}\right) \right] \right].
\end{align}

In summary,
\begin{align}
    \mathcal{A}_{\text{det}} &= \ln p_0 (\bm{X}_0)- i  \int\limits_{0}^T dt \bm{Q}(t)\cdot \left[\dot{\bm{X}}(t) - \overline{\overline{M}} \bm{\nabla} V(\bm{X}) \right], \label{eq:Adet_res} \\
    \mathcal{A}_{\text{add}} &= -D_0 \int\limits_{0}^T dt Q(t)^2, \label{eq:Aadd_res}\\
    \mathcal{A}_{\text{mul}} &= \int\limits_{0}^T dt \left(-\alpha \left[\overline{\overline{M}}\bm{\nabla}^2 V(\bm{X},t)\right] +\frac{1}{2}\left( i\bm{Q}(t) \cdot \overline{\overline{\Pi}} - \alpha \nabla \overline{\overline{\Pi}} \right)^2 - \frac{\alpha^2}{2}\left( (\nabla \overline{\overline{\Pi}})^2 + c \nabla \overline{\overline{\Pi}} {}^t\nabla \overline{\overline{\Pi}}\right)\right)\label{eq:Amul_res}
\end{align}

One may regroup the noise terms differently, to encode their physical meaning for the dynamics rather than their origin~\cite{ArnoulxdePirey2022,Aron2016}, as
\begin{align}
    \mathcal{A}_{\text{det}'} &= \ln p_0 (\bm{X}_0) + i  \int\limits_{0}^T dt \bm{Q}(t)\cdot \left[ \overline{\overline{M}} \bm{\nabla} V(\bm{X}) \right] \label{eq:Adet'_res} \\
    \mathcal{A}_{\text{diss}} &=  \int\limits_{0}^T dt \left(- i \bm{Q}(t)\cdot\dot{\bm{X}}(t) -D_0 Q(t)^2 -\frac{1}{2}\left( \bm{Q}(t) \cdot \overline{\overline{\Pi}} \right)^2 - \alpha \left( i\bm{Q}(t) \cdot \overline{\overline{\Pi}} \right)\cdot \nabla\overline{\overline{\Pi}} \right) \label{eq:Amul_diss} \\
    \mathcal{A}_{\text{jac}} &= -\alpha \int\limits_{0}^T dt \left( \overline{\overline{M}}\bm{\nabla}^2 V(\bm{X},t)  + \frac{\alpha c}{2} \nabla \overline{\overline{\Pi}} {}^t\nabla \overline{\overline{\Pi}}\right) \label{eq:Amul_jac}
\end{align}
where $\text{diss}$ stands for dissipation (and now groups up all terms that come from interactions with a bath) and $\text{jac}$ stands for Jacobian.

In these expression, without replacing $V$ by some effective potential, the multiplicative noise always leads to spurious components in the action, no matter the $\alpha$ (no choice of $\alpha$ leads to a Boltzmann distribution with the original potential).
In Eq.~\ref{eq:EffectiveForce_GenericLambda}, we found the effective force acting on the system for this choice of noise.
Another view~\cite{GonzalezArenas2012,ArnoulxdePirey2022} is that one may replace the deterministic force so as to cancel out the dependence on $\Lambda$ and $\alpha$ in the equilibrium distribution, namely 
\begin{align}
    \bm{F}(\bm{X}) \equiv \overline{\overline{M}}\bm{\nabla} V(\bm{X}) \mapsto \bm{F}_{\text{reg}}(\bm{X}) =\overline{\overline{D}} \bm{F} (\bm{X}) + \frac{1}{2}(1 - \alpha) \nabla\cdot \overline{\overline{\Pi}}{}^2 + c \left( \alpha - \frac{1}{2}\right) \nabla\cdot (\overline{\overline{\Pi}}{}^t \overline{\overline{\Pi}}) - \alpha c\overline{\overline{\Pi}}\cdot\nabla ^t\overline{\overline{\Pi}}. \label{eq:FPE_ForceReplacement_Rule}
\end{align}
Indeed, injecting this ``regularized'' force into Eq.~\ref{eq:EffectiveForce_GenericLambda} yields $\widetilde{\bm{F}}_{\text{reg}}(\bm{X}) = \bm{F}(\bm{X})$.
In other words, this replacement ensures that the equilibrium distribution is given by a Boltzmann distribution using the work of $F$ as its energy.

Replacing the deterministic part in the action yields the new expressions
\begin{align}
    \mathcal{A}_{\text{det}'} &= \ln p_0 (\bm{X}_0) + i  \int\limits_{0}^T dt \bm{Q}(t)\cdot \left[ \overline{\overline{D}}(\bm{X}) \,\overline{\overline{M}} \bm{\nabla} V(\bm{X}) \right] \\
    \mathcal{A}_{\text{diss}} &=  \int\limits_{0}^T dt \left(- i \bm{Q}(t)\cdot\dot{\bm{X}}(t) -D_0 Q(t)^2 -\frac{1}{2}\left( \bm{Q}(t) \cdot \overline{\overline{\Pi}} \right)^2 - \alpha \left( i\bm{Q}(t) \cdot \overline{\overline{\Pi}} \right)\cdot \nabla\overline{\overline{\Pi}}\right) \nonumber \\
    &\hphantom{aa}+ i \int\limits_{0}^T dt \bm{Q}(t) \cdot \left( \frac{1}{2}(1 - \alpha) \nabla\cdot \overline{\overline{\Pi}}{}^2 + c \left( \alpha - \frac{1}{2}\right) \nabla\cdot (\overline{\overline{\Pi}}{}^t \overline{\overline{\Pi}}) - \alpha c\overline{\overline{\Pi}}\cdot\nabla ^t\overline{\overline{\Pi}} \right) \\
    \mathcal{A}_{\text{jac}} &= -\alpha \int\limits_{0}^T dt \left( \bm{\nabla} \left(\overline{\overline{D}}(\bm{X}) \, \overline{\overline{M}}\bm{\nabla}  V(\bm{X},t)\right)  + \frac{\alpha c}{2} \nabla \overline{\overline{\Pi}} {}^t\nabla \overline{\overline{\Pi}}\right) \nonumber \\
    &\hphantom{aa} - \alpha \int\limits_{0}^T dt \bm{\nabla}\left( \frac{1}{2}(1 - \alpha) \nabla\cdot \overline{\overline{\Pi}}{}^2 + c \left( \alpha - \frac{1}{2}\right) \nabla\cdot (\overline{\overline{\Pi}}{}^t \overline{\overline{\Pi}}) - \alpha c\overline{\overline{\Pi}}\cdot\nabla ^t\overline{\overline{\Pi}}\right)
\end{align}
After some algebra, these equations can be rewritten as
\begin{align}
    \mathcal{A}_{\text{det}'} &= \ln p_0 (\bm{X}_0) + i  \int\limits_{0}^T dt \bm{Q}(t)\cdot \left[ \overline{\overline{D}}(\bm{X}) \,\overline{\overline{M}} \bm{\nabla} V(\bm{X}) \right] \label{eq:Adet'_res_replaced} \\
    \mathcal{A}_{\text{diss}} &=  \int\limits_{0}^T dt \left(- i \bm{Q}(t)\cdot\dot{\bm{X}}(t) -D_0 Q(t)^2 -\frac{1}{2}\left( \bm{Q}(t) \cdot \overline{\overline{\Pi}} \right)^2 +(1- 2\alpha) \left( i\bm{Q}(t) \cdot \overline{\overline{\Pi}} \right)\cdot \nabla\overline{\overline{\Pi}}\right) \nonumber \\
    &\hphantom{aa}+ i c \int\limits_{0}^T dt \bm{Q}(t) \cdot \left(  \left( \alpha - \frac{1}{2}\right) \nabla\cdot (\overline{\overline{\Pi}}{}^t \overline{\overline{\Pi}}) - \alpha \overline{\overline{\Pi}}\cdot\nabla ^t\overline{\overline{\Pi}} \right)\label{eq:Amul_diss_replaced} \\
    \mathcal{A}_{\text{jac}} &= -\alpha \int\limits_{0}^T dt \left( \bm{\nabla} \left(\overline{\overline{D}}(\bm{X}) \, \overline{\overline{M}}\bm{\nabla}  V(\bm{X},t)\right)  + (1-\alpha) \left(  (\nabla \overline{\overline{\Pi}})^2 +  \overline{\overline{\Pi}} \nabla^2 \overline{\overline{\Pi}}\right) \right) \nonumber \\
    &\hphantom{aa} - \alpha c \int\limits_{0}^T dt \left( (2 \alpha - 1) \left( \nabla \overline{\overline{\Pi}}\cdot\nabla ^t\overline{\overline{\Pi}}  - \frac{1}{2} {}^t\overline{\overline{\Pi}} \nabla^2 \overline{\overline{\Pi}}\right) - \frac{1}{2} \left(\alpha \nabla \overline{\overline{\Pi}}\cdot\nabla ^t\overline{\overline{\Pi}} + \overline{\overline{\Pi}} \nabla^2 {}^t\overline{\overline{\Pi}} \right) \right)\label{eq:Amul_jac_replaced}
\end{align}

A few comments on these expressions are in order.
First, note that in the general case of non-symmetric $\overline{\overline{\Pi}}$ and $c\neq 0$, no single choice of $\alpha$ will lead to a dissipative part of the action that reduces to the additive-noise part.
In other words, $c$ and the lack of reciprocity in the intensity of the pairwise multiplicative noise matrix seemingly lead to a breaking of the time-reversal symmetry.
Second, in the case $c = 0$, one recovers that the Stratonovich convention makes the dissipative part equal to that of the additive-noise case, as expected from Sec.~\ref{sec:Special_Spatial_Cases} and past work on multiplicative noise~\cite{GonzalezArenas2012,ArnoulxdePirey2022}.
Finally, in the case of symmetric $\overline{\overline{\Pi}}$, the expressions simplify considerably, leading to
\begin{align}
    \mathcal{A}_{\text{det}'} &= \ln p_0 (\bm{X}_0) + i  \int\limits_{0}^T dt \bm{Q}(t)\cdot \left[ \overline{\overline{D}}(\bm{X}) \,\overline{\overline{M}} \bm{\nabla} V(\bm{X}) \right] \label{eq:Adet'_res_symmetric} \\
    \mathcal{A}_{\text{diss}} &=  \int\limits_{0}^T dt \left(- i \bm{Q}(t)\cdot\dot{\bm{X}}(t) -D_0 Q(t)^2 -\frac{1}{2}\left( \bm{Q}(t) \cdot \overline{\overline{\Pi}} \right)^2 +(1- 2\alpha + c (\alpha - 1)) \left( i\bm{Q}(t) \cdot \overline{\overline{\Pi}} \right)\cdot \nabla\overline{\overline{\Pi}}\right)\label{eq:Amul_diss_symmetric} \\
    \mathcal{A}_{\text{jac}} &= -\alpha \int\limits_{0}^T dt \left( \bm{\nabla} \left(\overline{\overline{D}}(\bm{X}) \, \overline{\overline{M}}\bm{\nabla}  V(\bm{X},t)\right)  + \frac{1}{2}(1-\alpha)(1-c) \nabla\cdot  \overline{\overline{\Pi}}{}^2+ \frac{\alpha c}{2}  (\nabla \overline{\overline{\Pi}})^2 \right)\label{eq:Amul_jac_symmetric}
\end{align}
In particular, it is interesting to note that the extra dissipative term due to multiplicative noise can now be cancelled for $c \neq 0$ for a suitable $\alpha_c$ such that
\begin{align}
    1 - 2 \alpha_c + c (\alpha_c - 1) = 0
\end{align}
or, equivalently,
\begin{align}
    \alpha_c = \frac{c-1}{c-2}.
\end{align}
Note a few special cases,
\begin{itemize}
    \item [(i)] For $c = 0$, one recovers $\alpha_c = 1/2$ (Stratonovich convention), like for usual multiplicative noise,
    \item [(ii)] For $c=1$, one finds $\alpha_c = 0$ (It\={o} convention), 
    \item [(iii)] For $c = -1$, one finds $\alpha_c = 2/3$, which is not a named convention.
\end{itemize}
This highlights how $c$ modifies time-reversal properties of the dynamics independently of both the symmetry of $\overline{\overline{\Pi}}$ and $\alpha$.
We establish in Sec.~\ref{sec:TRS_Spatial} that \textit{no} joint choice of $\alpha$ and $c \neq 0$ results in a full restoration of time-reversal symmetry, so that these dynamics are always unequivocally irreversible.

Finally, note that the dynamical generating function $Z_d$ and the full MSRJD action are conventionally defined with extra source terms, so that
\begin{align}
    Z_d[\bm{J}_{\bm{X}}, \bm{J}_{\bm{Q}}] &= \frac{1}{\mathcal{N}} \int d^{dN}\bm{X}_0 \mathcal{D}\bm{X} \mathcal{D}\bm{Q} e^{\mathcal{A}_{MSRJD}[\bm{X}, \bm{Q}, \bm{J}_{\bm{X}}, \bm{J}_{\bm{Q}} | \bm{X}_0]} \label{eq:Dynamical_Generating_Function} \\
    \mathcal{A}_{MSRJD}[\bm{X}, \bm{Q}, \bm{J}_{\bm{X}}, \bm{J}_{\bm{Q}} | \bm{X}_0] &\equiv \mathcal{A}[\bm{X}, \bm{Q} | \bm{X}_0] + \int\limits_{0}^T dt' \left[ \bm{J}_{\bm{X}}( t') \cdot \bm{X}(t') + i \bm{J}_{\bm{Q}}(t')\cdot\bm{Q}(t') \right] . \label{eq:MSRJD_Action}
\end{align}
Including these extra source terms is in particular useful to define and compute response functions, just like in usual statistical mechanics.
In the case $c=0$, we show in Sec.~\ref{sec:TRS_Spatial} that this link lets us derive a generalized Fluctuation-Dissipation Theorem when $\Lambda_{ij} \neq 0$.

\subsubsection{OM Action and Entropy Production~\label{sec:Spatial_OM_EPR}}

We now establish the expression of the Onsager-Machlup (OM) action rather than the MSRJD one.
It is obtained by integrating over the response field in the path-integral, which may be achieved as the integral is always Gaussian.
Indeed, the relevant integral is of the form
\begin{align}
   &\int \mathcal{D}\bm{Q} \exp\left[\int\limits_{0}^T dt \left(-\bm{Q}_t \overline{\overline{D}} {}^t \bm{Q}_t + i \bm{Q}_t \cdot \bm{b}(\bm{X}_t,\dot{\bm{X}}_t,\alpha)\right)\right] \nonumber \\
   &\hphantom{a}= \prod\limits_{0\leq t \leq T} \int d\bm{Q}_t \exp\left[-\bm{Q}_t \overline{\overline{D}} {}^t \bm{Q}_t + i \bm{Q}_t \cdot\bm{b}(\bm{X}_t,\dot{\bm{X}}_t,\alpha) \right] \\
   &\hphantom{a}= \prod\limits_{0\leq t \leq T} \exp \left[ -\frac{1}{4} {}^{t}\bm{b}(\bm{X}_t,\dot{\bm{X}}_t,\alpha)\overline{\overline{D}}{}^{-1} \bm{b}(\bm{X}_t,\dot{\bm{X}}_t,\alpha)  + \frac{1}{2}\ln \det\left(\pi \overline{\overline{D}}{}^{-1}\right)\right]
\end{align}
This expression can be reinjected into the full generating function, Eqs.~\ref{eq:Adet'_res_replaced},~\ref{eq:Amul_diss_replaced},~\ref{eq:Amul_jac_replaced} to extract the Onsager-Machlup action,
\begin{align}
    \mathcal{A}_{\text{OM}} &= \ln p_0 (\bm{X}_0) +\mathcal{A}_{\text{jac}} + \int\limits_{0}^T dt \left[ -\frac{1}{4} {}^{t}\bm{b}_t^{(\alpha)}\overline{\overline{D}}{}^{-1} \bm{b}_t^{(\alpha)}  + \frac{1}{2}\ln \det\left(\pi \overline{\overline{D}}{}^{-1}\right)\right]\label{eq:AOM} \\
    \bm{b}_t^{(\alpha)} &\equiv \overline{\overline{D}}\,\overline{\overline{M}} \nabla V - \dot{\bm{X}}_t +(1-2\alpha + c(\alpha -1)) \overline{\overline{\Pi}}\nabla \overline{\overline{\Pi}} \label{eq:AOM_bDefinition}\\
    \mathcal{A}_{\text{jac}} &= -\alpha \int\limits_{0}^T dt \left( \bm{\nabla} \left(\overline{\overline{D}}(\bm{X}) \, \overline{\overline{M}}\bm{\nabla}  V(\bm{X},t)\right)  + (1-\alpha) \left(  (\nabla \overline{\overline{\Pi}})^2 +  \overline{\overline{\Pi}} \nabla^2 \overline{\overline{\Pi}}\right) \right) \nonumber \\
    &\hphantom{aa} - \alpha c \int\limits_{0}^T dt \left( (2 \alpha - 1) \left( \nabla \overline{\overline{\Pi}}\cdot\nabla ^t\overline{\overline{\Pi}}  - \frac{1}{2} {}^t\overline{\overline{\Pi}} \nabla^2 \overline{\overline{\Pi}}\right) - \frac{1}{2} \left(\alpha \nabla \overline{\overline{\Pi}}\cdot\nabla ^t\overline{\overline{\Pi}} + \overline{\overline{\Pi}} \nabla^2 {}^t\overline{\overline{\Pi}} \right) \right)\label{eq:Amul_jac_OM}
\end{align}
which verifies
\begin{align}
    1 = \frac{1}{\mathcal{N}} \int d^{dN}\bm{X}_0 \mathcal{D}\bm{X} e^{\mathcal{A}_{OM}[\bm{X}|\bm{X}_0]}
\end{align}
so that the probability of a path conditional on the initial condition is such that
\begin{align}
    \mathbb{P}[\bm{X} | \bm{X}_0] \propto e^{\mathcal{A}_{OM}[\bm{X}| \bm{X}_0]}.
\end{align}
In particular, the ratio between the probability of a path in the forward direction and that of the same path but time-reversed is readily expressed as
\begin{align}
    \frac{\mathbb{P}[\mathcal{T}\bm{X} | \bm{X}_T]}{\mathbb{P}[\bm{X} | \bm{X}_0]} = \exp\left[ \mathcal{T}\mathcal{A}_{OM}[\bm{X}|\bm{X}_0] - \mathcal{A}_{OM}[\bm{X}|\bm{X}_0]\right].
\end{align}

Keeping only the terms that do not trivially vanish by TRS and shifting/rescaling the time interval following $[0;T] \mapsto [-T, T]$ for convenience, the previous equation reads
\begin{align}
    \frac{\mathbb{P}[\mathcal{T}\bm{X} | \bm{X}_T]}{\mathbb{P}[\bm{X} | \bm{X}_{-T}]} &= \mathcal{J}\exp\left[ \ln p_0 (\bm{X}_T) - \ln p_0(\bm{X}_{-T}) + \int\limits_{T}^{-T} dt \left[ -\frac{1}{4} {}^{t}\bm{b}_{-t}^{(1-\alpha)}\overline{\overline{D}}{}^{-1} \bm{b}_{-t}^{(1-\alpha)} \right] - \int\limits_{-T}^T dt \left[ -\frac{1}{4} {}^{t}\bm{b}_{t}^{(\alpha)}\overline{\overline{D}}{}^{-1} \bm{b}_{t}^{(\alpha)} \right]\right].
\end{align}
In that expression, $\mathcal{J}$ is the part that comes from the Jacobian part of the action, and
\begin{align}
    b_{-t}^{(1-\alpha)} = \overline{\overline{D}}\,\overline{\overline{M}} \nabla V - d_t\bm{X}_{-t} -(1-2\alpha + \alpha c) \overline{\overline{\Pi}}\nabla \overline{\overline{\Pi}}
\end{align}
so that
\begin{align}
    \frac{\mathbb{P}[\mathcal{T}\bm{X} | \bm{X}_T]}{\mathbb{P}[\bm{X} | \bm{X}_{-T}]} &= \mathcal{J}\exp\left[ \ln p_0 (\bm{X}_T) - \ln p_0(\bm{X}_{-T}) + \int\limits_{-T}^{T} dt \left[ -\frac{1}{4} {}^{t}\widetilde{\bm{b}}_{t}^{(\alpha)}\overline{\overline{D}}{}^{-1} \widetilde{\bm{b}}_{t}^{(\alpha)} \right] - \int\limits_{-T}^T dt \left[ -\frac{1}{4} {}^{t}\bm{b}_{t}^{(\alpha)}\overline{\overline{D}}{}^{-1} \bm{b}_{t}^{(\alpha)} \right]\right]
\end{align}
where
\begin{align}
    \widetilde{\bm{b}}_{t}^{(\alpha)} &= \overline{\overline{D}}\,\overline{\overline{M}} \nabla V + \dot{\bm{X}}_{t} -(1-2\alpha + \alpha c) \overline{\overline{\Pi}}\nabla \overline{\overline{\Pi}}.
\end{align}
Choosing the Stratonovich convention for simplicity (which also makes $\mathcal{J} = 1$), one has
\begin{align}
    \widetilde{\bm{b}}_{t}^{(\frac{1}{2})} &= \overline{\overline{D}}\,\overline{\overline{M}} \nabla V  -\frac{c}{2} \overline{\overline{\Pi}}\nabla \overline{\overline{\Pi}}+ \dot{\bm{X}}_{t}, \\
    {\bm{b}}_{t}^{(\frac{1}{2})} &= \overline{\overline{D}}\,\overline{\overline{M}} \nabla V  -\frac{c}{2} \overline{\overline{\Pi}}\nabla \overline{\overline{\Pi}}- \dot{\bm{X}}_{t}.
\end{align}
As a result, only cross-terms mixing $\dot{X}$ with the rest of $b_t$ survive. 
As a result, the expression of the entropy production simplifies to
\begin{align}
    \frac{\mathbb{P}[\mathcal{T}\bm{X} | \bm{X}_T]}{\mathbb{P}[\bm{X} | \bm{X}_{-T}]} &= \exp\left[ \ln p_0 (\bm{X}_T) - \ln p_0(\bm{X}_{-T}) -\frac{1}{2} \int\limits_{-T}^{T} dt \left[ {}^t\hspace{-0.1cm}\left(\overline{\overline{D}}\,\overline{\overline{M}} \nabla V  -\frac{c}{2} \overline{\overline{\Pi}}\nabla \overline{\overline{\Pi}}\right)\overline{\overline{D}}{}^{-1}  \dot{\bm{X}}_t  + {}^t\dot{\bm{X}}_t\overline{\overline{D}}{}^{-1}  \hspace{-0.1cm}\left(\overline{\overline{D}}\,\overline{\overline{M}} \nabla V  -\frac{c}{2} \overline{\overline{\Pi}}\nabla \overline{\overline{\Pi}}\right)\right] \right].
\end{align}
Recalling that $\overline{\overline{D}}$, and therefore its inverse, are both symmetric within the current assumptions (see Eq.~\ref{eq:D_general_definition_RO}), this can be further simplified to
\begin{align}
    \frac{\mathbb{P}[\mathcal{T}\bm{X} | \bm{X}_T]}{\mathbb{P}[\bm{X} | \bm{X}_{-T}]} &= \exp\left[ \ln p_0 (\bm{X}_T) - \ln p_0(\bm{X}_{-T}) - \int\limits_{-T}^{T} dt \left[ {}^t\dot{\bm{X}}_t\overline{\overline{D}}{}^{-1}  \hspace{-0.1cm}\left(\overline{\overline{D}}\,\overline{\overline{M}} \nabla V  -\frac{c}{2} \overline{\overline{\Pi}}\nabla \overline{\overline{\Pi}}\right)\right] \right].
\end{align}
Assuming that the initial condition is drawn from the steady-state (Boltzmann) distribution, the first three terms cancel out, leaving only
\begin{align}
    \frac{\mathbb{P}[\mathcal{T}\bm{X} | \bm{X}_T]}{\mathbb{P}[\bm{X} | \bm{X}_{-T}]} &= \exp\left[\frac{c}{2} \int\limits_{-T}^{T} dt \left[ {}^t\dot{\bm{X}}_t\overline{\overline{D}}{}^{-1}  \overline{\overline{\Pi}}\nabla \overline{\overline{\Pi}}\right] \right].
\end{align}
Introducing the definition of entropy production $\Delta S$~\cite{Seifert2005,Seifert2012},
\begin{align}
\frac{\mathbb{P}[\mathcal{T}\bm{X} | \bm{X}_T]}{\mathbb{P}[\bm{X} | \bm{X}_{-T}]} &\equiv e^{-\Delta S(T)},
\end{align}
this expression yields an expression for an entropy production rate $\sigma(t)$ defined by
\begin{align}
    \Delta S(T) \equiv \int\limits_{-T}^T dt \sigma(t),
\end{align}
which here reads
\begin{align}
    \sigma(t) = -\frac{c}{2}{}^t\dot{\bm{X}}_t\overline{\overline{D}}{}^{-1}  \overline{\overline{\Pi}}\nabla \overline{\overline{\Pi}}.
\end{align}
In particular, time-reversal symmetry, corresponding to $\sigma = 0$, is recovered for $c = 0$.

The complexity of the entropy production rate is almost entirely hidden in the inverse of $\overline{\overline{D}}$.
In particular, consider a simple example given in Sec.~\ref{sec:Examples}, namely the case $\Lambda_{ij}(\bm{X}) = \Lambda_0(\bm{X})$ in $d=1$.
In that case, see Eq.~\ref{eq:InverseD_OneLambda0}, in the limit $D_0 = 0$, the elements of the inverse of $\overline{\overline{D}}{}^{-1}$ diverge as $c \to -1$, so that in the limit that was used to generate hyperuniformity in Ref.~\cite{Anand2025} the entropy production diverges.

More generally, consider the case of a symmetric $\Lambda_{ij}$ and no projector, given in Eq.~\ref{eq:D_general_definition_RO}, which we here remind the reader of,
\begin{align}
    D_{iiab} &= \delta_{ab} \left( D_0 + \frac{1}{2}\sum\limits_{j \neq i} \Lambda_{ij}(\bm{X})\right), \\
    D_{ijab} &\underset{i\neq j}{=} \delta_{ab} \frac{c}{2} \Lambda_{ij}(\bm{X}).
\end{align}
As previously noted, in the case $c = -1$,  $\overline{\overline{D}}/D_0$ is doubly stochastic.
As a result, in the limit $D_0 \to 0$, the sum of rows and columns of $\overline{\overline{D}}{}^{-1}$ diverges, so does the entropy production rate, and the matrix eventually becomes non-invertible.
Note however that all other $c\in (-1;1]\setminus\{0\}$ seem well-behaved at this level, even for $D_0 = 0$.
In particular, $c = 1$ is seemingly unremarkable -- except possibly in some very specific cases, like $D_0=0$, $N=2$ in Eq.~\ref{eq:InverseD_OneLambda0}.

Finally, in the most general case of a non-$ij$-symmetric $\Lambda_{ijab}$ with a projector, the same calculation but with more cumbersome terms eventually yields the result given in main text,
\begin{align}
    \sigma_0(t) &= -\frac{c}{2}{}^t\dot{\bm{X}}_t\overline{\overline{D}}{}^{-1}  \bm{\Phi}
\end{align}
with, as noted in Eq.~\ref{eq:D_general_definition_proj_simpler}
\begin{align}
    D_{iiab} &= D_0 \delta_{ab} + \frac{1}{2} \sum\limits_{k\neq i}  \Lambda_{ik} P_{ab} \nonumber \\
    D_{ijab} &=\frac{c}{2} \sqrt{\Lambda_{ij}\Lambda_{ji}} P_{ab} \hphantom{aa} \text{(if $j\neq i$)}. 
\end{align}
and where we introduced an effective force
\begin{align}
    \Phi_{ia} &\equiv \sum\limits_{j\neq i} \sum\limits_{b,p=1}^d \partial_{x_{j,b}} \left[\left(\sqrt{\Lambda_{ijap}\Lambda_{jibp}}\right)  \right].
\end{align}

\subsection{Temporal noise \label{sec:Temporal_EP_Analytical}}

We consider interacting passive particles immersed in an active bath following discrete-time dynamics (see Methods section in the main text). It was recently shown that the discrete-time dynamics of such systems can be approximated by a continuous-time stochastic differential equation (SDE), which is our starting point \cite{Anand2026}.

\subsubsection{Dynamics}

We now consider the temporal dynamics described by the continuous SDE
\begin{align}
    \dot{\bm{r}}_i(t) &= -\mu_i \sum\limits_{j \neq i} \bm{\nabla}_i V(\bm{r}_{ij}) + \sqrt{2 D_0} \bm{\eta}_i(\bm{\theta},t) \equiv U_i\left[\left\{ \bm{r}_{ij}\right\}, \bm{\eta}_i\right], \label{eq:PGD_dynamics}
\end{align}
where $D_0$ is a diffusion constant associated to a thermal bath, and the noise terms $\bm{\eta}_i$ verify
\begin{align}
    \left\langle \eta_{i}^a (\bm{\theta},t)  \right\rangle &= 0  \\
    \left\langle \eta_{i}^a (\bm{\theta},t) \eta_{j}^b (\bm{\theta},t') \right\rangle &= \delta_{ab} \delta_{ij} \Gamma(\bm{\theta},t - t'), \label{eq:NoiseCorrelations}
\end{align}
with $\Gamma$ a memory kernel that here takes the form
\begin{align}
    \Gamma(\bm{\theta},t-t') &= \delta(t-t') + \frac{1}{\theta^2} \sum\limits_{m = 1}^M\sum\limits_{n = 0}^{M-m}\theta_{n}\theta_{n+m} \left(\delta(t-t' - m \tau) + \delta(t - t'+m\tau) \right) \label{eq:MemoryKernel}
\end{align}
where the memory consists of $M$ off-zero Dirac-delta peaks and $\theta_0 = 1$.
Each peak at $m>0$ has an amplitude $\theta_m \in\mathbb{R}$, and the sum is normalized by 
\begin{align}
    \theta^2 = 1 +\sum\limits_{m=1}^M \theta_m^2. \label{eq:ThetaNormalization}
\end{align}
If $\theta_k = 0$ for all $m$, the memory reduces to the usual Dirac delta associated with Gaussian white noise.
It is convenient to introduce the coefficients $c_m$ defined for $1\leq m \leq M$ by
\begin{align}
    c_m \equiv \frac{1}{\theta^2} \sum\limits_{n=0}^{M-m} \theta_{n} \theta_{n+m}.
\end{align}
The memory kernel may then be re-expressed in terms of the list of $c_m$,
\begin{align}
    \Gamma(\bm{\theta},t-t') &= \delta(t-t') +  \sum\limits_{m = 1}^M c_m(\bm{\theta}) \left(\delta(t-t' - m \tau) + \delta(t - t'+m\tau) \right) \label{eq:MemoryKernel_cm}
\end{align}
A case of particular interest within the context of Perturbed Gradient Descent (PGD) is that of $M = 1$ where
\begin{align}
    \Gamma(\theta_1,t-t') &= \delta(t-t') + c_1(\theta_1) \left(\delta(t-t' - \tau) + \delta(t - t'+\tau) \right).
\end{align}
For instance, $\theta_1 = \pm1$ yields the simple example 
\begin{align}
    \Gamma(\pm1,t-t') &= \delta(t-t') \pm \frac{1}{2} \left(\delta(t-t' - \tau) + \delta(t - t'+\tau) \right),
\end{align}
where $c_1$ saturates its minimal and maximal values $\pm 1/2$, which may be used as a simple illustrative example.

Since the process is non-Markovian, note that one may not establish a Fokker-Planck equation. 
Alternatives exist for simple enough choices of Gaussian but non-Markovian processes~\cite{Hänggi1995} but they do not easily extend to this case.

\subsubsection{MSRJD Action and Time-Reversal Symmetry}

By symmetry with the case of spatial noise, we here explicitly specify a discretization convention $\alpha$ for the discretized SDE.
Following the same first steps as in the case of spatial noise, we can again write an action $\mathcal{A}$ that is split into three components -- a deterministic part, a part due to noise, and a part due to the Jacobian
\begin{align}
    \mathcal{A}\left[\bm{X}, \dot{\bm{X}}, \bm{Q}; \bm{X}_0 \right] &= \mathcal{A}_{det}\left[\bm{X}, \dot{\bm{X}}, \bm{Q}; \bm{X}_0 \right] + \mathcal{A}_{\text{add}}\left[\bm{X}, \dot{\bm{X}}, \bm{Q} \right] + \mathcal{A}_{\text{jac}}\left[\bm{X}, \dot{\bm{X}}, \bm{Q} \right]\label{eq:Action_decomp_temporal}
\end{align}
with
\begin{align}
    e^{\mathcal{A}_{\text{det}}} &= \exp\left[\ln p_0 (\bm{X}_0)- i  \int\limits_{0}^T dt \bm{Q}(t)\cdot \left[\dot{\bm{X}}(t) - \overline{\overline{M}} \bm{\nabla} V(\bm{X}) \right)\right], \label{eq:Adet_def_temporal} \\
    e^{\mathcal{A}_{\text{add}}} &= \int  \prod\limits_{n=1}^N\left[\mathcal{D}\bm{\eta}_n p_\eta(\bm{\eta}_n) \right] \exp \left[  i \sqrt{2 D_0}\int\limits_0^T dt   \bm{Q}(t) \cdot  \bm{\eta}(t) \right], \label{eq:Aadd_def_temporal}\\
    e^{\mathcal{A}_{\text{jac}}} &=  \exp \left[ - \int\limits_0^T dt \left[ \alpha\text{Tr}\frac{\delta \bm{U}(t) }{\delta \bm{X}(t)}   \right]\right]. \label{eq:Ajac_def_temporal}
\end{align}

To progress further, one must express the distribution of the noise that appears in the path integral.
Here, it is crucial to note that it is non-Markovian yet Gaussian, so that the functional probability of a noise history for one particle and one cartesian component of the noise reads
\begin{align}
    p_\eta(\eta_n^a) &= \frac{1}{\mathcal{N}} \exp\left[-\frac{1}{2}\iint\limits_{[0,T]^2} dt dt' \eta_{n}^a(t) \Gamma^{-1}(\bm{\theta},t - t')\eta_{n}^a(t') \right]
\end{align}
where we introduce the inverse kernel $\Gamma^{-1}$ such that
\begin{align}
    \int\limits dt'' \Gamma(\bm{\theta},t-t'')\Gamma^{-1}(\bm{\theta}, t''- t') = \delta(t - t').
\end{align}
Note that, using the expression of the memory kernel in Eq.~\ref{eq:MemoryKernel}, the inverse has to verify
\begin{align}
        \delta(t - t') &= \int\limits dt'' \left[ \delta(t - t'') + \sum\limits_{m = 1}^M c_m(\bm{\theta}) \left(\delta(t - t'' - m \tau) + \delta(t - t'' + m \tau)\right) \right]\Gamma^{-1}(\bm{\theta}, t''- t') \\
        &= \Gamma^{-1}(\bm{\theta},t - t') + \sum\limits_{m=1}^M c_m (\bm{\theta}) (\Gamma^{-1}(\bm{\theta}, t - t'- m \tau ) + \Gamma^{-1}(\bm{\theta}, t - t'+ m \tau )). \label{eq:GammaInverse_Condition}
\end{align}
The inverse can be assumed to be expressed following the ansatz
\begin{align}
    \Gamma^{-1}(\bm{\theta}, t - t') = \sum\limits_{n\in \mathbb{Z}} b_n \delta(t - t'- n\tau).
\end{align}
Injecting this ansatz into Eq.~\ref{eq:GammaInverse_Condition} yields
\begin{align}
    \delta(t - t') = \sum\limits_{n\in \mathbb{Z}} b_n \delta(t - t'- n\tau) + \sum\limits_{m=1}^M\sum\limits_{n = -\infty}^\infty c_m (\bm{\theta}) b_n (\bm{\theta}) \left(\delta(t-t'-(n+ m)\tau) +\delta(t-t'-(n- m)\tau)\right) \label{eq:InverseGamma_ExplicitSum}
\end{align}
This expression is akin to an identity between polynomials: the only way to verify it is to enforce that the prefactor of the $\delta(t-t')$ on the right-hand-side is 1, while all others must be zero.
Thus, it is useful to group up terms according to the delay in the $\delta$,
\begin{align}
     \delta(t - t') = \left(b_0 + \sum\limits_{m=1}^M c_m (b_{-m} + b_{m}) \right)\delta(t - t') + \sum\limits_{p\in \mathbb{Z}^\star}\left[ b_p + \sum\limits_{m=1}^{M} c_m (b_{p-m} + b_{p+m})\right]  \delta(t - t'- p\tau),
\end{align}
which yields a first condition from the first term
\begin{align}
    b_0 + \sum\limits_{m=1}^M c_m (b_{-m} + b_{m}) &= 1 \label{eq:InverseCondition1}
\end{align}
then extra conditions that apply to all $p \in \mathbb{Z}^\star$,
\begin{align}
    b_p +\sum\limits_{m = 1}^M c_m (b_{p-m} + b_{p+m}) &= 0.\label{eq:InverseCondition2}
\end{align}
Note that both conditions can be written in a compact form as
\begin{align}
    b_p +\sum\limits_{m = 1}^M c_m (b_{p-m} + b_{p+m}) &= \delta_{p,0}.
\end{align}
In Sec.~\ref{sec:InvertingGamma} we establish more general expressions for the inverse of $\Gamma$.

The integral over noise is then a Gaussian integral,
\begin{align}
    e^{\mathcal{A}_{\text{add}}} &= \prod\limits_{n=1}^N\prod\limits_{a=1}^d \int  \left[\prod\limits_{t = 0}^T d\eta^a_n(t)\right] \exp \left[ -\frac{1}{2} \iint\limits_{[0;T]^2} dt dt'\eta_{n}^a(t) \Gamma^{-1}(\bm{\theta},t - t')\eta_{n}^a(t')  + i \sqrt{2 D_0}\int\limits_0^T dt   q_{n,a}(t) \cdot  \eta_n^a(t) \right].
\end{align}
Using the fact that
\begin{align}
    \int d\bm{X} e^{-\frac{1}{2}\bm{X}\overline{\overline{A}}{}^t\bm{X} + i{}^t\bm{B}\cdot\bm{X}} = \sqrt{\det(2\pi \overline{\overline{A}}{}^{-1} )}\,e^{-\frac{1}{2}^t\bm{B}\overline{\overline{A}}{}^{-1}\bm{B} },
\end{align}
this yields
\begin{align}
        e^{\mathcal{A}_{\text{add}}} &= \prod\limits_{n=1}^N\prod\limits_{a=1}^d \sqrt{\det_{tt'} (2\pi \Gamma(\bm{\theta}, t-t'))} \exp \left[ -D_0 \iint\limits_{[0;T]^2} dt dt'q_{n}^a(t) \Gamma(\bm{\theta},t - t')q_{n}^a(t') \right].
\end{align}
The determinant may be computed by noticing that
\begin{align}
    \det A = \exp \text{Tr} \ln A
\end{align}
and that 
\begin{align}
    \Gamma = \mathbb{1} + \Delta
\end{align}
(where here the identity means $\delta(t - t')$), so that
\begin{align}
    \det \Gamma \approx \exp \text{Tr} \Delta.
\end{align}
By definition, $\Delta$ is traceless, so that $\det \Gamma \approx 1$ to leading order (the next order yields $\text{Tr}\Delta^2 /2 = \theta^2/2$, so that following orders are simply constants anyway).
As a result, the determinant may be absorbed into the normalization to leading order, so that
\begin{align}
        e^{\mathcal{A}_{\text{add}}} &=  \exp \left[ -D_0 \iint\limits_{[0;T]^2} dt dt'\bm{Q}(t) \Gamma(\bm{\theta},t - t')\bm{Q}(t') \right]. \label{eq:Noise_Action_Result}
\end{align}

In summary,
\begin{align}
    \mathcal{A}_{\text{det}} &= \ln p_0 (\bm{X}_0)- i  \int\limits_{0}^T dt \bm{Q}(t)\cdot \left[\dot{\bm{X}}(t) - \overline{\overline{M}} \bm{\nabla} V(\bm{X}) \right], \label{eq:Adet_res_temporal} \\
    \mathcal{A}_{\text{add}} &=  -D_0 \iint\limits_{[0;T]^2} dt dt'\bm{Q}(t) \Gamma(\bm{\theta},t - t')\bm{Q}(t') \label{eq:Aadd_res_temporal}\\
    \mathcal{A}_{\text{jac}} &= -\alpha\int\limits_{0}^T dt \overline{\overline{M}}\bm{\nabla}^2 V(\bm{X},t)\label{eq:Ajac_res_temporal}
\end{align}
Conventionally, one generally regroups terms differently by defining the dissipative part of the action (that contains all terms that relate to interactions with the bath) so that
\begin{align}
    \mathcal{A}_{\text{det}+\text{jac}} &= \ln p_0 (\bm{X}_0) + i  \int\limits_{0}^T dt \bm{Q}(t)\cdot \left[ \overline{\overline{M}} \bm{\nabla} V(\bm{X}) \right] -\alpha\int\limits_{0}^T dt \overline{\overline{M}}\bm{\nabla}^2 V(\bm{X},t), \label{eq:Adetjac_res_temporal} \\
    \mathcal{A}_{\text{diss}} &=   \int\limits_{0}^T dt (-i \bm{Q}(t))\left[ \dot{\bm{X}}(t) + D_0 \int\limits_{0}^T dt' \Gamma(\bm{\theta},t - t')(-i\bm{Q}(t')) \right]\label{eq:Adiss_res_temporal}.
\end{align}
In particular, for the memory kernel considered in these notes, which may be written compactly as
\begin{align}
    \Gamma(\bm{\theta}, t -t') = \delta(t - t') + \sum\limits_{m\in \mathbb{Z}^\star} c_{|m|}(\bm{\theta}) \delta(t - t'-m\tau),
\end{align}
the dissipative action may be explicitly written as
\begin{align}
    \mathcal{A}_{\text{diss}} &=   \int\limits_{0}^T dt (-i \bm{Q}(t))\left[ \dot{\bm{X}}(t) + D_0 (-i\bm{Q}(t)) + D_0\sum\limits_{m \in \mathbb{Z}^\star} c_{|m|}(\bm{\theta})(-i\bm{Q}(t - m\tau)) \right]\label{eq:Adiss_kernel}.
\end{align}
Because of the fact that dynamics are overdamped, no transformation of the usual type
\begin{align}
    i \bm{Q}_t \mapsto i \bm{Q}_{-t} + D_0^{-1} d_t\bm{X}_{-t}
\end{align}
(where the time-dependence is written as an index) can leave the dissipative part of the action unchanged.
As a result, these dynamics are not time-reversal symmetric, unless $c_m = 0$.
Note that time-reversal symmetry could be recovered in underdamped dynamics with the same noise correlations and a suitably chosen friction kernel, but these (clearly different) dynamics are not the focus of our work.

\subsubsection{OM Action and Entropy Production}

In this section, we establish the expression of the Onsager-Machlup (OM) action rather than the MSRJD one.
It is once again obtained by integrating over the response field in the path-integral, which may be achieved as the integral is always Gaussian.
In the context of coloured additive noise, the relevant integral is of the form
\begin{align}
   &\int \mathcal{D}\bm{Q} \exp\left[\iint\limits_{[0;T]^2} dt dt' \left(-\frac{1}{2}\bm{Q}_t A_{tt'}(\bm{\theta}) \bm{Q}_{t'} + i \bm{Q}_t \cdot \bm{B}_t\right)\right] \nonumber \\
   &\hphantom{a}= \prod\limits_{0\leq t,t' \leq T} \int \mathcal{D}\bm{Q} \exp\left[-\frac{1}{2}\bm{Q}_t A_{tt'}(\bm{\theta}) \bm{Q}_{t'} + i \bm{Q}_t \cdot \bm{B}_t\right] \\
   &\hphantom{a}= \prod\limits_{0\leq t,t' \leq T} \exp \left[ -\frac{1}{2} \bm{B}_tA_{tt'}^{-1} \bm{B}_{t'}  + \frac{1}{2}\ln \det_{tt'}\left(2\pi A_{tt'}^{-1}\right)\right] \\
   &\hphantom{a}= \exp\left[\iint\limits_{[0;T]^2} dt dt' \left( -\frac{1}{2} \bm{B}_tA_{tt'}^{-1} \bm{B}_{t'}\right)+ \frac{1}{2}\ln \det_{tt'}\left(2\pi A_{tt'}^{-1}\right)\right]
\end{align}
Identifying, in the overdamped case and assuming that a single damping coefficient $\gamma_0$ is relevant, 
\begin{align}
    A_{tt'} &= 2 D_0 \Gamma_{tt'} \\
    A_{tt'}^{-1} &= \frac{1}{2D_0} \Gamma_{tt'}^{-1} \\
    \bm{B}_t &= -\frac{1}{\gamma_0}\nabla V(\bm{X})  -\dot{\bm{X}}_t
\end{align}
this expression can be reinjected into the full generating function to extract the Onsager-Machlup action,
\begin{align}
    \mathcal{A}_{OM} &= \ln p_0 (\bm{X}_0) +\frac{\alpha}{\gamma_0} \int\limits_{0}^T dt \nabla^2 V(\bm{X}) +\frac{1}{2}\ln \det\left(\frac{\pi}{D_0} \Gamma_{tt'}^{-1}(\bm{\theta})\right)\nonumber \\
    &\hphantom{aa}+\iint\limits_{[0;T]^2} dt dt'\left[ - \frac{1}{4D_0} \left( \dot{\bm{X}}_t + \frac{1}{\gamma_0}\nabla V(\bm{X}_t) \right)  \Gamma_{tt'}^{-1}(\bm{\theta})\left( \dot{\bm{X}}_{t'} + \frac{1}{\gamma_0}\nabla V(\bm{X}_{t'}) \right) \right],
\end{align}
which verifies
\begin{align}
    1 = \frac{1}{\mathcal{N}} \int d^{dN}\bm{X}_0 \mathcal{D}\bm{X} e^{\mathcal{A}_{OM}[\bm{X}|\bm{X}_0]}
\end{align}
so that the probability of a path conditional on the initial condition is such that
\begin{align}
    \mathbb{P}[\bm{X} | \bm{X}_0] \propto e^{\mathcal{A}_{OM}[\bm{X}| \bm{X}_0]}.
\end{align}
In particular, the ratio between the probability of a path in the forward direction and that of the same path but time-reversed is readily expressed as
\begin{align}
    \frac{\mathbb{P}[\mathcal{T}\bm{X} | \bm{X}_T]}{\mathbb{P}[\bm{X} | \bm{X}_0]} = \exp\left[ \mathcal{T}\mathcal{A}_{OM}[\bm{X}|\bm{X}_0] - \mathcal{A}_{OM}[\bm{X}|\bm{X}_0]\right].
\end{align}

Choosing the Stratonovich convention $\alpha = 1/2$, keeping only the terms that do not trivially vanish by TRS and shifting/rescaling the time interval following $[0;T] \mapsto [-T, T]$ for convenience, the previous equation reads
\begin{align}
    \frac{\mathbb{P}[\mathcal{T}\bm{X} | \bm{X}_T]}{\mathbb{P}[\bm{X} | \bm{X}_{-T}]} &= \exp\left[ \ln p_0 (\bm{X}_T) - \ln p_0(\bm{X}_{-T}) + \iint\limits_{[-T;T]^2} dt dt'\left[ - \frac{1}{4D_0} \left( -\dot{\bm{X}}_t + \frac{1}{\gamma_0}\nabla V(\bm{X}_t) \right)  \Gamma_{tt'}^{-1}(\bm{\theta})\left( - \dot{\bm{X}}_{t'} + \frac{1}{\gamma_0}\nabla V(\bm{X}_{t'}) \right) \right] \right. \nonumber \\
    &\hphantom{aaaaaaaaaaaaaaaaaaaaaaaaaa}- \left.\iint\limits_{[-T;T]^2} dt dt'\left[ - \frac{1}{4D_0} \left( \dot{\bm{X}}_t + \frac{1}{\gamma_0}\nabla V(\bm{X}_t) \right)  \Gamma_{tt'}^{-1}(\bm{\theta})\left( \dot{\bm{X}}_{t'} + \frac{1}{\gamma_0}\nabla V(\bm{X}_{t'}) \right) \right]\right]. \\
    &= \exp\left[ \ln p_0 (\bm{X}_T) - \ln p_0(\bm{X}_{-T}) + \iint\limits_{[-T;T]^2} dt dt'\left[ \frac{1}{2D_0\gamma_0} \left( \dot{\bm{X}}_t \Gamma_{tt'}^{-1}(\bm{\theta}) \nabla V(\bm{X}_{t'}) +   \nabla V(\bm{X}_{t})\Gamma_{tt'}^{-1}(\bm{\theta}) \dot{\bm{X}}_{t'}\right) \right]\right].
\end{align}
Identifying $\beta = 1/(D_0 \gamma_0)$, this reads
\begin{align}
    \frac{\mathbb{P}[\mathcal{T}\bm{X} | \bm{X}_T]}{\mathbb{P}[\bm{X} | \bm{X}_{-T}]} 
    &= \exp\left[ \ln p_0 (\bm{X}_T) - \ln p_0(\bm{X}_{-T}) + \iint\limits_{[-T;T]^2} dt dt'\left[ \frac{\beta}{2} \left( \dot{\bm{X}}_t \Gamma_{tt'}^{-1}(\bm{\theta}) \nabla V(\bm{X}_{t'}) +   \nabla V(\bm{X}_{t})\Gamma_{tt'}^{-1}(\bm{\theta}) \dot{\bm{X}}_{t'}\right) \right]\right].
\end{align}
In particular, as a special case, if $\Gamma = \Gamma^{-1} = \mathbb{1}$ and $p_0(\bm{X}) = e^{-\beta V(\bm{X})}/Z$, which corresponds to usual thermal equilibrium with Gaussian white noise the above expression yields
\begin{align}
    \frac{\mathbb{P}[\mathcal{T}\bm{X} | \bm{X}_T]}{\mathbb{P}[\bm{X} | \bm{X}_{-T}]} 
    &= \exp\left[ \ln p_0 (\bm{X}_T) - \ln p_0(\bm{X}_{-T}) + \int\limits_{[-T;T]} dt \left[ \beta \dot{\bm{X}}_t \nabla V(\bm{X}_{t}) \right]\right] =1.
\end{align}
In the more general case of correlated noise, one may decompose $\Gamma^{-1}$ as $\Gamma^{-1} = \mathbb{1} + \Upsilon$ and assume that the system still displays a Boltzmnann distribution in steady state, so that
\begin{align}
    \frac{\mathbb{P}[\mathcal{T}\bm{X} | \bm{X}_T]}{\mathbb{P}[\bm{X} | \bm{X}_{-T}]} 
    &= \exp\left[\hphantom{a} \iint\limits_{[-T;T]^2} dt dt'\left[ \frac{\beta \Upsilon_{tt'}(\bm{\theta}) }{2} \left( \dot{\bm{X}}_t \cdot \nabla V(\bm{X}_{t'}) +   \nabla V(\bm{X}_{t})\cdot \dot{\bm{X}}_{t'}\right) \right]\right],
\end{align}
which one may identify as a net dissipated power over a cycle going from $\bm{X}_{-T}$ to $\bm{X}_{T}$ then back.
Using the fact that $\Upsilon_{tt'}$ should be symmetric, the sum may be simplified using a $t \leftrightarrow t'$ transformation in one of the terms, yielding
\begin{align}
    \frac{\mathbb{P}[\mathcal{T}\bm{X} | \bm{X}_T]}{\mathbb{P}[\bm{X} | \bm{X}_{-T}]} 
    &= \exp\left[\hphantom{a} \iint\limits_{[-T;T]^2} dt dt'\left[ \beta \Upsilon_{tt'}(\bm{\theta}) \dot{\bm{X}}_t \cdot \nabla V(\bm{X}_{t'}) \right]\right], \label{eq:EPR_Final}
\end{align}
so that the EP reads
\begin{align}
    \Delta S
    &=-\iint\limits_{[-T;T]^2} dt dt'\left[ \beta \Upsilon_{tt'}(\bm{\theta}) \dot{\bm{X}}_t \cdot \nabla V(\bm{X}_{t'}) \right]. \label{eq:GenericDeltaS_Temporal}
\end{align}
This expression generically does not reduce to a single integral over time of an EPR.

Explicit expressions of the entropy production are derived in some simple cases in Sec.~\ref{sec:Examples}.
In particular, we shall show that the entropy production diverges as the criterion to observe DHU~\cite{Anand2026}, $\sum_m c_m = -1/2$ regardless of $M$, is met.
We present some mathematical insight into that observation in Secs.~\ref{sec:M=1} and~\ref{sec:M=2}.
We also show in Sec.~\ref{sec:Temporal_Thermostat} that adding a second, standard thermostat to the system once again regularizes EP and makes it maximal, not infinite as this criterion is met.

\section{Simple examples of noise designs\label{sec:Examples}}

\subsection{Spatial Noise}

\subsubsection{Homogeneous pairwise noise}

As a simple example, let us consider the case $\Lambda_{ijab}(\bm{X}) = \Lambda_0(\bm{X})$ in $d=1$.
In that case, the $a$ and $b$ indices disappear ($d=1$) and the $D$ matrix simply reads 
\begin{align}
    D_{ii} &= D_0 + \frac{N-1}{2} \Lambda_0 (\bm{X}) \\
    D_{ij} &\underset{i\neq j}{=} \frac{c}{2} \Lambda_{0} (\bm{X}),  
\end{align}
so that
\begin{align}
    \overline{\overline{D}} = \left(D_0 + \frac{N-1-c}{2} \Lambda_0\right) \overline{\overline{I}} +  \frac{c}{2} \Lambda_{0} (\bm{X}) \overline{\overline{J}}
\end{align}
with $\overline{\overline{I}}$ the identity and $\overline{\overline{J}}$ a matrix of ones.
This matrix is amenable to inversion by the Sherman-Morrison formula~\cite{Sherman1950}.
The latter states that for $\overline{\overline{M}}$ an invertible $n\times n$ matrix, and $\bm{u}, \bm{v}$ two $n\times 1$ vectors, the matrix defined by $\overline{\overline{M}} + \bm{u}{}^t\bm{v}$ is invertible if and only if $1+ {}^t\bm{v}\overline{\overline{M}}{}^{-1}\bm{u} \neq 0$ and that, in this case,
\begin{align}
    \left(\overline{\overline{M}} + \bm{u}{}^t\bm{v}\right)^{-1} = \overline{\overline{M}}{}^{-1} - \frac{\overline{\overline{M}}{}^{-1}\bm{u}{}^t\bm{v}\overline{\overline{M}}{}^{-1}}{1+ {}^t\bm{v}\overline{\overline{M}}{}^{-1}\bm{u}}.
\end{align}
Choosing 
\begin{align}
    \bm{u} &= \sqrt{|c| \Lambda_0 (\bm{X}) / 2} \left(1\, 1\, \ldots\, 1 \right)^t, \quad {}^t\bm{v} = \operatorname{sign}(c)\sqrt{|c| \Lambda_0 (\bm{X}) / 2} \left(1\, 1\, \ldots\, 1 \right), \\
    \overline{\overline{M}} &=  \left(D_0 + \frac{N-1-c}{2} \Lambda_0\right) \overline{\overline{I}},
\end{align}
(where isolating the sign function avoids defining intermediate complex values if $c < 0$), this formula yields
\begin{align}
    \overline{\overline{D}}{}^{-1} &= \frac{1}{D_0 + (N-1-c) \Lambda_0 /2}\left(\overline{\overline{I}} - \frac{c \Lambda_0 /2}{D_0 + (N-1-c)\Lambda_0/2+N c \Lambda_0 /2}\overline{\overline{J}}\right), \\
    &= \frac{2}{2D_0 + (N-1-c) \Lambda_0 }\left(\overline{\overline{I}} -  \frac{c \Lambda_0}{2D_0 + (c+1)(N-1)\Lambda_0 }\overline{\overline{J}}\right). \label{eq:InverseD_OneLambda0}
\end{align}

\subsubsection{Asymmetric, non-projector noise}

As a simple example of an $i\leftrightarrow j$ asymmetric case but with no projector component, $P_{ab} =\delta_{ab}$, consider the case of $\Lambda_{ij} = \Lambda_-$ if $i < j$ and $\Lambda_{ij} = \Lambda_+$ if $i > j$.
Then, one has
\begin{align}
    D_{iiab} &= \delta_{ab} \left( D_0 + \frac{i-1}{2}\Lambda_+ + \frac{N-i}{2}\Lambda_- \right), \\
    D_{ijab} &\underset{i\neq j}{=} \delta_{ab} \frac{c}{2} \sqrt{\Lambda_-\Lambda_+}.
\end{align}
In the simple example of a $1d$ system, the matrix can thus be decomposed as
\begin{align}
    \overline{\overline{D}} = \text{Diag}\left( D_0 + \frac{i-1}{2}\Lambda_+ + \frac{N-i}{2}\Lambda_- - \frac{c}{2}\sqrt{\Lambda_- \Lambda_+}\right) + \frac{c}{2}\sqrt{\Lambda_- \Lambda_+} \, \overline{\overline{J}}.
\end{align}
This matrix can still be inverted using the Sherman-Morrison formula~\cite{Sherman1950}, where now
\begin{align}
    ^t\bm{u} &= \sqrt{c \sqrt{ \Lambda_-} / 2} \left(1\, 1\, \ldots\, 1 \right),\\
    {}^t\bm{v} &= \sqrt{c \sqrt{ \Lambda_+} / 2} \left(1\, 1\, \ldots\, 1 \right), \\
    \overline{\overline{M}} &= \text{Diag}\left( D_0 + \frac{i-1}{2}\Lambda_+ + \frac{N-i}{2}\Lambda_- - \frac{c}{2}\sqrt{\Lambda_- \Lambda_+}\right).
\end{align}
Without getting into the fine details of the solution, one may at least check that
\begin{align}
    1 + {}^t\bm{v} \overline{\overline{M}}{}^{-1}\bm{u} &= 1 + \frac{c}{2}\sqrt{\Lambda_- \Lambda_+} \sum\limits_{i=1}^N \frac{1}{D_0 + \frac{i-1}{2}\Lambda_+ + \frac{N-i}{2}\Lambda_- - \frac{c}{2}\sqrt{\Lambda_- \Lambda_+}} \\
    &= 1 + \frac{c}{2}\sqrt{\Lambda_- \Lambda_+} \sum\limits_{i=1}^N \frac{1}{D_0 + \frac{1}{2}(N \Lambda_- - \Lambda_+) - \frac{c}{2}\sqrt{\Lambda_- \Lambda_+} + \frac{i}{2}(\Lambda_+ - \Lambda_-)}.
\end{align}
If $\Lambda_+ \neq \Lambda_-$, the sum may be computed by introducing the polygamma function $\psi(z)$ such that
\begin{align}
    \sum\limits_{i=1}^N\frac{1}{a + i b} = \frac{\psi(1+a/b + N) - \psi(1 + a/b)}{b},
\end{align}
by defining
\begin{align}
    a &= D_0 + \frac{1}{2}(N \Lambda_- - \Lambda_+) - \frac{c}{2}\sqrt{\Lambda_- \Lambda_+}, \\
    b &= \frac{\Lambda_+ - \Lambda_-}{2},
\end{align}
so that
\begin{align}
    1 + {}^t\bm{v} \overline{\overline{M}}{}^{-1}\bm{u} &= 1 + c\frac{\sqrt{\Lambda_- \Lambda_+}}{\Lambda_+ - \Lambda_-} \left( \psi\left( \frac{2D_0 + N \Lambda_+ - \Lambda_-  - c\sqrt{\Lambda_- \Lambda_+}}{\Lambda_+ - \Lambda_-}   \right) - \psi\left( \frac{2D_0 + (N-1) \Lambda_- - c\sqrt{\Lambda_- \Lambda_+}}{\Lambda_+ - \Lambda_-}  \right)\right).
\end{align}
Let's assume (without any loss of generality) that $\Lambda_+ > \Lambda_-$.
The polygamma function is a strictly growing, concave function, so that the factor multiplying $c$ is always positive: the inverse is always well-defined for $c \geq 0$.
For $c<0$, only very specific choices of $\Lambda_{\pm}$ would be problematic.

\subsubsection{Projector noise}

Consider the minimal example of $N=2$ particles in $d=2$, with a projection along the center-of-mass component and an otherwise constant $\Lambda$, so that
\begin{align}
    \sqrt{\Lambda_{ijab}} = \sqrt{\Lambda_0} \bm{\hat{e}}_a \cdot \bm{\hat{r}}_{ij} \otimes \bm{\hat{r}}_{ij} \cdot \bm{\hat{e}}_b.
\end{align}
Parametrising the unit vector by an angle $\theta$ such that $\bm{\hat{r}}_{ij} = (\cos\theta, \sin\theta)$, and the short-hand notation
\begin{align}
    R(\theta) = \begin{pmatrix}
        \cos^2\theta & \cos\theta\sin\theta \\
        \cos\theta\sin\theta & \sin^2\theta
    \end{pmatrix}
\end{align}
the tensor thus reads
\begin{align}
\overline{\overline{D}} = D_0 \overline{\overline{I}} + \frac{\Lambda_0}{2}\begin{pmatrix} R(\theta) & cR(\theta)  \\
c R(\theta)  & R(\theta)   \end{pmatrix}. \label{eq:proj_example}
\end{align}
Interestingly, 
\begin{align}
    \det R(\theta) = 0,
\end{align}
so that $R$ is not invertible and, therefore, neither is the matrix multiplying $\Lambda_0$ in Eq.~\ref{eq:proj_example}.
As a result, $\overline{\overline{D}}$ is not invertible in the case $D_0 = 0$ for this choice of projection matrix, irrespective of $c$, and the EPR will diverge as $D_0 \to 0$.
More generally, assume that $\sqrt{\Lambda_{ijab}} = P_{ab}(\bm{r}_{ij}) \sqrt{\Lambda_0}$ with $P_{ab}$ generated by a projection matrix.
In this more general setting, noting that due to projection operators being symmetric, and since they also verify $P^2 = P$
\begin{align}
    \sum\limits_{p=1}^d P_{ap}P_{bp} = (P^2)_{ab} = P_{ab},
\end{align}
so that
\begin{align}
\overline{\overline{D}} = D_0 \overline{\overline{I}} + \frac{\Lambda_0 }{2}\begin{pmatrix} P(\bm{r}_{12}) & c P(\bm{r}_{12}) \\
c P(\bm{r}_{12})  & P(\bm{r}_{12})   \end{pmatrix}. \label{eq:proj_example_genericP}
\end{align}
The inverse is thus well defined in the limit $D_0 \to 0$ if and only if $P$ itself is invertible.
Since $P$ is a projection matrix, 
\begin{align}
    P^2 = P.
\end{align}
Assuming that $P$ admits an inverse $P^{-1}$, multiplying the equation above by this inverse yields
\begin{align}
    P = I
\end{align}
so that the only invertible projection matrix is the identity.
Thus, the only dynamics with finite EPR in the limit $D_0 \to 0$ is that with $\sqrt{\Lambda_{ijab}} = \delta_{ab}\sqrt{\Lambda_{ij}}$.
However, any amount of thermal noise $D_0$ regularizes the matrix and lets it admit an inverse.
The reasoning remains unchanged for $N=2$ in arbitrary $d$.
For $N>2$ and arbitrary $d$, the argument does not get much more complicated as long as $\Lambda_0$ remains a constant, the matrix just becomes
\begin{align}
\overline{\overline{D}} = D_0 \overline{\overline{I}} + \frac{\Lambda_0 }{2}\begin{pmatrix} \sum\limits_{k\neq 1} P(\bm{r}_{1k}) & cP(\bm{r}_{12})& \ldots& c P(\bm{r}_{1N}) \\
cP(\bm{r}_{21})& \sum\limits_{k\neq 2} P(\bm{r}_{2k})& \ldots& c P(\bm{r}_{2N}) \\
\vdots & \vdots & \ddots & \vdots \\
cP(\bm{r}_{N1}) & cP(\bm{r}_{N2})& \ldots & \sum\limits_{k\neq N} P(\bm{r}_{Nk})
\end{pmatrix}. \label{eq:proj_example_genericP_N}
\end{align}
Thus, generically, $P$ not being invertible precludes inverting $\overline{\overline{D}}$ in the case $D_0 =0$.

Going back to Eq.~\ref{eq:proj_example}, in the case $D_0>0$, the matrix may be invertible.
It is then convenient to rewrite it as
\begin{align}
\overline{\overline{D}} =\begin{pmatrix} M & c \Lambda_0 R(\theta)/2  \\
c \Lambda_0 R(\theta)/2  & M   \end{pmatrix}. \label{eq:proj_example_block}
\end{align}
with $M$ the matrix given by
\begin{align}
    M = \begin{pmatrix}
        D_0 + \Lambda_0 \cos^2\theta/2 & \Lambda_0 \cos\theta\sin\theta /2 \\
        \Lambda_0 \cos\theta\sin\theta /2  & D_0 + \Lambda_0 \sin^2\theta/2
    \end{pmatrix}.
\end{align}
The inverse of this matrix is always well-defined for $D_0>0$ and reads
\begin{align}
    M^{-1} = \frac{1}{D_0 (\Lambda_0 + 2D_0)}\begin{pmatrix}
        2D_0 + \Lambda_0 \sin^2 \theta & -\Lambda_0 \cos\theta \sin\theta \\
        -\Lambda_0 \cos\theta \sin\theta & 2D_0 + \Lambda_0 \cos^2 \theta
    \end{pmatrix}.
\end{align}
As a result, the inverse of $\overline{\overline{D}}$ itself may be written explicitly.
To do so, recall that for a $2\times2$ block matrix
\begin{align}
    \overline{\overline{M}}_B = \begin{pmatrix}
        A & B \\
        C & D
    \end{pmatrix},
\end{align}
if $A$ is invertible and that the Schur complement 
\begin{align}
    S = D - C A^{-1}B
\end{align}
is invertible, then the inverse can be written explicitly as
\begin{align}
        \overline{\overline{M}}_B^{-1} &= \begin{pmatrix}
        E & F \\
        G & H
    \end{pmatrix}, \\
    E &= A^{-1} + A^{-1} B S^{-1} C A^{-1}\\
    F &= - A^{-1} B S^{-1}\\
    G &= - S^{-1} C A^{-1}\\
    H &= S^{-1}.
\end{align}
Identifying 
\begin{align}
    A &= D = M, \\
    B &= C = c\Lambda_0 R(\theta)/2,
\end{align}
it is useful to write
\begin{align}
    R M^{-1}R = \frac{2}{2D_0 + \Lambda_0}R
\end{align}
so that the Schur complement reads
\begin{align}
    S &= M - \frac{c^2\Lambda_0^2}{4} R M^{-1} R \\
    &= M - \frac{c^2 \Lambda_0^2}{2(\Lambda_0 + 2 D_0)} R \\
    &= D_0 I + \frac{\Lambda_0}{2}\left( 1 - c^2 \frac{\Lambda_0}{\Lambda_0 + 2 D_0} \right) R.
\end{align}
This matrix takes the same form as $M$ but with a renormalized $\Lambda_0$, according to the mapping
\begin{align}
    \Lambda_0 \mapsto \Lambda_0 \left( 1 - c^2 \frac{\Lambda_0}{\Lambda_0 + 2 D_0} \right),
\end{align}
so that it is invertible as long as $D_0 >0$.
In particular, for $c = 0$, $S = M$ and, for $c = \pm 1$,
\begin{align}
    \Lambda_0 \mapsto \frac{2 D_0\Lambda_0}{\Lambda_0 + 2 D_0}.
\end{align}
After some algebra, the inverse of $S$ can be injected into the various submatrices to yield
\begin{align}
    E &= H = \frac{1}{2D_0} \begin{pmatrix}
        \frac{8D_0^2 + 6D_0 \Lambda_0 + \Lambda_0^2 (1-c^2) - \Lambda_0(2D_0 + \Lambda_0 (1-c^2)) \cos2\theta}{(2D_0 + (1-c)\Lambda_0 ) (2D_0 + (1+c)\Lambda_0 )} & -2\Lambda_0 \frac{(2 D_0 + \Lambda_0 (1-c^2) )\cos\theta\sin\theta}{(2D_0 + (1-c)\Lambda_0 ) (2D_0 + (1+c)\Lambda_0 )} \\
        -2\Lambda_0 \frac{(2 D_0 + \Lambda_0 (1-c^2) )\cos\theta\sin\theta}{(2D_0 + (1-c)\Lambda_0 ) (2D_0 + (1+c)\Lambda_0 )} &  \frac{8D_0^2 + 6D_0 \Lambda_0 + \Lambda_0^2 (1-c^2) + \Lambda_0(2D_0 + \Lambda_0 (1-c^2)) \cos2\theta}{(2D_0 + (1-c)\Lambda_0 ) (2D_0 + (1+c)\Lambda_0 )}
    \end{pmatrix} \\
    F &= G =-2 c \Lambda_0\begin{pmatrix}
        \frac{\cos^2 \theta}{(2D_0 + (1-c)\Lambda_0 ) (2D_0 + (1+c)\Lambda_0 ) }  & \frac{\cos \theta \sin\theta}{4D_0^2 + 4D_0\Lambda_0+\Lambda_0^2 (1-c^2)} \\
        \frac{\cos \theta \sin\theta}{4D_0^2 + 4D_0\Lambda_0+\Lambda_0^2 (1-c^2)} & \frac{\sin^2 \theta}{(2D_0 + (1-c)\Lambda_0 ) (2D_0 + (1+c)\Lambda_0 ) }
    \end{pmatrix} 
\end{align}
It is then interesting to expand the expressions above in the limit $D_0 \to 0$, keeping only terms up to $O(1)$, which yields
\begin{align}
    E &\approx \frac{1}{D_0} R\left(\theta+ \frac{\pi}{2}\right) + \frac{2}{\Lambda_0 (1+c)(1-c)} R(\theta) + O(D_0) \\
    F &\approx -\frac{2c}{\Lambda_0 (1+c)(1-c)} R(\theta) + O(D_0).
\end{align}
Thus, for $D_0 \to 0^{+}$, the inverse of $\overline{\overline{D}}$ diverges for specific choices of $c$, namely $c \to \pm 1$ -- like in the simple example leading to Eq.~\ref{eq:InverseD_OneLambda0}.

\subsection{Temporal Noise}

We now focus on tractable dynamics with temporal noise.
In order to discuss entropy production, we must once again focus on the inverse of an object, but it is this time the kernel $\Gamma$.
We start by writing general considerations on inverting $\Gamma$ in Sec.~\ref{sec:InvertingGamma}, before considering cases of growing complexity in the next subsections -- namely, $M=1$ in Sec.~\ref{sec:M=1}, $M=2$ in Sec.~\ref{sec:M=2}, then a few general considerations in Sec.~\ref{sec:InvGamma_General}.

\subsubsection{Inverting Gamma \label{sec:InvertingGamma}}

We here focus on inverting $\Gamma$ for a generic set of parameters $\bm{\theta}$.
To do so, recall the definition of the inverse,
\begin{align}
    \int d\tau \Gamma(\bm{\theta};t, \tau) \Gamma^{-1}(\bm{\theta}; \tau , t') = \delta(t - t'). \label{eq:GammaInverseDefinition}
\end{align}
To solve for the inverse, it is useful to define the Fourier transform of the dyadic kernel $\bm{\Gamma}$ as
\begin{align}
    \widehat{\Gamma}(\bm{\theta}; \omega, \omega') &\equiv \int dt dt' e^{i (\omega't' - \omega t)} \Gamma(\bm{\theta}; t, t') \\
    \Gamma(\bm{\theta}; t,t') &= \frac{1}{4\pi^2}\int d\omega d\omega' e^{i(\omega t - \omega't')}.
\end{align}

Using these definitions, as well as the definition of the Fourier transform of the Dirac $\delta$,
\begin{align}
    \delta(x-x_0) = \frac{1}{2\pi} \int dq e^{iq(x-x_0)}
\end{align}
the definition of $\Gamma^{-1}$ can be rewritten as
\begin{align}
     \int d\tau \Gamma(\bm{\theta};t, \tau) \Gamma^{-1}(\bm{\theta}; \tau , t')&= \frac{1}{(2\pi)^4}\int d\tau d\omega d\omega' dk dk' e^{i (\omega t - \omega'\tau + k\tau - k't')} \widehat{\Gamma}(\bm{\theta}; \omega, \omega') \widehat{\Gamma}^{-1}(\bm{\theta}; k, k') \\
     &= \frac{1}{(2\pi)^3}\int d\omega d\omega' dk dk' \delta(k - \omega')e^{i (\omega t - k't')} \widehat{\Gamma}(\bm{\theta}; \omega, \omega') \widehat{\Gamma}^{-1}(\bm{\theta}; k, k') \\
     &=\frac{1}{(2\pi)^3}\int d\omega dk dk'e^{i (\omega t - k't')} \widehat{\Gamma}(\bm{\theta}; \omega, k) \widehat{\Gamma}^{-1}(\bm{\theta}; k, k').
\end{align}
Thus, one may compute the Fourier transform of Eq.~\ref{eq:GammaInverseDefinition}, in which the left-hand side reads
\begin{align}
    \int dt dt' e^{i(q't' - qt)} \int d\tau \Gamma(\bm{\theta};t, \tau) \Gamma^{-1}(\bm{\theta}; \tau , t') &=\frac{1}{(2\pi)^3}\int dt dt'  d\omega dk dk'e^{i(\omega-q) t} e^{i ( q'- k')t'} \widehat{\Gamma}(\bm{\theta}; \omega, k) \widehat{\Gamma}^{-1}(\bm{\theta}; k, k') \\
    &=\frac{1}{2\pi}\int d\omega dk dk'\delta(\omega-q) \delta( q'- k') \widehat{\Gamma}(\bm{\theta}; \omega, k) \widehat{\Gamma}^{-1}(\bm{\theta}; k, k') \\
    &= \frac{1}{2\pi}\int  dk \widehat{\Gamma}(\bm{\theta}; q, k) \widehat{\Gamma}^{-1}(\bm{\theta}; k, q').
\end{align}
Writing the Fourier transform of the right-hand-side,
\begin{align}
    \int dt dt'e^{i(q't'- qt)} \delta(t -t') &= \int dt e^{it(q'- q)} =2\pi \delta(q-q')
\end{align}
then yields the Fourier-space definition of the inverse,
\begin{align}
    \int  dk \widehat{\Gamma}(\bm{\theta}; q, k) \widehat{\Gamma}^{-1}(\bm{\theta}; k, q') &= 4\pi^2 \delta(q-q'). \label{eq:FourierGammaInverseDefinition}
\end{align}

It is then useful to introduce the definition of $\Gamma$ as
\begin{align}
    \Gamma(\bm{\theta}; t,t') = \delta(t-t') + \sum\limits_{m\in\mathbb{N}^\star} c_m \left(\delta(t-t'-m\tau)+\delta(t-t'+m\tau)\right),
\end{align}
which yields
\begin{align}
    \widehat{\Gamma}(\bm{\theta}; \omega, \omega')&= \int dt e^{it(\omega'- \omega )} + \sum\limits_{m\in \mathbb{N}^\star} c_m\int dt  e^{it(\omega'- \omega )} \left( e^{i m\omega'\tau} + e^{-i m\omega'\tau}\right) \\
    &= 2\pi \delta(\omega - \omega') + 4\pi\delta(\omega-\omega') \sum\limits_{m\in \mathbb{N}^\star} c_m \cos ( m\omega\tau )\label{eq:FourierGamma}
\end{align}
Injecting Eq.~\ref{eq:FourierGamma} into Eq.~\ref{eq:FourierGammaInverseDefinition} yields
\begin{align}
    2\pi \delta(q-q') &= \int dk \left[\delta(q - k) + 2\delta(q-k) \sum\limits_{m\in \mathbb{N}^\star} c_m \cos ( m q \tau )\right]\widehat{\Gamma}^{-1}(\bm{\theta}; k, q') \\
    &= \left[1 + 2 \sum\limits_{m\in \mathbb{N}^\star} c_m \cos ( m q \tau )\right]\widehat{\Gamma}^{-1}(\bm{\theta}; q, q')
\end{align}
which can be inverted to yield
\begin{align}
    \widehat{\Gamma}^{-1}(\bm{\theta}; q, q') = \frac{2\pi \delta(q-q')}{1 + 2 \sum\limits_{m\in \mathbb{N}^\star} c_m \cos ( m q \tau )}.
\end{align}
Finally, this general expression can be inverse Fourier-transformed to yield the inverse,
\begin{align}
    {\Gamma}^{-1}(\bm{\theta}; t, t') &= \frac{1}{2\pi} \int dq  \frac{e^{i q (t -t')}}{1+ 2\sum\limits_{m\in \mathbb{N}^\star} c_m \cos mq\tau } . \label{eq:invGamma_generic_Fourier_inverse}
\end{align}
This is the most compact form of the general expression of the inverse of $\Gamma$.
It is however useful to rewrite it in a slightly different way, as
\begin{align}
    {\Gamma}^{-1}(\bm{\theta}; t, t') &=  \frac{1}{2\pi} \sum\limits_{n=0}^\infty (-2)^n\int dq e^{i q (t-t')} \left(\sum\limits_{m=1}^\infty c_m \cos m q \tau\right)^n \label{eq:GammaInv_SumCosines} \\
    &= \delta(t-t') + \frac{1}{2\pi} \sum\limits_{n=1}^\infty (-2)^n \sum\limits_{m_1,m_2, \ldots, m_n =1}^\infty  \int dq e^{i q (t-t')} \prod\limits_{p=1}^n \left[ c_{m_p} \cos m_p q \tau \right]. \label{eq:GeneralInverseGamma_IntegralForm}
\end{align}
Rewriting 
\begin{align}
   2^n\prod\limits_{p=1}^n \left[ c_{m_p} \cos m_p q \tau \right] &= \prod\limits_{p=1}^n \left[ c_{m_p} (e^{i m_p q \tau} +e^{-i m_p q \tau})  \right] \\
   &= \prod\limits_{p=1}^n \left[ c_{m_p} e^{ i  m_p q \tau}(1 +e^{-2i m_p q \tau})  \right] \\
   &= \prod\limits_{p=1}^n \left[ c_{m_p} e^{ i  m_p q \tau}\right]\prod\limits_{p=1}^n \left[1 +e^{-2i m_p q \tau}  \right] \\
   &= \prod\limits_{p=1}^n \left[ c_{m_p} e^{ i  m_p q \tau}\right]\sum\limits_{s=0}^n \sum\limits_{\pi^s \in P_s(n)}  \prod\limits_{u = 1}^s e^{-2i m_{\pi^s_u} q \tau} 
\end{align}
where in the last equation we introduced a sum over all permutations of $s$ elements among $n$.
In other words, in the last line,
\begin{align}
    \sum\limits_{s=0}^n \sum\limits_{\pi^s \in P_s(n)}  \prod\limits_{u = 1}^s e^{-2i m_{\pi^s_u} q \tau} &= 1 + \sum\limits_{a=1}^n e^{-2im_a q \tau} + \left(e^{-2i(m_1+m_2) q \tau} + e^{-2i(m_1+m_3) q \tau} + e^{-2i(m_1+m_4) q \tau} +\ldots\right) \nonumber \\
    &\hphantom{aa}+ \left(e^{-2i(m_1+m_2+m_3) q \tau} + e^{-2i(m_1+m_2+m_4) q \tau} + e^{-2i(m_1+m_2+m_5) q \tau} +\ldots\right) \nonumber \\
    & \hphantom{aa}\cdots \nonumber \\
    &\hphantom{aa} + e^{-2i (m_1 + \ldots +m_n)q\tau}.
\end{align}
Altogether,
\begin{align}
   {\Gamma}^{-1}(\bm{\theta}; t, t') &=  \delta(t-t') + \frac{1}{2\pi} \sum\limits_{n=1}^\infty (-1)^n \sum\limits_{m_1,m_2, \ldots, m_n =1}^\infty \prod\limits_{p=1}^n \left[ c_{m_p} \right]  \sum\limits_{s=0}^n \sum\limits_{\pi^s \in P_s(n)}\int dq e^{i q (t-t' + (\sum\limits_{p=1}^n m_p - 2\sum\limits_{u=1}^sm_{\pi^s_u})\tau)} \\ 
   &=\delta(t-t') +  \sum\limits_{n=1}^\infty (-1)^n \sum\limits_{m_1,m_2, \ldots, m_n =1}^\infty \prod\limits_{p=1}^n \left[ c_{m_p} \right]  \sum\limits_{s=0}^n \sum\limits_{\pi^s \in P_s(n)} \delta \left[t-t' + \left(\sum\limits_{p=1}^n m_p - 2\sum\limits_{u=1}^sm_{\pi^s_u}\right)\tau\right].
\end{align}
This rather unwieldy expression at least has the advantage that it proves that $\Gamma^{-1}$ is indeed a sum of Dirac $\delta$'s delayed by multiples of $\tau$, as assumed within an ansatz before.

The expression of the inverse may be made simpler by introducing a multinomial expansion.
Consider Eq.~\ref{eq:GammaInv_SumCosines} and write the multinomial expansion for the power of the sum of cosines,
\begin{align}
    {\Gamma}^{-1}(\bm{\theta}; t, t') &=  \frac{1}{2\pi} \sum\limits_{n=0}^\infty (-1)^n\int dq e^{i q (t-t')} \left(\sum\limits_{m=1}^\infty c_m  \left(e^{ i m q \tau} + e^{ -i m q \tau}\right)\right)^n \\
    &=  \frac{1}{2\pi} \sum\limits_{n=0}^\infty (-1)^n\int dq e^{i q (t-t')} \sum\limits_{k_{1,1}+k_{1,2}+\ldots+k_{M,2} = n} \begin{pmatrix}
        n \\
        k_{1,1},\ldots, k_{{M,2}}
    \end{pmatrix} \prod_{m=1}^\infty c_m^{k_{m,1}+k_{m,2}} e^{i mq\tau (k_{m,1} - k_{m,2})} \\
    &= \sum\limits_{n=0}^\infty (-1)^n \sum\limits_{k_{1,1}+k_{1,2}+\ldots+k_{M,2} = n} \begin{pmatrix}
        n \\
        k_{1,1},\ldots, k_{{M,2}}
    \end{pmatrix} \left( \prod_{m=1}^\infty c_m^{k_{m,1}+k_{m,2}}\right) \delta\left(t-t'+\sum\limits_{m=1}^\infty m(k_{m,1}-k_{m,2})\tau\right)
\end{align}
With this writing, the $b_p$ coefficient in front of $\delta(t-t'+p\tau)$ in this inverse can be expressed explicitly as
\begin{align}
    b_p = \sum\limits_{n=0}^\infty (-1)^n \sum\limits_{\substack{k_{1,1}+k_{1,2}+\ldots+k_{M,2} = n \\
    \sum\limits_{m} m(k_{m,1} - k_{m,2}) = p }}\begin{pmatrix}
        n \\
        k_{1,1},\ldots, k_{{M,2}}
    \end{pmatrix} \left( \prod_{m=1}^\infty c_m^{k_{m,1}+k_{m,2}}\right).
\end{align}
Representing the Kronecker delta that selects $p$ as its Fourier transform,
\begin{align}
    \delta_{p, \sum\limits_{m} m(k_{m,1} - k_{m,2})} = \frac{1}{2\pi} \int\limits_{-\pi}^{\pi} du e^{iu (\sum\limits_{m}m(k_{m,1} - k_{m,2})-p)},
\end{align}
one may write
\begin{align}
    b_p &= \frac{1}{2\pi}\int\limits_{-\pi}^\pi du e^{-ipu}\sum\limits_{n=0}^\infty (-1)^n \sum\limits_{\substack{k_{1,1}+k_{1,2}+\ldots+k_{M,2} = n }}\begin{pmatrix}
        n \\
        k_{1,1},\ldots, k_{{M,2}}
    \end{pmatrix} \left( \prod_{m=1}^\infty (c_m e^{i u m})^ {k_{m,1}} (c_m e^{-i u m})^ {k_{m,2}}\right) \\
    &= \frac{1}{2\pi}\int\limits_{-\pi}^\pi du e^{-ipu}\sum\limits_{n=0}^\infty  \left(- 2\sum\limits_{m=1}^\infty c_m \cos m u \right)^n \\
    &= \frac{1}{2\pi}\int\limits_{-\pi}^\pi du \frac{e^{-ipu}}{1 + 2\sum\limits_{m=1}^\infty c_m \cos m u}.
\end{align}
In particular, by parity, the imaginary part of the integral is always zero and
\begin{align}
    b_p     &= \frac{1}{2\pi}\int\limits_{-\pi}^\pi du \frac{\cos p u}{1 + 2\sum\limits_{m=1}^\infty c_m \cos m u}, \label{eq:bp_integralrep}
\end{align}
which immediately yields $b_p = b_{-p}$.
To compute this integral, it is useful to introduce the Chebyshev polynomials $T_n(x)$ such that
\begin{align}
    \cos nx = T_n(\cos x),
\end{align}
so that
\begin{align}
    b_p     &= \frac{1}{2\pi}\int\limits_{-\pi}^\pi du \frac{T_p(\cos u)}{1 + 2\sum\limits_{m=1}^\infty c_m T_m(\cos u)}, \label{eq:bp_integralrep_Chebyshev}
\end{align}
then to perform Weierstrass's change of variable $u \mapsto v = \tan (u/2)$, which yields
\begin{align}
    b_p     &= \frac{2}{\pi}\int\limits_{0}^\infty \frac{dv}{1+v^2} \frac{T_p\!\left(\frac{1-v^2}{1+v^2}\right)}{1 + 2\sum\limits_{m=1}^\infty c_m T_m\!\left(\frac{1-v^2}{1+v^2}\right)}. \label{eq:bp_integralrep_ChebyshevWeierstrass}
\end{align}
This expression is the integral of a rational function and may in general be computed in terms of the roots of its denominator.

\subsubsection{Case $M=1$ \label{sec:M=1}}

Consider the simple case $M = 1$ so that $c_1 = c$ and $c_{n>1}=0$.
Then, it is best to go back to Eq.~\ref{eq:GeneralInverseGamma_IntegralForm} and to rewrite it as
\begin{align}
    {\Gamma}^{-1}(\bm{\theta}; t, t') &= \delta(t-t') + \frac{1}{2\pi} \sum\limits_{n=1}^\infty (-2)^n c^n   \int dq e^{i q (t-t')}  \cos^n  q \tau \\
    &= \delta(t-t') + \frac{1}{2\pi} \sum\limits_{n=1}^\infty (-1)^n c^n   \int dq e^{i q (t-t')} e^{inq\tau} \left(1 + e^{-2iq\tau} \right)^n \\
    &= \delta(t-t') + \frac{1}{2\pi} \sum\limits_{n=1}^\infty (-1)^n c^n   \int dq e^{i q (t-t')} e^{inq\tau} \sum\limits_{s=0}^n \begin{pmatrix} n \\s \end{pmatrix} e^{-2siq\tau} \\
    &= \delta(t-t') + \frac{1}{2\pi} \sum\limits_{n=1}^\infty\sum\limits_{s=0}^n  (-1)^n c^n \begin{pmatrix} n \\s \end{pmatrix}  \int dq e^{i q (t-t' + (n-2s)\tau)} \\
    &= \delta(t-t') + \sum\limits_{n=1}^\infty\sum\limits_{s=0}^n  (-1)^n c^n \begin{pmatrix} n \\s \end{pmatrix}  \delta(t-t' + (n-2s)\tau).
\end{align}
To progress further, it is useful to split the sum over $n$ into evens and odds, leading to
\begin{align}
    {\Gamma}^{-1}(\bm{\theta}; t, t')
    &= \delta(t-t') + \sum\limits_{p=1}^\infty\sum\limits_{s=0}^{2p}  c^{2p} \begin{pmatrix} 2p \\s \end{pmatrix}  \delta(t-t' + 2(p-s)\tau) - c\sum\limits_{p=0}^\infty\sum\limits_{s=0}^{2p+1}  c^{2p} \begin{pmatrix} 2p +1\\s \end{pmatrix}  \delta(t-t' + 2(p-s)\tau + \tau).
\end{align}
Finally, it is useful to define $m = p-s$ so that
\begin{align}
    {\Gamma}^{-1}(\bm{\theta}; t, t')
    &= \delta(t-t') + \sum\limits_{p=1}^\infty\sum\limits_{m=-p}^{p}  c^{2p} \begin{pmatrix} 2p \\p-m \end{pmatrix}  \delta(t-t' + 2m\tau) - c\sum\limits_{p=0}^\infty\sum\limits_{m=-p-1}^{p}  c^{2p} \begin{pmatrix} 2p +1\\p-m \end{pmatrix}  \delta(t-t' + (2m+1)\tau ).
\end{align}
Finally, one may group all terms with an equal delay, recalling that the odd and even delays are easier to treat separately.
Defining
\begin{align}
    {\Gamma}^{-1}(\bm{\theta}; t, t')
    &= \sum\limits_{n\in \mathbb{Z}} b_n \delta(t-t'+ n\tau),
\end{align}
in this simple case one has:
\begin{align}
    b_0 &= 1 + \sum\limits_{p=1}^\infty c^{2p} \begin{pmatrix} 2p \\p \end{pmatrix} = \frac{1}{\sqrt{1-4c^2}} \\
    b_{2m+1} &= -\sum\limits_{p\geq \max(-m-1,m)} c^{2p+1} \begin{pmatrix} 2p +1\\p-m \end{pmatrix} \\
    &= - c^{|2m+1|}  {}_2F_1\left(\left\{ \frac{3}{4}+\left|m+\frac{1}{4}\right|,\frac{3}{4}+\left|m+\frac{3}{4}\right|\right\}, \left\{1+|2m+1|\right\}, 4c^2 \right), \\
    b_{2m} &= \sum\limits_{p\geq |m|} c^{2p} \begin{pmatrix} 2p \\p-m \end{pmatrix} \\
    &= c^{2|m|}  {}_2F_1\left(\left\{ \frac{1}{2}+|m|,1+|m|\right\}, \left\{1+2|m|\right\}, 4c^2 \right).
\end{align}

Notice that $b_{2p} =  b_{-2p}$ and  $b_{2p+1} = b_{-(2p+1)} $ (this second case is a bit harder to see, notice that the sign switch is equivalent to $p \mapsto p'= -p-1$, so that the transformation switches the first two arguments of ${}_2F_1$, the order of which is arbitrary, and leaves the rest unchanged).
In particular, one may work out the first few values,
\begin{align}
    b_0 &= \frac{1}{\sqrt{1-4c^2}} \label{eq:b0_singlec} \\
    b_1 &= \frac{1}{2c}\left(1 - \frac{1}{\sqrt{1-4c^2}}\right) \label{eq:b1_singlec} \\
    b_2 &= \frac{1-2c^2 -\sqrt{1-4c^2}}{2c^2 \sqrt{1-4c^2}}.
\end{align}
There are two special choices of $c$ that are worth noting, $c= 0$ and $c = \pm 1/2$.
For $c = 0$, $b_n = \delta_{n,0}$ as expected.
For $c = \pm1/2$, the expressions above diverge. 

To check these expressions, consider the explicit formulation of the definition of the inverse in Eq.~\ref{eq:InverseGamma_ExplicitSum} and the resulting conditions given in Eqs.~\ref{eq:InverseCondition1} and~\ref{eq:InverseCondition2}.
For $M=1$ and a single $c$ value, and using the even-ness of the $b_p$, these conditions become
\begin{align}
    b_0 + 2 c\, b_{1} &= 1, \\
    b_p +c (b_{p-1} + b_{p+1}) &= 0,\,  \forall p >0.\label{eq:InverseCondition2M=1}
\end{align}
For $c = 0$, both conditions are immediately fulfilled.
For $c\neq 0$, the first condition is verified by the expressions given in Eqs.~\ref{eq:b0_singlec} and~\ref{eq:b1_singlec}.
Condition~\ref{eq:InverseCondition2M=1} in the case $c\neq 0$ is a bit trickier to check.
Consider the even case, $p = 2m >0$.
In that situation, one has to evaluate
\begin{align}
    C_{2m} &= b_{2m} + c(b_{2m+1} + b_{2m -1}).
\end{align}
To check the inequality, it is easier to use the expressions of the coefficients written as sums of binomial coefficients, so that
\begin{align}
    C_{2m} = \sum\limits_{p\geq m} c^{2p} \begin{pmatrix} 2p \\p-m \end{pmatrix} - c\left[\sum\limits_{q \geq m} c^{2q+1}\begin{pmatrix} 2q +1 \\q-m \end{pmatrix}  + \sum\limits_{r \geq m-1} c^{2r+1}\begin{pmatrix} 2r +1 \\r-m +1 \end{pmatrix}\right].
\end{align}
To progress further, the term proportional to $c$ can be rewritten as a single sum and one may use Pascal's triangle equality, which states that
\begin{align}
    \begin{pmatrix} 2q +1 \\q-m \end{pmatrix} + \begin{pmatrix} 2q +1 \\q-m+1 \end{pmatrix} = \begin{pmatrix} 2q +2 \\q-m +1\end{pmatrix}.
\end{align}
Using this equality,
\begin{align}
    C_{2m} = \sum\limits_{p\geq m} c^{2p} \begin{pmatrix} 2p \\p-m \end{pmatrix} -\left[ c^{2m} +\sum\limits_{q \geq m} c^{2q+2}\begin{pmatrix} 2q +2 \\q-m+1 \end{pmatrix}  \right].
\end{align}
A re-indexation of the right-most sum using $q+1 \mapsto p$ then yields the desired condition, $C_{2m} = 0$.
The same proof can be reproduced for $C_{2m+1}$ by a simple permutation.
This concludes the proof that we found the correct inverse $\Gamma^{-1}$ for $M=1$.

A simpler solution comes from the expression in Eq.~\ref{eq:bp_integralrep}.
Using that expression in the case $M=1$ immediately yields a definition of an hypergeometric function
\begin{align}
    b_p = \frac{1}{1-2c} {}_3\widetilde{F}_2\left(\left\{ \frac{1}{2},1,1 \right\}, \left\{1-p,1+p \right\}; \frac{4c}{2c -1}\right). \label{eq:M=1_bn}
\end{align}

Thus, in the simple case $M=1$ with a coefficient $c$, the entropy production $\Delta S$ in the exponent of Eq.~\ref{eq:EPR_Final} may be written explicitly,
\begin{align}
    \Delta S 
    &= -\iint\limits_{[-T;T]^2} dt dt'\left[ \beta \left[ (b_0 -1)\delta_{t,t'} + \sum\limits_{n \neq 0}b_n\delta_{t,t'-n\tau}\right]\dot{\bm{X}}_t \cdot \nabla V(\bm{X}_{t'}) \right], \\
    &= -\int\limits_{[-T;T]} dt \left[ \beta  (b_0 -1)\dot{\bm{X}}_t \cdot \nabla V(\bm{X}_{t}) + \beta \sum\limits_{n\neq 0}b_n\dot{\bm{X}}_{t-n\tau} \cdot \nabla V(\bm{X}_{t}) \right].\label{eq:EPR_Final_M=1}
\end{align}
Interestingly, this expression contains a non-zero instantaneous entropy production rate $\sigma_0(t)$ that reads
\begin{align}
    \sigma_0(t) &\equiv -\beta  (b_0 -1)\dot{\bm{X}}_t \cdot \nabla V(\bm{X}_{t}), \\
    &= -\beta \frac{1-\sqrt{1-4c^2}}{\sqrt{1-4c^2}}\dot{\bm{X}}_t \cdot \nabla V(\bm{X}_{t}), \label{eq:c=1_sametime_EPR}
\end{align}
that does not involve a delayed velocity-force product.
This term vanishes in the limit $c\to 0$ as expected, and diverges in the limits $c \to \pm 1/2$.
Note that the $b_n(c)$ series at a fixed $|c| \in (0;1/2)$ decays exponentially with $n$, so that the number of delayed dissipation terms in the entropy production is typically finite.
In practice, since it depends on functions of $\sqrt{1-4c^2}$, the number of terms to keep will increase as $c \to \pm 1/2$.

\subsubsection{Case $M=2$\label{sec:M=2}}

Now, consider the case $M = 2$, so that $c_{1,2} \neq 0$ and $c_{n>2}=0$.
Starting from Eq.~\ref{eq:bp_integralrep}, the expression of $b_p$ reads
\begin{align}
    b_p     &= \frac{1}{2\pi}\int\limits_{-\pi}^\pi du \frac{\cos p u}{1 + 2 c_1 \cos  u + 2 c_2 \cos 2 u}.
\end{align}
This expression can be written as a rational function of $\cos u$ when introducing Chebyshev polynomials,
\begin{align}
    b_p     &= \frac{1}{2\pi}\int\limits_{-\pi}^\pi du \frac{T_p(\cos u)}{1-2c_2 + 2 c_1 \cos  u + 4 c_2 \cos^2 u} \equiv \frac{1}{2\pi}\int\limits_{-\pi}^\pi du R_p(\cos u)
\end{align}
with
\begin{align}
    R_p(X) \equiv \frac{T_p(X)}{1-2c_2 + 2 c_1 X + 4 c_2 X^2}.
\end{align}
This is a subcase of a broader family of integrals that are easily solved using Cauchy's residue theorem.
The broader case is that of a rational function $R$ of both $\cos u$ and $\sin u$,
\begin{align}
    I = \int\limits_{-\pi}^\pi du R(\cos u, \sin u).
\end{align}
In that case, the trick is to compute a contour integral of a function $f(z)$ defined as
\begin{align}
    f(z) \equiv \frac{1}{iz}R\left(\frac{z+z^{-1}}{2}, \frac{z-z^{-1}}{2i}\right)
\end{align}
over the unit circle $\mathcal{C}$ in the complex plane, as a simple change of variables $z \mapsto e^{i\theta}$ leading to $dz = i e^{i\theta} d\theta = iz d\theta$ yields
\begin{align}
    \oint_\mathcal{C} dz f(z) = \int\limits_{-\pi}^\pi\frac{1}{ie^{i \theta} }R(\cos\theta,\sin \theta) ie^{i\theta }d\theta = I.
\end{align}
Then, invoking the residue formula, this contour integral can be written as the sum of the residues of the poles of $f$ that are contained in the unit circle.
More specifically, assuming that $f$ only has a finite number of poles $z_p$ that do not belong to $\mathcal{C}$,
\begin{align}
    \oint_\mathcal{C} dz f(z) =2i\pi \sum\limits_{|z_p|<1} \text{Res}\left[f, z_p \right],
\end{align}
with the residue of $f$ at $z_p$ is given, for a pole of order $n$, by
\begin{align}
    \text{Res}\left[f, z_p \right] \equiv \frac{1}{(n-1)!} \lim_{z\to z_p} \partial_z^{(n-1)} \left[(z-z_p)^n f(z) \right].
\end{align}
In particular, for simple poles,
\begin{align}
    \text{Res}\left[f, z_p \right] = \lim_{z\to z_p} \left[(z-z_p) f(z)\right].
\end{align}
If \textit{simple} poles exist \textit{on} the unit circle, the integral is no longer strictly defined, but its Cauchy principal value may be evaluated instead.
To do so, the contour has to be modified to become the unit circle except around poles $z_c$ such that $|z_c|=1$, where the contour must instead follow a half-circle with vanishing radius $z_c + r e^{i\phi}$ with $\phi \in [0;\pi]$ and $r \to 0^+$.
With that modification, calling $\mathcal{C}'$ the modified circle in the limit $r \to 0$,
\begin{align}
    \text{PV}\oint_\mathcal{C'} dz f(z) =2i\pi \sum\limits_{|z_p|<1} \text{Res}\left[f, z_p \right] + i\pi \sum\limits_{|z_c|=1} \text{Res}\left[f, z_c \right].
\end{align}
Here, injecting the definition of $R_p$ into that of $f$ yields
\begin{align}
    f_p(z) &\equiv \frac{1}{iz} \frac{T_p\left[ (z+1/z)/2\right]}{1-2c_2 +c_1(z +1/z) + c_2 (z+1/z)^2} \\
    &= \frac{z}{i} \frac{T_p\left[ (z+1/z)/2\right]}{c_2 z^4 + c_1 z^3 +  z^2 + c_1 z +c_2}.
\end{align}
Since $T_p$ is of degree $p$, this can be rewritten as
\begin{align}
    f_p(z) &=\frac{z}{iz^{p}} \frac{z^pT_p\left[ (z+1/z)/2\right]}{c_2 z^4 + c_1 z^3 + z^2 + c_1 z +c_2},
\end{align}
where $z^pT_p\left[ (z+1/z)/2\right] \in P_{2p}[\mathbb{C}]$ is a polynomial.
Thus, in practice, the non-trivial poles (read: the poles away from the origin) are given by the roots of a polynomial $Q$ given by
\begin{align}
    Q(Z) = c_2 Z^4 + c_1 Z^3 +  Z^2 + c_1 Z + c_2.
\end{align}
This is a symmetric quartic polynomial: its roots may actually be written explicitly from the roots of the polynomial $\widetilde{Q}$ on $X = (Z +Z^{-1})/2$ -- which is exactly the denominator of $R_p$,
\begin{align}
    \widetilde{Q}(X) =  4c_2 X^2 + 2c_1 X + 1-2c_2.
\end{align}
This quadratic polynomial has a discriminant
\begin{align}
    \Delta = 4c_1^2 - 16 c_ 2(1-2c_ 2).
\end{align}
It does not matter whether the roots of this polynomial are real at this stage, only whether they are simple roots.
A double root occurs when $\Delta = 0$, or 
\begin{align}
    C_{1,X}^{\pm} = \pm 2\sqrt{c_2(1-2c_2)}.
\end{align}
For either of the two following inequalities
\begin{align}
    0\leq &c_2 \leq \frac{2-\sqrt{2}}{8} \equiv C_X\\
\frac{1}{2}-C_X \leq &c_2 \leq \frac{1}{2},
\end{align}
this equation has 2 real solutions for $-1/2< c_1 <1/2$.
The roots of $\widetilde{Q}$ are given by
\begin{align}
    X_{1,2} = \frac{-c_1 \pm \sqrt{\Delta}}{4c_2} = \frac{-c_1 \pm 2\sqrt{c_1^2 - 4 c_2 (1-2c_2)}}{4c_2}
\end{align}
where the square root is here understood as being extended to negative numbers via $\sqrt{-1} = i$ and $X_{1,2} \in \mathbb{C}$.
In particular, the double root takes the value $X_0 = -c_1 / 4c_2$.

The roots of $Q$ are found by requiring that $X_{1,2} = (Z + Z^{-1})/2$ at the roots, which implies
\begin{align}
    Z^2 - 2X_{1,2}Z + 1 = 0,
\end{align}
a quadratic polynomial with a discriminant
\begin{align}
    \Delta' = 4(X_{1,2}^2 - 1) = 4(X_{1,2} - 1)(X_{1,2} + 1).
\end{align}
The four roots of $Q$ are thus given by
\begin{align}
    Z_{1,2,3,4} =  X_{1,2}\pm  \sqrt{X_{1,2}^2 - 1} 
\end{align}
where the four indices are obtained by combining the choice of $X_{1,2}$ and the choice of $\pm$, and in general $Z_k \in \mathbb{C}$.

One may thus write, assuming simple poles only,
\begin{align}
    b_0 &=  i \sum\limits_{|Z_k|<1} \text{Res}\left[ f_0, Z_k\right], \\
    \text{PV\,} b_0 &=  i \sum\limits_{|Z_k|<1} \text{Res}\left[ f_0, Z_k\right] + \frac{i}{2} \sum\limits_{|Z_k|=1} \text{Res}\left[ f_0, Z_k\right],
\end{align}
with
\begin{align}
    f_0(z) &=\frac{z}{i} \frac{1}{c_2 z^4 + c_1 z^3 + z^2 + c_1 z +c_2} \\
    &= \frac{z/i}{c_2\prod\limits_{k=1}^4(z-Z_k)},
\end{align}
or, assuming simple poles,
\begin{align}
    i \text{Res}\left[ f_0, Z_k\right] = \frac{Z_k}{c_2\prod\limits_{j\neq k}(Z_k - Z_j)}.
\end{align}
If poles become multiple on the circle, the integral diverges, as the vanishing-radius semi-circle formula breaks down.

Computing the integral using residues then boils down to identifying when roots lie (strictly) in the unit disk, or more concretely when the inequality
\begin{align}
|Z_{1,2,3,4}|^2 = \left|X_{1,2} \pm \sqrt{X_{1,2}^2 - 1}  \right|^2 \leq 1
\end{align}
is true, and when it saturates.
Some simple but cumbersome algebra yields, using the convention 
\begin{align}
    Z_1 &= X_1+ \sqrt{X_1^2 -1}, \\
    Z_2 &=  X_2+ \sqrt{X_2^2 -1}, \\
    Z_3 &=  X_1- \sqrt{X_1^2 -1}, \\
    Z_4 &= X_2- \sqrt{X_2^2 -1},
\end{align}
 the following conditions for \textit{equality}, i.e. roots belonging to the unit \textit{circle},
 \begin{align}
   \left( \forall c_1,\frac{1}{4}+\frac{1}{4}\sqrt{1-2c_1^2}\leq c_2 \right)\cup\left(c_1>0, c_2\leq \frac{1}{2}(2c_1 -1)\right)&\Rightarrow |Z_1| = |Z_3| =  1 \\
   \left( \forall c_1,\frac{1}{4}+\frac{1}{4}\sqrt{1-2c_1^2}\leq c_2 \right)\cup\left(c_1<0, c_2\leq -\frac{1}{2}(2c_1 +1)\right)&\Rightarrow |Z_2| = |Z_4| =  1.
 \end{align}
 The conditions for strict inequalities, on the other hand, are
 \begin{align}
   \left( \forall c_1,\frac{1}{4}+\frac{1}{4}\sqrt{1-2c_1^2}> c_2 \right)\cap\left(c_1>0, c_2> \frac{1}{2}(2c_1 -1)\right)\cap\left(c_1>0 \,\cup\,\vphantom{\int}c_2<0\right)&\Rightarrow |Z_1| <  1 \\
   \left( \forall c_1,\frac{1}{4}+\frac{1}{4}\sqrt{1-2c_1^2}> c_2 \right)\cap\left(c_1>0 \,\cap\,\vphantom{\int}c_2>0\right)&\Rightarrow |Z_2| <  1 \\
   \left( \forall c_1,\frac{1}{4}+\frac{1}{4}\sqrt{1-2c_1^2}> c_2 \right)\cap\left(c_1<0 \,\cap\,\vphantom{\int}c_2>0\right)&\Rightarrow |Z_3| <  1 \\
   \left( \forall c_1,\frac{1}{4}+\frac{1}{4}\sqrt{1-2c_1^2}> c_2 \right)\cap\left(c_1>0, c_2> -\frac{1}{2}(2c_1 +1)\right)\cap\left(c_1<0 \,\cup\,\vphantom{\int}c_2>0\right)&\Rightarrow |Z_4| <  1.
 \end{align}
 
\begin{figure}
    \centering
    \includegraphics[width=0.24\linewidth]{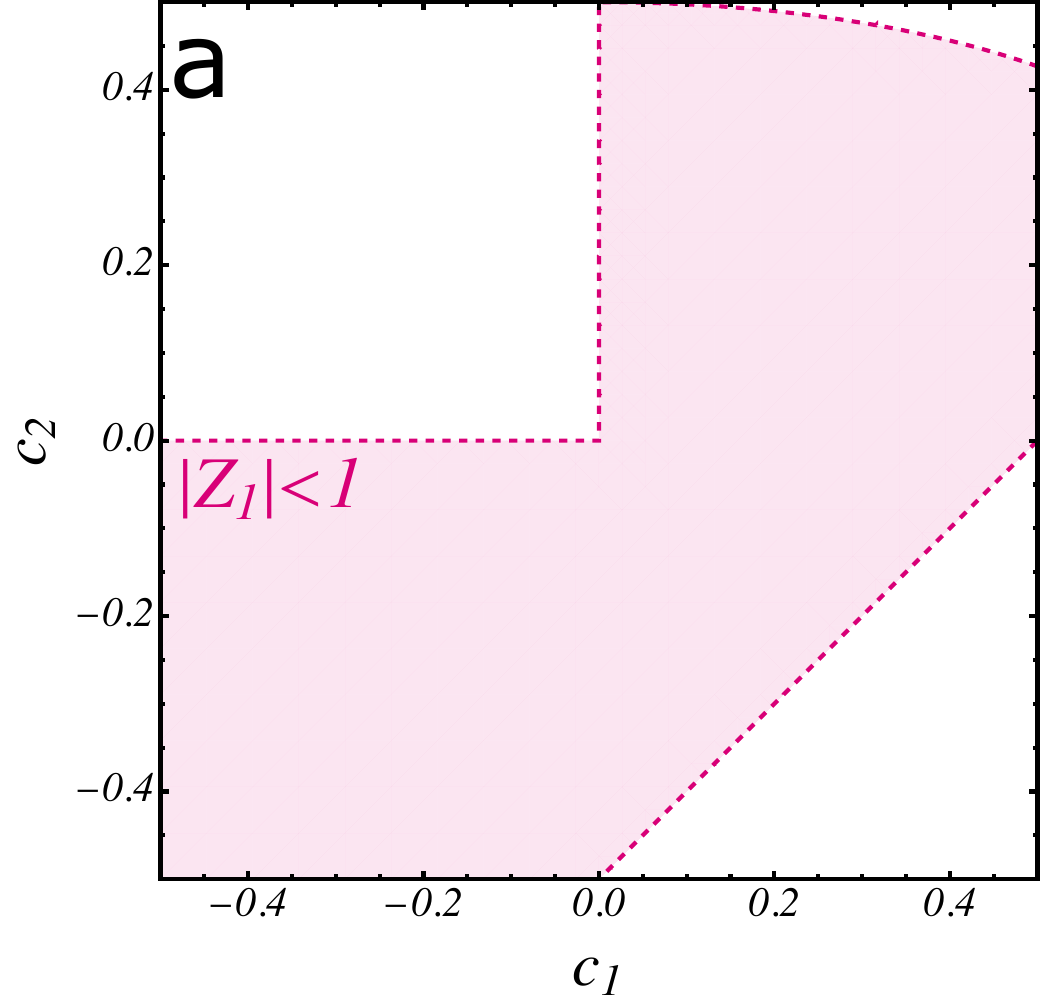}
    \includegraphics[width=0.24\linewidth]{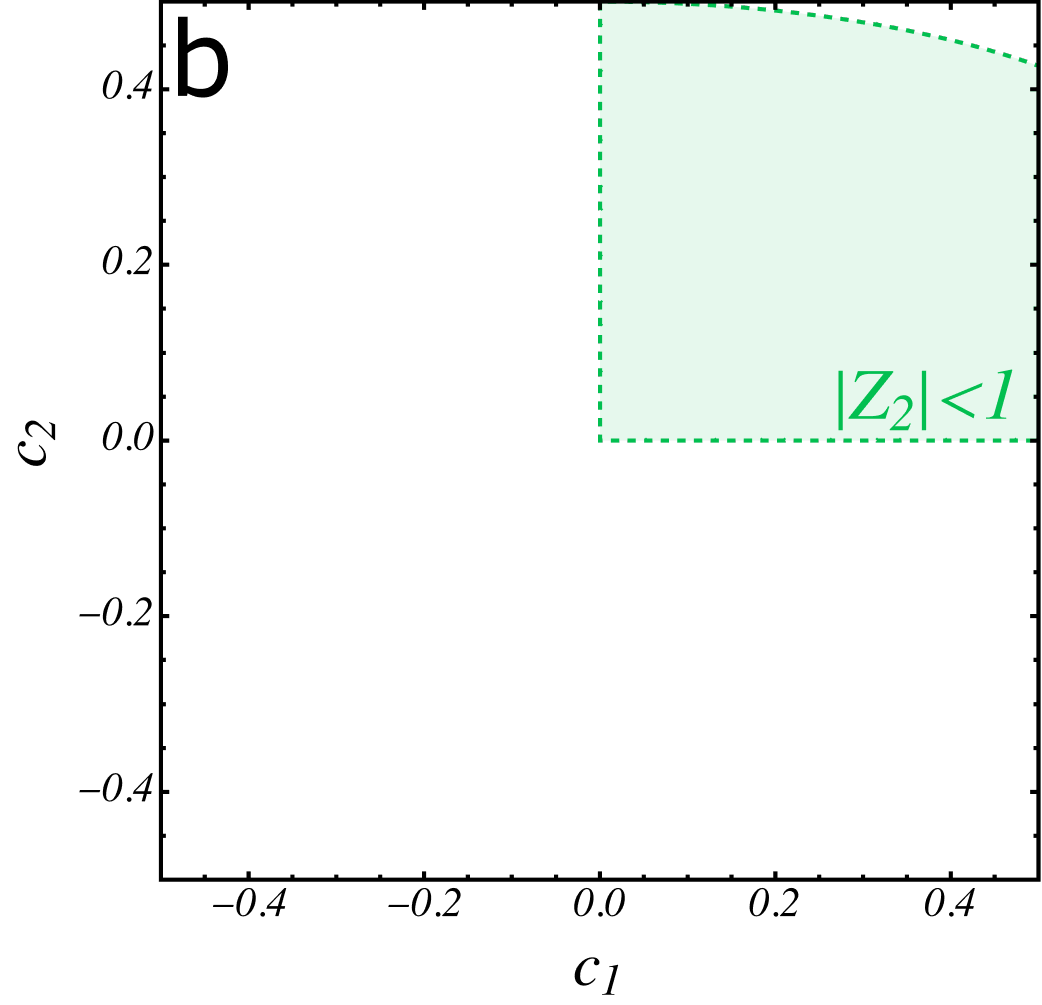}
    \includegraphics[width=0.24\linewidth]{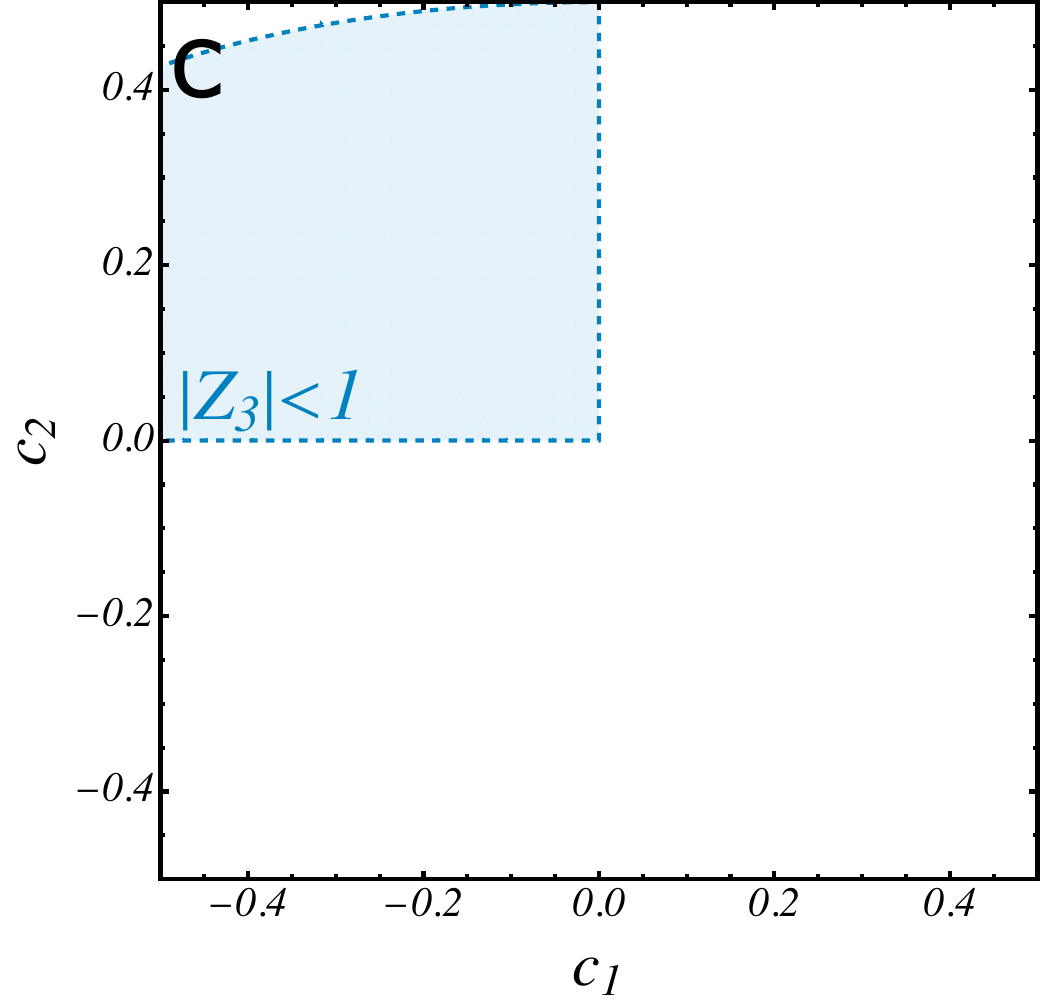}
    \includegraphics[width=0.24\linewidth]{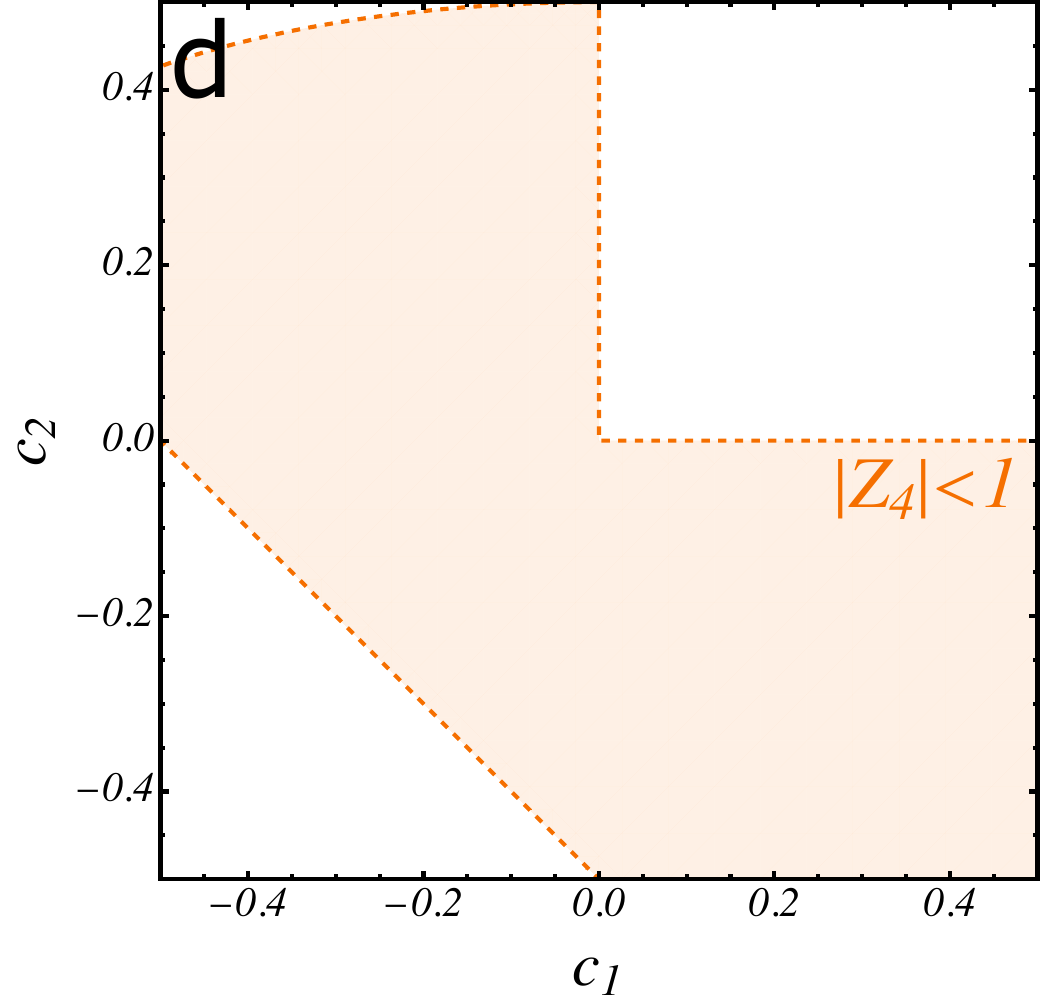}
    \caption{\textbf{Location of roots $Z_k$}.
    The domains of $(c_1,c_2)$ in which $(a)$ $Z_1$,$(b)$ $Z_2$,$(c)$ $Z_3$,$(d)$ $Z_4$ lie in the unit disk are highlighted as colored areas.}
    \label{fig:Zdomains}
\end{figure}

Altogether, in the absence of a Cauchy principal value extension, the integral diverges in 3 cases:
\begin{align}
    \left(1+\sqrt{1-2c_1^2}\leq 4c_2 \right)\cup\left( c_1+c_2\leq -\frac{1}{2}\right)\cup\left( c_2-c_1\leq -\frac{1}{2}\right)\Rightarrow b_0 \text{ diverges}. \label{eq:DVCriterion_M2}
\end{align}
In the convergent case, namely when none of the conditions~\ref{eq:DVCriterion_M2} are fulfilled, it's easier describe the values of the integrals splitting the $(c_1,c_2)$ plane into quadrants.
Defining $g_k(c_1,c_2)$ for $k\in\{1,2,3,4\}$ as
\begin{align}
    g_k(c_1,c_2) = \frac{Z_k/c_2}{\prod_{j\neq k}Z_k - Z_j},
\end{align}
one finds
\begin{align}
    b_0 = \begin{cases}
        g_1 + g_4 \text{ if } c_2 <0, \\
        g_3 + g_4 \text{ if }c_2>0\text{ and } c_1 <0, \\
        g_1+g_2 \text{ if }c_2>0\text{ and } c_1 >0.
    \end{cases}
\end{align}
These expressions do not simplify much.
However, it is worth noting that, in the last two cases, the individual $g$'s can evaluate to complex numbers but the two components of $b_0$ are then complex-conjugate, so that the resulting value is real.

Back to the meaning of $b_0$, this calculation shows that the same-time term in the entropy production rate for $M=2$ also diverges as $2\sum_m c_m \to -1/2$.
However, it also diverges at less trivial lines, namely $c_2-c_1 = -1/2$ and $4 c_2 = 1 + \sqrt{1-2c_1^2}$, which should be investigated numerically.

\begin{figure}
    \centering
    \includegraphics[width=0.32\linewidth]{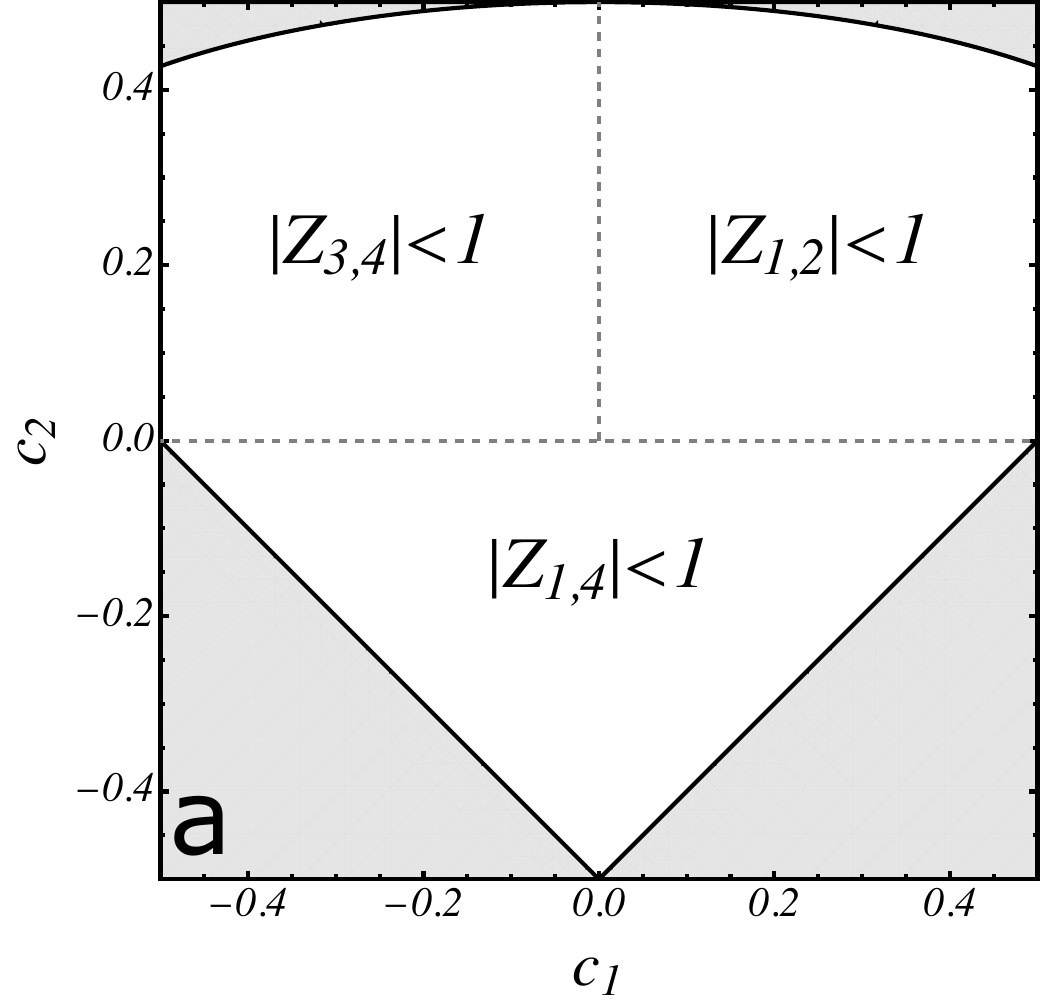}
    \includegraphics[width=0.32\linewidth]{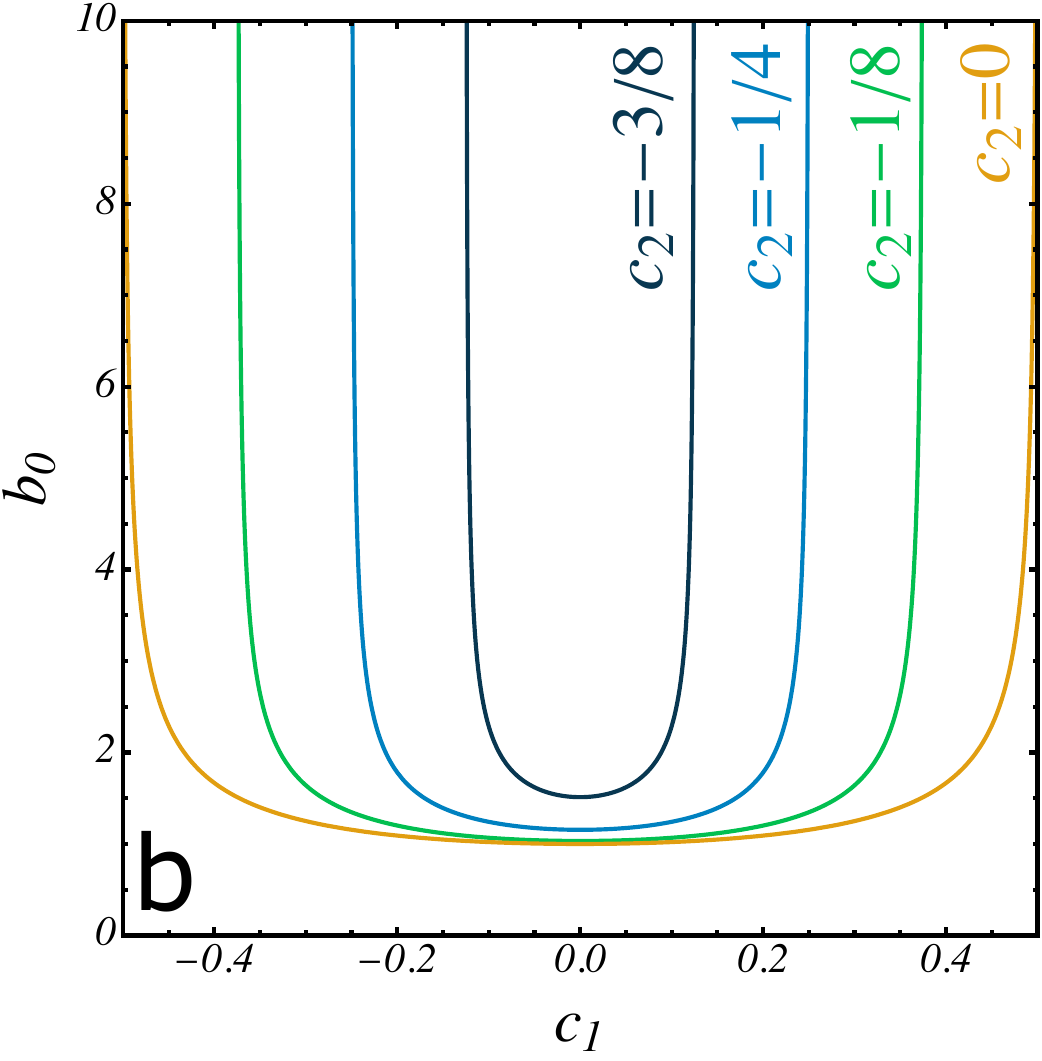}
    \includegraphics[width=0.32\linewidth]{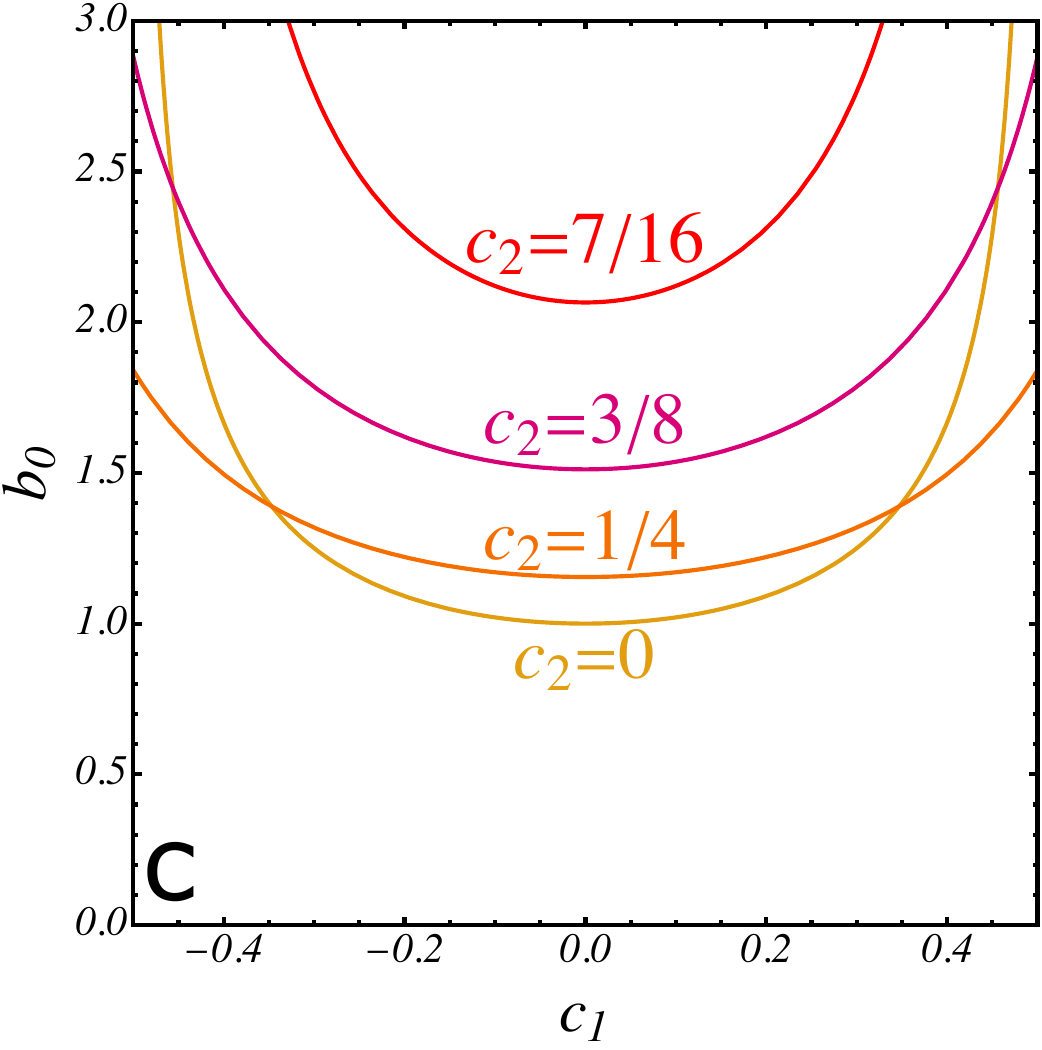}
    
    \caption{\textbf{Results for $b_0$.}
    $(a)$ Combinations of residues yielding the value of $b_0$ in the $(c_1,c_2)$ plane.
    In the gray areas, the integral diverges.
    $(b)$ Values of $b_0$ against $c_1$ for various choices of $c_2<0$ (colored lines).
    $(b)$ Values of $b_0$ against $c_1$ for various choices of $c_2>0$ (colored lines).
    }
    \label{fig:b0M2}
\end{figure}

It is also worth noting that, although their meaning in this system is unclear as of yet, the Cauchy principal values of $b_0$ are well-behaved in most of the cases in which $b_0$ diverges.
The simplest case is that of
\begin{align}
     c_1\neq 0 \text{ and }\frac{1}{4}\left(1 + \sqrt{1-2c_1^2}\right)\leq c_2<\frac{1}{2}.
\end{align}
In that case, all four poles lie on the unit circle, so that
\begin{align}
    \text{PV } b_0 = \frac{1}{2}\sum\limits_{k=1}^4 g_k(c_1,c_2) = 0.
\end{align}

The other two diverging regions are symmetric: in each of them, a pair of poles hits the unit circle via the real axis at the boundary of the domain.
More specifically, $Z_2$ and $Z_4$ hit the circle at $+1$ while $Z_1$ is in the disk for $c_1 <0$ and $Z_1$ and $Z_3$ hit the circle at $-1$ while $Z_4$ is in the disk for $c_1 >0$.
Thus, for $c_2 < 0$, the principal values in each regime assume the values
\begin{align}
    \text{PV } b_0 = \begin{cases}
        g_1(c_1,c_2) + \frac{1}{2}(g_2(c_1,c_2) +g_4(c_1,c_2)) \text{ if $c_1 <0$} \\
        g_4(c_1,c_2) + \frac{1}{2}(g_1(c_1,c_2) +g_3(c_1,c_2)) \text{ if $c_1 >0$}
    \end{cases}.
\end{align}
Both of these values are finite and non-zero.


\subsubsection{General case\label{sec:InvGamma_General}}

The strategy highlighted in the case $M=2$ is in principle usable for any $M \in \mathbb{N}^\star$, but it is difficult to reach generic expressions for an arbitrary $M$.
We here outline the steps for which conclusions can be drawn.

Recalling Eq.~\ref{eq:bp_integralrep}, the general expression of $b_0$ at any finite $M$ is 
\begin{align}
    b_0     &= \frac{1}{2\pi}\int\limits_{-\pi}^\pi du \frac{1}{1 + 2\sum\limits_{m=1}^M c_m \cos m u} = \frac{1}{2\pi} \int\limits_{-\pi}^\pi du R_0(\cos u)
\end{align}
where the rational function $R_0$ can be expressed as a function of the Chebyshev polynomials with degrees smaller than, or equal to, $M$,
\begin{align}
    R_0(X) \equiv \frac{1}{1+2\sum\limits_{m=1}^M c_m T_m(X)}.
\end{align}
To prove the divergence of $b_0$ at a given set of $c_m$ coefficients, it is sufficient to show that the corresponding function
\begin{align}
    f_0(z) = \frac{1}{iz} R_0(\frac{z + z^{-1}}{2})
\end{align}
has a pole on the unit circle for these parameters.
The poles are the roots of a symmetric polynomial with degree $2M$,
\begin{align}
    Q(Z) = c_M (Z^{2M}+1) + c_{M-1}(Z^{2M -1} + Z) + \ldots + c_1(Z^{M+1}+Z^{M-1}) + Z^{M}.
\end{align}
It is not possible to write an analytical expression for the roots of these polynomial in the general case.
However, an easy subcase of interest is that of $c_1 = c_2 =\ldots=c_M = c$.
In that case,
\begin{align}
    Q(Z) = (1-c) Z^M+ c \sum\limits_{n=0}^{2M} Z^n =(1-c)Z^M + c\frac{1-Z^{2M+1}}{1-Z}.
\end{align}
In the case $Z = 1$, this polynomial yields
$Q(1) = 1+2M c$, so that $c = -1/(2M)$, such that $\sum_m c_m = -1/2$, is a root.
Thus, generically, there is a pole on the unit circle for $c=-1/(2M)$ in the case of equal $c$.
This suggests that there is still a diverging entropy production for $\sum_mc_m = -1/2$ even in the fully general case $M>1$ with distinct $c$'s.

More generally, taking $Z=1$ in the original equation for $Q$ and requiring $Q(1) = 0$ yields the condition for divergence
\begin{align}
    1 + 2\sum\limits_{m=1}^M c_m =0.
\end{align}

Likewise, imposing $Q(-1)=0$ reveals another less trivial condition,
\begin{align}
    (-1)^M + 2\sum\limits_{m=1}^M (-1)^m c_m =0,
\end{align}
which is the generalization of the condition $c_2 - c_1 = -1/2$ in the case $M = 2$.

\section{MSD of the center of mass for temporal noise}

Here, it will be interesting to consider the MSD of the center of mass of the system described by our dynamics with temporal noise.
To do so, consider Eq.~\ref{eq:PGD_dynamics} and integrate it from time $0$ to time $t$,
\begin{align}
    \bm{r}_i (t) = \bm{r}_i (0) - \mu_i \sum\limits_{j \neq i} \int\limits_{0}^t dt'\bm{\nabla}_i V(\bm{r}_{ij}(t')) + \sqrt{2 D_0} \int\limits_{0}^t dt' \bm{\eta}_i (t').
\end{align}
Introducing the position of the center of mass as 
\begin{align}
    \bm{r}_G(t) = \frac{1}{N} \sum\limits_{i=1}^N \bm{r}_i(t),
\end{align}
and assuming that interactions are reciprocal, the dynamics of the center of mass simply obey
\begin{align}
    \bm{r}_G (t) = \bm{r}_G (0)  + \frac{\sqrt{2 D_0}}{N} \sum\limits_{i=1}^N\int\limits_{0}^t dt' \bm{\eta}_i (t').
\end{align}
As a result, one may introduce the MSD of the center of mass as
\begin{align}
    \Delta r_G^2 (t) &\equiv \left\langle \left(\bm{r}_G(t) - \bm{r}_G(0)\right)^2 \right\rangle \\
    &= \frac{2 D_0}{N^2} \sum\limits_{i,j=1}^N \iint\limits_{[0;t]^2} dt'dt'' \left\langle \bm{\eta}_i (t') \cdot \bm{\eta}_j(t'') \right\rangle \\
    &= \frac{2 d D_0}{N^2} \sum\limits_{i,j=1}^N \iint\limits_{[0;t]^2} dt'dt'' \left\langle \eta_{i,a} (t') \cdot \eta_{j,a}(t'') \right\rangle \\
    &= \frac{2 d D_0}{N}  \iint\limits_{[0;t]^2} dt'dt'' \Gamma(\bm{\theta}, t'- t'').
\end{align}
Using the expression of $\Gamma$, Eq.\ref{eq:MemoryKernel_cm}, this yields
\begin{align}
    \Delta r_G^2 (t) &= \frac{2 d D_0}{N}  \iint\limits_{[0;t]^2} dt'dt'' \left [ \delta(t'- t'')  + \sum\limits_{m=1}^M c_m \left(  \delta(t'- t'' - m\tau) +  \delta(t'- t''+m\tau) \right) \right] \\
    &= \frac{2 d D_0}{N} \left[t + \sum\limits_{m=1}^M c_m \left( \int\limits_{0}^{t-m\tau} dt' + \int\limits_{m\tau}^{t} dt' \right) \right] \\
    &= \frac{2 d D_0}{N} \left[t + 2 \sum\limits_{m=1}^M c_m \left( t - m\tau \right) \right] \\
    &= \frac{2 d D_0}{N} \left[\left(1 + 2 \sum\limits_{m=1}^M c_m \right)t - 2 \sum\limits_{m=1}^M m c_m  \tau \right].
\end{align}
Thus, the center of mass displays diffusive motion with an effective diffusion constant
\begin{align}
    D_G = \frac{2 d D_0}{N} \left(1 + 2 \sum\limits_{m=1}^M c_m \right).
\end{align}
In particular, this diffusion constant vanishes when
\begin{align}
    \sum\limits_{m=1}^M c_m = -\frac{1}{2}. \label{eq:Static_COM_Condition}
\end{align}
This condition thus ensures that the center of mass is conserved at the level of the MSD.


\section{Adding a standard thermostat to temporal noise \label{sec:Temporal_Thermostat}}

It is interesting to consider what happens in the presence of an additional source of noise that follows from a standard thermal bath in the case of temporal noise, so that the equation of motion, Eq.~\ref{eq:PGD_dynamics}, is replaced by
\begin{align}
    \dot{\bm{r}}_i(t) &= -\mu_i \sum\limits_{j \neq i} \bm{\nabla}_i V(\bm{r}_{ij}) + \sqrt{2 D_0} \bm{\eta}_i(\bm{\theta},t) + \sqrt{2 D_{T}} \bm{\zeta}_i(t)\label{eq:PGD_dynamics_thermal}
\end{align}
with $\bm{\eta}$ following the same correlation pattern as before and 
\begin{align}
    \left\langle \zeta_{i,a}(t) \right\rangle &= 0, \\
    \left\langle \zeta_{i,a}(t) \eta_{j,b}(t') \right\rangle &= 0, \\
    \left\langle \zeta_{i,a}(t) \zeta_{j,b}(t') \right\rangle &= \delta_{ij} \delta_{ab} \delta(t-t') .
\end{align}
In short, this addition leads to an additional factor in the path integral, of the form
\begin{align}
    I_{T} &= \int \prod\limits_{n=1}^N\left[\mathcal{D}\bm{\zeta}_n p_{\zeta}(\bm{\zeta}_n)\right] \exp\left[i \sqrt{2 D_T} \int\limits_{0}^T dt \bm{Q}(t) \cdot \bm{\zeta}(t) \right], \\
    &= \exp\left[ - D_T \int\limits_{0}^T dt Q(t)^2 \right].
\end{align}
Adding this factor to the definition of the dissipative part of the action, Eq.~\ref{eq:Adiss_res_temporal}, one may notice that adding this extra source of noise is exactly equivalent, at the level of the path-integral formulation to replacing $\Gamma$ by $\Gamma_T$ defined as
\begin{align}
    \Gamma_T(\bm{\theta}, t,t') &\equiv \Gamma(\bm{\theta}, t,t') +  \frac{D_T}{D_0} \delta(t - t') .
\end{align}

The expression of the inverse of $\Gamma_T$ is given by a modified version of Eq.~\ref{eq:invGamma_generic_Fourier_inverse},
\begin{align}
    {\Gamma}_T^{-1}(\bm{\theta}; t, t') &= \frac{1}{2\pi} \int dq  \frac{e^{i q (t -t')}}{1+ \rho_T + 2\sum\limits_{m\in \mathbb{N}^\star} c_m \cos mq\tau } , \label{eq:invGamma_thermal_generic_Fourier_inverse}
\end{align}
where we define $\rho_T \equiv D_T/D_0$ for brevity.
Consequently, the coefficients $b_p$ of the inverse are defined through
\begin{align}
    b_p     &= \frac{1}{2\pi}\int\limits_{-\pi}^\pi du \frac{\cos p u}{1 + \rho_T + 2\sum\limits_{m=1}^\infty c_m \cos m u}.\label{eq:bp_integralrep_thermal}
\end{align}
This integral representation may be factored as
\begin{align}
    b_p     &= \frac{1}{1+\rho_T}\frac{1}{2\pi}\int\limits_{-\pi}^\pi du \frac{\cos p u}{1 + 2\sum\limits_{m=1}^\infty \frac{c_m}{1+\rho_T} \cos m u},
\end{align}
which yields the scaling relation
\begin{align}
    b_p(c_1,\ldots,c_M; \rho_T) = \frac{1}{1+\rho_T}\, b_p\!\left(\frac{c_1}{1+\rho_T},\ldots,\frac{c_M}{1+\rho_T}; 0\right), \label{eq:bp_thermal_scaling}
\end{align}
relating the coefficients with thermostat to those without, evaluated at rescaled noise correlations $\widetilde{c}_m \equiv c_m/(1+\rho_T)$.
One may verify this relation by checking the recurrence conditions: defining $\Gamma_{T}^{-1} = \sum_n b_n \delta(t - t' + n\tau)$, the condition $\Gamma_T \cdot \Gamma_T^{-1} = \mathbb{1}$ reads
\begin{align}
    (1+\rho_T)b_0 + 2\sum\limits_{m=1}^M c_m\, b_{m} &= 1, \\
    (1+\rho_T)b_p +\sum\limits_{m = 1}^M c_m (b_{p-m} + b_{p+m}) &= 0, \quad \forall p >0,
\end{align}
both of which are satisfied by Eq.~\ref{eq:bp_thermal_scaling}.

Intuitively, the thermostat dilutes the colored-noise correlations: the effective correlation coefficients shrink from $c_m$ to $\widetilde{c}_m = c_m/(1+\rho_T)$, and the overall weight of $\Gamma_T^{-1}$ is reduced by a factor $1/(1+\rho_T)$.
In particular, the divergence condition $\sum_m c_m = -1/2$ is regularized, since $\sum_m \widetilde{c}_m = \frac{\sum_m c_m}{1+\rho_T} > -1/2$ for any $D_T > 0$.

The derivation of the entropy production itself follows the same steps as in Sec.~\ref{sec:Temporal_EP_Analytical}, but with $\Gamma$ replaced by $\Gamma_T$.
One must however heed the fact that the equilibrium reference changes.
When $c_m = 0$, the combined noise is white with effective diffusion $D_0 + D_T$, so that the steady-state Boltzmann distribution has inverse temperature
\begin{align}
    \beta_T \equiv \frac{1}{(D_0 + D_T)\gamma_0} = \frac{\beta}{1+\rho_T}.
\end{align}
The equilibrium value of $\Gamma_T^{-1}$ at $c_m = 0$ is $\frac{1}{1+\rho_T}\delta(t-t')$, not $\delta(t-t')$.
Consequently, the non-equilibrium part of $\Gamma_T^{-1}$ that drives entropy production is
\begin{align}
    \widetilde{\Upsilon}_{T}(t,t') \equiv \Gamma_T^{-1}(t,t') - \frac{1}{1+\rho_T}\delta(t-t'),
\end{align}
and the entropy production reads
\begin{align}
    \Delta S = -\beta \iint\limits_{[-T;T]^2} dt\,dt'\left[\widetilde{\Upsilon}_{T}(t,t') \dot{\bm{X}}_t \cdot \nabla V(\bm{X}_{t'}) \right]. \label{eq:EPR_Final_thermal}
\end{align}
Using the scaling relation, Eq.~\ref{eq:bp_thermal_scaling}, the components of $\widetilde{\Upsilon}_T$ are $\widetilde{\Upsilon}_{T,n} = \Upsilon_n(\widetilde{c})/(1+\rho_T)$ where $\Upsilon(\widetilde{c})$ denotes the $\Gamma^{-1}(\widetilde{c}) - \mathbb{1}$ matrix at the rescaled coefficients, so that Eq.~\ref{eq:EPR_Final_thermal} simplifies to
\begin{align}
    \Delta S = -\beta_T \iint\limits_{[-T;T]^2} dt\,dt'\left[\Upsilon_{tt'}(\widetilde{\bm{c}}) \dot{\bm{X}}_t \cdot \nabla V(\bm{X}_{t'}) \right]. \label{eq:EPR_thermal_rescaled}
\end{align}
In other words, the entropy production with an additional thermostat is exactly given by the same expression as that without an additional thermostat, but with rescaled coefficients $\widetilde{c}_m = c_m/(1+\rho_T)$ and $\beta \to \beta_T = \beta/(1+\rho_T)$.

In the case $M=1$, the scaling relation together with Eqs.~\ref{eq:b0_singlec}--\ref{eq:b1_singlec} yields
\begin{align}
    b_0 &= \frac{1}{\sqrt{(1 + \rho_T)^2 - 4 c^2}}, \label{eq:b0_thermal}\\
    b_1 &= \frac{1}{2c}\left(1 - \frac{1+\rho_T}{\sqrt{(1 + \rho_T)^2 - 4 c^2}}\right), \label{eq:b1_thermal}\\
    b_2 &= \frac{(1+\rho_T)^2 - 2c^2 - (1+\rho_T)\sqrt{(1+\rho_T)^2 - 4c^2}}{2c^2\sqrt{(1+\rho_T)^2 - 4c^2}}, \label{eq:b2_thermal}
\end{align}
and the general coefficient is given by the hypergeometric formula (cf.\ Eq.~\ref{eq:M=1_bn})
\begin{align}
    b_p = \frac{1}{(1+\rho_T)-2c} {}_3\widetilde{F}_2\left(\left\{ \frac{1}{2},1,1 \right\}, \left\{1-p,1+p \right\}; \frac{4c}{2c -(1+\rho_T)}\right). \label{eq:M=1_bn_thermal}
\end{align}
In particular, for $c = \pm 1/2$,
\begin{align}
    b_0(c = \pm 1/2) = \frac{1}{\sqrt{(1 + \rho_T)^2 - 1}},
\end{align}
which is always finite for $D_T > 0$, with limiting behaviors
\begin{align}
    b_0(c = \pm 1/2) &\underset{\rho_T\to 0}{\sim} \frac{1}{\sqrt{2\rho_T} }, \\
    b_0(c = \pm 1/2) &\underset{\rho_T\to \infty}{\sim}  \frac{1}{\rho_T}.
\end{align}
The instantaneous entropy production rate reads
\begin{align}
    \sigma_0(t) &= -\beta\left(b_0 - \frac{1}{1+\rho_T}\right)\dot{\bm{X}}_t\cdot\nabla V(\bm{X}_t) = \beta_T (b_0(\widetilde{c}) - 1)\dot{W}_t, \label{eq:sigma0_thermal}
\end{align}
which vanishes for $c = 0$ at any $D_T$, as required by equilibrium.

\section{TRS Checks for Spatial noise\label{sec:TRS_Spatial}}

We here focus on the time-reversal properties of the dynamical action of spatial noise dynamics, and how it affects the relation between correlation and response.
First, from Eqs.~\ref{eq:Dynamical_Generating_Function},~\ref{eq:MSRJD_Action} and the definition of the dynamical average of an observable in Eq.~\ref{eq:Dynamical_Average}, one may notice that the correlation between any two components of the state vector can be expressed as
\begin{align}
    C_{ijab}(t,t') \equiv \left\langle x_{i,a}(t)x_{j,b}(t') \right\rangle = \left.\frac{\delta^2 Z_d\left[\bm{J}_{\bm{X}}, \bm{J}_{\bm{Q}}\right]}{\delta J_{x_{i,a}}(t) \delta J_{x_{j,b}}(t')}\right|_{\bm{J}_{\bm{X}} = \bm{0}, \bm{J}_{\bm{Q}} = \bm{0}}.
\end{align}
Likewise, a response is the change in the dynamical variable induced by the change $\bm{F}(\bm{X}) \mapsto \bm{F}(\bm{X}) + \bm{f}(t)$ at time $t$, so that it can be written as
\begin{align}
    R_{mnab}(t,t') &\equiv \left.\frac{\delta \left\langle x_{m,a} (t)\right\rangle_f}{\delta f_{n,b}(t')} \right|_{\bm{f} = \bm{0}}.
\end{align}
Using the definition of the dynamical average, Eq.~\ref{eq:Dynamical_Average}, and defining $\mathcal{A}_f$ the dynamical action with the addition of $\bm{f}$, one then has
\begin{align}
    R_{mnab}(t,t') &= \left.\frac{\delta}{\delta f_{n,b}(t')}\left[ \int \mathcal{D}\bm{X} \mathcal{D}\bm{Q} d\bm{X}_0 x_{m,a}(t) e^{\mathcal{A}_f[\bm{X}, \bm{Q} | \bm{X}_0]} \right]\right|_{\bm{f} = \bm{0}} \\
    &= \left\langle x_{m,a}(t) \left.\frac{\delta \mathcal{A}_f[\bm{X},\bm{Q}|\bm{X}_0]}{\delta f_{n,b}(t')}\right|_{\bm{f} = \bm{0}} \right\rangle \label{eq:Response_ActionDerivative}
\end{align}
and one may notice that adding $f_{n,b}$ to the deterministic part will simply end up in a $i q_{n,b}f_{n,b}$ term in the action, so that the response is the correlator of $\bm{X}$ with the response field $i \bm{Q}$,
\begin{align}
    R_{mnab}(t,t') &= \left\langle x_{m,a}(t) i q_{n,b} (t') \right\rangle \label{eq:Response_as_Average} \\
    &= \left.\frac{\delta^2 Z_d[\bm{J}_{\bm{X}}, \bm{J}_{\bm{Q}}]}{\delta J_{x_{m,a}}(t)\delta J_{q_{n,b}}(t')} \right|_{\bm{J}_{\bm{X}} = \bm{J}_{\bm{Q}} = \bm{0}}
\end{align}

In the case of the action of Eqs.~\ref{eq:Adet'_res_replaced},~\ref{eq:Amul_diss_replaced},~\ref{eq:Amul_jac_replaced}, if $\Lambda$ is independent of the one-body part of $\bm{F}$, one has
\begin{align}
    \mathcal{A}_f[\bm{X}, \bm{Q} | \bm{X}_0] &= \mathcal{A}[\bm{X}, \bm{Q} | \bm{X}_0] +  \int\limits_{0}^T dt \left[ i\bm{Q}(t)\cdot\overline{\overline{D}}\bm{f}(t)- \alpha \bm{\nabla}\overline{\overline{D}}\bm{f}(t)\right] \\
    &= \mathcal{A}[\bm{X}, \bm{Q} | \bm{X}_0] + \sum\limits_{m,n=1}^N\sum\limits_{a,b=1}^d\int\limits_{0}^T dt \left[ iq_{m,a}(t) D_{manb}(\bm{X})f_{nb}(t)- \alpha \partial_{x_{m,a}}D_{manb}(\bm{X})f_{nb}(t)\right]
\end{align}
so that, using Eq.~\ref{eq:Response_ActionDerivative}
\begin{align}
    R_{mnab}(t,t') = \left\langle x_{m,a}(t) \left(i\bm{Q}(t') - \alpha\bm{\nabla}\right)\overline{\overline{D}}(\bm{X}(t'))\hat{\bm{e}}_{n,b}\right\rangle
\end{align}
as usual with multiplicative noise~\cite{ArnoulxdePirey2022}.

The correlation and response functions are usually linked by fluctuation-dissipation relations.
These are established from the MSRJD action by establishing how to transform $\bm{Q}$ under time-reversal so as to leave the action unchanged -- the FDR is then given as the conservation law associated to that symmetry~\cite{ArnoulxdePirey2022}.
Here, consider the action as given by Eqs.~\ref{eq:Adet'_res_replaced},~\ref{eq:Amul_diss_replaced},~\ref{eq:Amul_jac_replaced}.
Then, consider two trajectories: one running forward from $\bm{X}_{-T}$ at time $-T$ for a time $2T$, and one running from the final state of the first trajectory $\bm{X}_T$ at time $-T$, with dynamics modified by the time-reversal transformation $\mathcal{T}$, for a time $2T$.
Note the introduction of short-hand notations and of a symmetrized time interval to lighten the procedure.
The real-space variables transform as:
\begin{align}
    \mathcal{T} \bm{X}_t = \bm{X}_{-t}.
\end{align}
The objective is to find how the response field and the parameters of the dynamics ($\alpha$, $c$) need to transform into $(\overline{\alpha},\overline{c})$ to leave the action invariant under $\mathcal{T}$.
It is convenient to temporarily define $\bm{\chi}_{-t}$ such that
\begin{align}
    \mathcal{T} \left[ i \bm{Q}_t\right] = i \bm{Q}_{-t} + \bm{\chi}_{-t}
\end{align}
where, in the usual additive noise case, for instance, one needs $\bm{\chi}_{-t} = \beta d_t \bm{X}_{-t}$.
We here treat simpler cases first and build up to the most generic case.

\subsection{Case of $c = 0$}

First, we consider the simple case of $c = 0$, which resembles conventional multiplicative noise the most.
Then, the various parts of the action read:
\begin{align}
    \mathcal{A}_{\text{det}'} &= \ln p_0 (\bm{X}_{-T}) + i  \int\limits_{-T}^T dt \bm{Q}(t)\cdot \left[ \overline{\overline{D}}(\bm{X}) \,\overline{\overline{M}} \bm{\nabla} V(\bm{X}) \right] \label{eq:Adet'_c0} \\
    \mathcal{A}_{\text{diss}} &=  \int\limits_{-T}^T dt \left(- i \bm{Q}(t)\cdot\dot{\bm{X}}(t) -D_0 Q(t)^2 -\frac{1}{2}\left( \bm{Q}(t) \cdot \overline{\overline{\Pi}} \right)^2 +(1- 2\alpha) \left( i\bm{Q}(t) \cdot \overline{\overline{\Pi}} \right)\cdot \nabla\overline{\overline{\Pi}}\right)\label{eq:Amul_diss_c0} \\
    \mathcal{A}_{\text{jac}} &= -\alpha \int\limits_{-T}^T dt \left( \bm{\nabla} \left(\overline{\overline{D}}(\bm{X}) \, \overline{\overline{M}}\bm{\nabla}  V(\bm{X},t)\right)  + (1-\alpha) \left(  (\nabla \overline{\overline{\Pi}})^2 +  \overline{\overline{\Pi}} \nabla^2 \overline{\overline{\Pi}}\right) \right)\label{eq:Amul_jac_c0}
\end{align}
We first focus on the time-reversed version of the dissipative part of the action.
It is more convenient to rewrite the latter as
\begin{align}
    \mathcal{A}_{\text{diss}} &=  \int\limits_{-T}^T dt (- i \bm{Q}_t)\cdot\left(  \overline{\overline{D}}(- i \bm{Q}_t) + d_t^{(\alpha)}\bm{X}_t \right)
\end{align}
where 
\begin{align}
    d_t^{(\alpha)}\bm{X}_t \equiv d_t \bm{X}_t - (1-2\alpha) \overline{\overline{\Pi}}_t\nabla \overline{\overline{\Pi}}_t.
\end{align}
In order for the latter to transform like a usual time derivative, one needs
\begin{align}
    \mathcal{T} d_t^{(\alpha)}\bm{X}_t = d_t^{(\alpha)} \bm{X}_{-t} = - d_t^{(\alpha)} \bm{X}_t,
\end{align}
which is achieved if $\overline{\alpha} = 1-\alpha$.
With that convention, the correct time-reversal operation to keep the dissipative part constant is
\begin{align}
    \mathcal{T}\bm{X}_t &= \bm{X}_{-t} \label{eq:TRS_X_c=0} \\
    \mathcal{T} i \bm{Q}_t &= i \bm{Q}_{-t} + \overline{\overline{D}}{}^{-1}d_t^{(\alpha)} \bm{X}_{-t} \label{eq:TRS_Q_c=0}
\end{align}
where $\overline{\overline{D}}$ was already shown to be invertible.
Indeed, injecting this expression into the time-reversed dissipative action yields
\begin{align}
    \mathcal{T}\mathcal{A}_{\text{diss}} &= \int\limits_{-T}^T dt \left(- i \bm{Q}_{-t} - \overline{\overline{D}}{}^{-1}d_t^{(\alpha)} \bm{X}_{-t}\right)\cdot\left( \overline{\overline{D}}\left(- i \bm{Q}_{-t}- \overline{\overline{D}}{}^{-1}d_t^{(\alpha)} \bm{X}_{-t} \right) + d_t^{(\alpha)}\bm{X}_{-t} \right) \\
    &= \int\limits_{-T}^T dt \left(- i \bm{Q}_{-t} \right)\cdot\left( \overline{\overline{D}}\left(- i \bm{Q}_{-t}\right) - d_t^{(\alpha)}\bm{X}_{-t} \right) \\
    &= \int\limits_{-T}^T dt \left(- i \bm{Q}_{-t} \right)\cdot\left( \overline{\overline{D}}\left(- i \bm{Q}_{-t}\right) + d_{-t}^{(\alpha)}\bm{X}_{-t} \right) \\
    &= \mathcal{A}_{\text{diss}}
\end{align}
The same expression can be used on the remainder of the action, here grouped up into a single term
\begin{align}
    \mathcal{T}\mathcal{A}_{\text{det}+\text{jac}} &= \ln p_0 (\bm{X}_{T}) + \int\limits_{-T}^T dt \left(i\bm{Q}_{-t} + \overline{\overline{D}}{}^{-1}d_t^{(\alpha)} \bm{X}_{-t} \right)\cdot \left[ \overline{\overline{D}} \,\overline{\overline{M}} \bm{\nabla} V \right] -(1-\alpha) \int\limits_{-T}^T dt \left( \bm{\nabla} \left(\overline{\overline{D}} \, \overline{\overline{M}}\bm{\nabla}  V\right)  + \alpha \left(  (\nabla \overline{\overline{\Pi}})^2 +  \overline{\overline{\Pi}} \nabla^2 \overline{\overline{\Pi}}\right) \right) \\
    &= \mathcal{A}_{\text{det}+\text{jac}} +\ln p_0 (\bm{X}_{T}) - \ln p_0 (\bm{X}_{-T}) - \int\limits_{-T}^T dt  \overline{\overline{D}}{}^{-1}d_t^{(\alpha)} \bm{X}_{t} \cdot \left[ \overline{\overline{D}} \,\overline{\overline{M}} \bm{\nabla} V \right] - (1 - 2 \alpha)\int\limits_{-T}^T dt\bm{\nabla} \left(\overline{\overline{D}} \, \overline{\overline{M}}\bm{\nabla}  V\right) \\
    &= \mathcal{A}_{\text{det}+\text{jac}} +\ln p_0 (\bm{X}_{T}) - \ln p_0 (\bm{X}_{-T}) - \int\limits_{-T}^T dt  \overline{\overline{D}}{}^{-1}d_t \bm{X}_{t} \cdot \left[ \overline{\overline{D}} \,\overline{\overline{M}} \bm{\nabla} V \right]  \nonumber \\
    &\hphantom{aaaaaa} + (1 - 2 \alpha)\int\limits_{-T}^T dt  \overline{\overline{D}}{}^{-1}\overline{\overline{\Pi}}\nabla \overline{\overline{\Pi}} \cdot \left[ \overline{\overline{D}} \,\overline{\overline{M}} \bm{\nabla} V \right] - (1 - 2 \alpha)\int\limits_{-T}^T dt \bm{\nabla} \left(\overline{\overline{D}} \, \overline{\overline{M}}\bm{\nabla}  V\right).
\end{align}
Using the diagonal nature of $\overline{\overline{D}}$ and the expression of its gradient, $\nabla \overline{\overline{D}} = \nabla \overline{\overline{\Pi}}{}^2 / 2$, this expression simplifies to
\begin{align}
    \mathcal{T}\mathcal{A}_{\text{det}+\text{jac}} &= \mathcal{A}_{\text{det}+\text{jac}} +\ln p_0 (\bm{X}_{T}) - \ln p_0 (\bm{X}_{-T}) - \int\limits_{-T}^T dt\, d_t \bm{X}_{t} \cdot \left[  \,\overline{\overline{M}} \bm{\nabla} V \right] - (1 - 2 \alpha)\int\limits_{-T}^T dt \overline{\overline{D}} \, \overline{\overline{M}}\bm{\nabla}^2  V.
\end{align}
Finally, one may assume that all mobilities are identical and use the chain rule as modified in the context of the $\alpha$-discretization,
\begin{align}
    d_t V_t = d_t \bm{X}_t \cdot \bm{\nabla} V_t + (1-2\alpha) \overline{\overline{D}} \bm{\nabla}^2  V
\end{align}
so that (taking the scale $D_0$ out of the potential),
\begin{align}
    \mathcal{T}\mathcal{A}_{\text{det}+\text{jac}} &= \mathcal{A}_{\text{det}+\text{jac}} +\ln p_0 (\bm{X}_{T}) - \ln p_0 (\bm{X}_{-T}) +\frac{\mu}{D_0} \int\limits_{-T}^T dt\, \,d_t V_{t} .
\end{align}
Since $\ln p_0 = - \beta V$, the integral then exactly compensates the difference of log-measures, and
\begin{align}
    \mathcal{T}\mathcal{A}_{\text{det}+\text{jac}} &= \mathcal{A}_{\text{det}+\text{jac}}.
\end{align}
Thus, the time-reversal transformation does leave the whole action invariant.

A direct consequence is that, defining the response function from Eq.~\ref{eq:Response_as_Average}, and using time-reversal symmetry as defined by Eqs.~\ref{eq:TRS_X_c=0} and~\ref{eq:TRS_Q_c=0}, one finds
\begin{align}
    R_{mnab}(t,t'; \alpha) &= \mathcal{T}\left\langle x_{m,a}(t) i q_{n,b} (t') \right\rangle \\
    &= \left\langle x_{m,a}(-t) i \left( q_{n,b}(-t') + \sum\limits_{p=1}^N \sum\limits_{c = 1}^d\left(\overline{\overline{D}}{}^{-1}\right)_{nbpc} d_{t'} ^{(\alpha)} x_{p,c}(-t')\right)_{n,b}  \right\rangle \\
    &= R_{mnab}(-t,-t'; 1- \alpha) +  \sum\limits_{p=1}^N \sum\limits_{c = 1}^d \left(\overline{\overline{D}}{}^{-1}\right)_{nbpc} d_{t'}^{(\alpha)} C_{mpac}(-t,-t'; 1-\alpha),
\end{align}
where the value of $\alpha$ is written explicitly to remind that it is also modified by the time-reversal operator.
To simplify this expression, recall (see Sec.~\ref{sec:Special_Spatial_Cases}) that $\overline{\overline{D}}$ is diagonal in the case $c=0$, so that the expression simplifies to
\begin{align}
    R_{mnab}(t,t'; \alpha) &= R_{mnab}(-t,-t'; 1-\alpha) + D_{nb}^{-1}(t') d_{t'}^{(\alpha)} C_{mnab}(-t,-t'; 1-\alpha).
\end{align}
Since the physics cannot depend on the choice of $\alpha$, one usually drops the $\alpha$ dependence entirely from this expression~\cite{ArnoulxdePirey2022}, leading to 
\begin{align}
    R_{mnab}(t,t') &= R_{mnab}(-t,-t') + D_{nb}^{-1}(t') d_{t'} C_{mnab}(-t,-t').
\end{align}
The time-reversal operation can also be applied to $C$, leading to the much simpler relation
\begin{align}
    C_{mnab}(t,t'; \alpha) &= C_{mnab}(-t,-t'; 1-\alpha)
\end{align}
which, dropping the $\alpha$ dependence once again, yields
\begin{align}
    R_{mnab}(t,t') &= R_{mnab}(-t,-t') + D_{nb}^{-1}(t') d_{t'} C_{mnab}(t,t').
\end{align}
Recalling that (see Sec.~\ref{sec:Special_Spatial_Cases}) the expression of the diagonal elements of $D$ read, in the case $c = 0$,
\begin{align}
    D_{nb} = D_0 + \frac{1}{2}\sum\limits_{p\neq n} \Lambda_{np}
\end{align}
this expression can be made more explicit,
\begin{align}
    R_{mnab}(t,t') &= R_{mnab}(-t,-t') + \left( D_0 + \frac{1}{2}\sum\limits_{p\neq n} \Lambda_{np} (\bm{X}_{t'}) \right)^{-1} d_{t'} C_{mnab}(t,t'). \label{eq:c=0_FDR}
\end{align}
Finally, one may notice that causality imposes $R(t,t') = 0$ if $t < t'$ (recall that $t'$ is the time at which a perturbation is applied to the system) so that the expression above can be simplified to read
\begin{align}
    R_{mnab}(t,t') &= \Theta(t- t') \left( D_0 + \frac{1}{2}\sum\limits_{p\neq n} \Lambda_{np} (\bm{X}_{t'}) \right)^{-1} d_{t'} C_{mnab}(t,t'). \label{eq:c=0_FDR_causal}
\end{align}
with $\Theta$ a Heaviside step.
Finally, owing to the fact that the symmetry is here written for the equilibrium distribution one may invoke time-translational invariance of the response, so that all functions save $\Lambda$ become dependent only on $\tau = t'-t$ and
\begin{align}
    R_{mnab}(\tau) &= \Theta(\tau)\left( D_0 + \frac{1}{2}\sum\limits_{p\neq n} \Lambda_{np} (\bm{X}_{t'}) \right)^{-1} d_{\tau} C_{mnab}(\tau). \label{eq:c=0_FDR_TTI_causal}
\end{align}
This last expression is the Fluctuation-Dissipation Relation for independent multiplicative pairwise noise.
It shows that the effect of the noise, in this case, it to turn the usual temperature into an effective, state-dependent one.

\subsection{Case of a symmetric $\Lambda$}

We now turn our attention to the more general case of a symmetric $\overline{\overline{\Pi}}$ with an arbitrary $c$.
Then, 
\begin{align}
    \mathcal{A}_{\text{det}'} &= \ln p_0 (\bm{X}_{-T}) + i  \int\limits_{-T}^T dt \bm{Q}_t\cdot \left[ \overline{\overline{D}}(\bm{X}) \,\overline{\overline{M}} \bm{\nabla} V(\bm{X}) \right] \\
    \mathcal{A}_{\text{diss}} &=  \int\limits_{-T}^T dt \left(- i \bm{Q}_t\cdot\dot{\bm{X}}_t -D_0 Q_t^2 -\frac{1}{2}\left( \bm{Q}_t \cdot \overline{\overline{\Pi}} \right)^2 +(1- 2\alpha + c (\alpha - 1)) \left( i\bm{Q}_t \cdot \overline{\overline{\Pi}} \right)\cdot \nabla\overline{\overline{\Pi}}\right) \\
    \mathcal{A}_{\text{jac}} &= -\alpha \int\limits_{-T}^T dt \left( \bm{\nabla} \left(\overline{\overline{D}}\, \overline{\overline{M}}\bm{\nabla}  V\right)  + \frac{1}{2}(1-\alpha)(1-c) \nabla\cdot  \overline{\overline{\Pi}}{}^2+ \frac{\alpha c}{2}  (\nabla \overline{\overline{\Pi}})^2 \right)
\end{align}

\subsubsection{Flipping $\alpha$ and keeping $c$ unchanged does not work}

As a first naïve attempt, one may write the dissipative action with $c$ as a fixed parameter,
\begin{align}
    \mathcal{A}_{\text{diss}} &=  \int\limits_{-T}^T dt (- i \bm{Q}_t)\cdot\left(  \overline{\overline{D}}(- i \bm{Q}_t) + d_t^{(\alpha)}\bm{X}_t + c(1 -\alpha)\overline{\overline{\Pi}}_t\nabla \overline{\overline{\Pi}}_t \right)
\end{align}
with the usual definition for the effective time derivative,
\begin{align}
    d_t^{(\alpha,c)}\bm{X}_t \equiv d_t \bm{X}_t - (1-2\alpha) \overline{\overline{\Pi}}_t\nabla \overline{\overline{\Pi}}_t.
\end{align}
With this definition, the only time-reversal operation that will cancel out all terms proportional to $d_t \bm{X}_{-t}$ is
\begin{align}
    \mathcal{T}\bm{X}_t &= \bm{X}_{-t} \label{eq:TRS_X_Symm_Lambda} \\
    \mathcal{T} i \bm{Q}_t &= i \bm{Q}_{-t} + \overline{\overline{D}}{}^{-1}d_t^{(\alpha)} \bm{X}_{-t} + c\alpha \overline{\overline{D}}{}^{-1} \overline{\overline{\Pi}}\nabla \overline{\overline{\Pi}} .\label{eq:TRS_Q_Symm_Lambda}
\end{align}
With that definition, using the results from the $c=0$ case, all the $c$-independent terms will be vanish in $(\mathcal{T} - \mathcal{I})\mathcal{A}_{\text{diss}}$, leaving only the $c$-dependent terms.
\begin{align}
    \mathcal{T}\mathcal{A}_{\text{diss}} &= \int\limits_{-T}^T dt \left(- i \bm{Q}_{-t} - \overline{\overline{D}}{}^{-1}d_t^{(\alpha)} \bm{X}_{-t} - c\alpha \overline{\overline{D}}{}^{-1} \overline{\overline{\Pi}}\nabla \overline{\overline{\Pi}}\right)\cdot\left( \overline{\overline{D}}\left(- i \bm{Q}_{-t}- \overline{\overline{D}}{}^{-1}d_t^{(\alpha)} \bm{X}_{-t} - c\alpha \overline{\overline{D}}{}^{-1} \overline{\overline{\Pi}}\nabla \overline{\overline{\Pi}}\right) + d_t^{(\alpha)}\bm{X}_{-t} + c\alpha\overline{\overline{\Pi}}_t\nabla \overline{\overline{\Pi}}_t\right) \\
    &= \mathcal{A}_{\text{diss}}(c=0)- c \alpha\int\limits_{-T}^T dt  \overline{\overline{D}}{}^{-1} \overline{\overline{\Pi}}\nabla \overline{\overline{\Pi}}\cdot\left( \overline{\overline{D}}\left(- i \bm{Q}_{-t}- \overline{\overline{D}}{}^{-1}d_t^{(\alpha)} \bm{X}_{-t} - c\alpha \overline{\overline{D}}{}^{-1} \overline{\overline{\Pi}}\nabla \overline{\overline{\Pi}}\right) + d_t^{(\alpha)}\bm{X}_{-t} + c\alpha\overline{\overline{\Pi}}_t\nabla \overline{\overline{\Pi}}_t\right) \nonumber \\
    &\hphantom{aa} + c\alpha \int\limits_{-T}^T dt \left(- i \bm{Q}_{-t} - \overline{\overline{D}}{}^{-1}d_t^{(\alpha)} \bm{X}_{-t}\right)\cdot \left( \overline{\overline{D}}\left(- \overline{\overline{D}}{}^{-1} \overline{\overline{\Pi}}\nabla \overline{\overline{\Pi}}\right) + \overline{\overline{\Pi}}_t\nabla \overline{\overline{\Pi}}_t\right) \\
    &= \mathcal{A}_{\text{diss}}(c=0) - \int\limits_{-T}^T dt \left(- i \bm{Q}_{-t} \right)\cdot\left(c \alpha \overline{\overline{\Pi}}_t\nabla \overline{\overline{\Pi}}_t \right) \\
    &= \mathcal{A}_{\text{diss}} - c \int\limits_{-T}^T dt \left(- i \bm{Q}_{-t} \right)\cdot\left( \overline{\overline{\Pi}}_t\nabla \overline{\overline{\Pi}}_t \right)
\end{align}
Thus, keeping $c$ outside of the derivative and simply performing the replacement $\alpha \mapsto 1 - \alpha$ is not enough to ensure time-reversal symmetry.

\subsubsection{Changing $c$ and flipping $\alpha$ does not work}

To solve this conundrum, notice that the dissipative part of the action is relatively similar to that of the $c = 0$ case.
It is thus tempting to write it as
\begin{align}
    \mathcal{A}_{\text{diss}} &=  \int\limits_{-T}^T dt (- i \bm{Q}_t)\cdot\left(  \overline{\overline{D}}(- i \bm{Q}_t) + d_t^{(\alpha,c)}\bm{X}_t \right)
\end{align}
where 
\begin{align}
    d_t^{(\alpha,c)}\bm{X}_t \equiv d_t \bm{X}_t - (1-2\alpha + c(\alpha - 1)) \overline{\overline{\Pi}}_t\nabla \overline{\overline{\Pi}}_t.
\end{align}
The time-reversed version of this modified derivative reads
\begin{align}
    \mathcal{T}d_t^{(\alpha,c)}\bm{X}_t \equiv d_t \bm{X}_{-t} - (1-2\overline{\alpha} + \overline{c}(\overline{\alpha} - 1)) \overline{\overline{\Pi}}_t\nabla \overline{\overline{\Pi}}_t.
\end{align}
The key question is then whether one wants to allow $\overline{c} \neq c$.

\begin{figure}
    \centering
    \includegraphics[width=0.5\linewidth]{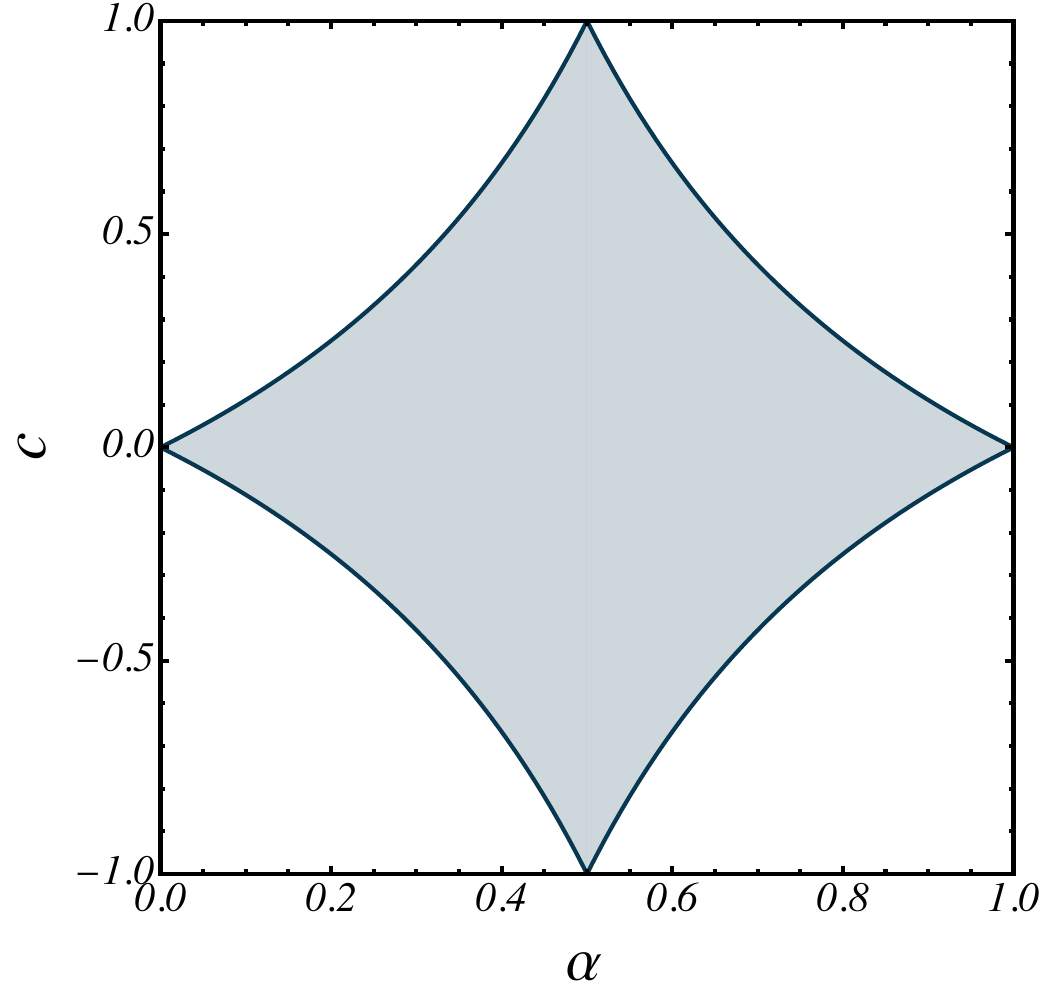}
    \caption{\textbf{Allowed $c$ choices at a given $\alpha$ with $c$ changed by TR.}
    The safe choices are shown as a shaded region.
    The solid lines correspond to all four sign combinations of $\pm[\alpha/(1-\alpha)]^{\pm1}$.
    }
    \label{fig:safe_cs}
\end{figure}

If one allows $\overline{c}$ to be transformed with time-reversal to make sure that $d_t^{(\alpha,c)}\bm{X}_t$ transforms like a time derivative, \textit{i.e.} $1-2\overline{\alpha} + \overline{c}(\overline{\alpha} - 1) = 2\alpha - 1 + c (1-\alpha)$ and uses the standard transformation for $\alpha$, the needed transformations are
\begin{align}
    \overline{\alpha} &= 1 - \alpha, \\
    \overline{c} &= c \frac{\alpha - 1}{\alpha}. 
\end{align}
In particular, note that $\alpha = 0$ and $\alpha = 1$ lead to problematic time-reversed $c$, while $\alpha = 1/2$ flips the sign of $c$.
Also note that this expression is problematic for a whole range of $\alpha$ for which $c$ gets out of the $[-1;1]$ interval -- if $c$ is defined as a Pearson correlation, this is not allowed.
It is easy to work out the set of $(c,\alpha)$ that are unproblematic, by requiring that $c$ remains in $[-1;1]$ under both one and two successive time reversal operations.
In the second time reversal operation, $\alpha \mapsto 1-\alpha$ in both the action and the replacement rule, symmetrizing the constraints around $\alpha = 1/2$.
The result is shown in Fig.~\ref{fig:safe_cs}
Note in particular that the Stratonovich convention allows for all $c$ values, while $c = 0$ allows for all discretization conventions, but that any other choice is constrained.
Assuming that one may choose such combinations,  using these rules and the time-reversal operation defined by 
\begin{align}
    \mathcal{T}\bm{X}_t &= \bm{X}_{-t} \label{eq:TRS_X_Symm_Lambda_cshift} \\
    \mathcal{T} i \bm{Q}_t &= i \bm{Q}_{-t} + \overline{\overline{D}}{}^{-1}d_t^{(\alpha,c)} \bm{X}_{-t}\label{eq:TRS_Q_Symm_Lambda_cshift}
\end{align}
will leave the dissipative part of the action unchanged following the same steps as in the $c=0$ case.

Next, one needs to check whether the other parts of the action are invariant under this transformation.
Recalling the expression of the sum of the remaining two terms, 
\begin{align}
    \mathcal{A}_{\text{det}+\text{jac}} &= \ln p_0 (\bm{X}_{-T}) + i  \int\limits_{-T}^T dt \bm{Q}_t\cdot \left[ \overline{\overline{D}}(\bm{X}) \,\overline{\overline{M}} \bm{\nabla} V(\bm{X}) \right]  -\alpha \int\limits_{-T}^T dt \left( \bm{\nabla} \left(\overline{\overline{D}}\, \overline{\overline{M}}\bm{\nabla}  V\right)  + \frac{1}{2}(1-\alpha)(1-c) \nabla\cdot  \overline{\overline{\Pi}}{}^2+ \frac{\alpha c}{2}  (\nabla \overline{\overline{\Pi}})^2 \right)
\end{align}

\begin{align}
    \mathcal{T}\mathcal{A}_{\text{det}+\text{jac}} &= \ln p_0 (\bm{X}_{T}) +  \int\limits_{-T}^T dt \left( i\bm{Q}_{-t} + \overline{\overline{D}}{}^{-1} d_t^{(\alpha,c)}\bm{X}_{-t} \right)\cdot \left[ \overline{\overline{D}}(\bm{X}) \,\overline{\overline{M}} \bm{\nabla} V(\bm{X}) \right] \nonumber \\
    &\hphantom{aa} -(1-\alpha) \int\limits_{0}^T dt \left( \bm{\nabla} \left(\overline{\overline{D}}\, \overline{\overline{M}}\bm{\nabla}  V\right)  + \frac{1}{2}\alpha(1-c\frac{\alpha - 1}{\alpha}) \nabla\cdot  \overline{\overline{\Pi}}{}^2+ \frac{(1-\alpha) c (\alpha - 1)}{2\alpha}  (\nabla \overline{\overline{\Pi}})^2 \right) \\
    &= \mathcal{A}_{\text{det}+\text{jac}}  + \ln p_0 (\bm{X}_{T}) - \ln p_0(\bm{X}_{-T}) +  \int\limits_{-T}^T dt \left( \overline{\overline{D}}{}^{-1} d_t^{(\alpha,c)}\bm{X}_{-t} \right)\cdot \left[ \overline{\overline{D}}(\bm{X}) \,\overline{\overline{M}} \bm{\nabla} V(\bm{X}) \right] \nonumber \\
    &\hphantom{aa} + \alpha \int\limits_{-T}^T dt \left( \bm{\nabla} \left(\overline{\overline{D}}\, \overline{\overline{M}}\bm{\nabla}  V\right)  + \frac{1}{2}(1-\alpha)(1-c) \nabla\cdot  \overline{\overline{\Pi}}{}^2+ \frac{\alpha c}{2}  (\nabla \overline{\overline{\Pi}})^2 \right) \nonumber \\
    &\hphantom{aa} -(1-\alpha) \int\limits_{0}^T dt \left( \bm{\nabla} \left(\overline{\overline{D}}\, \overline{\overline{M}}\bm{\nabla}  V\right)  + \frac{1}{2}(\alpha-c (\alpha - 1)) \nabla\cdot  \overline{\overline{\Pi}}{}^2 - c \frac{(1-\alpha)^2}{2\alpha}  (\nabla \overline{\overline{\Pi}})^2 \right).
\end{align}
Replacing the effective time derivative by its expression, this then yields
\begin{align}
    \mathcal{T}\mathcal{A}_{\text{det}+\text{jac}} 
    &= \mathcal{A}_{\text{det}+\text{jac}}  + \ln p_0 (\bm{X}_{T}) - \ln p_0(\bm{X}_{-T}) -  \int\limits_{-T}^T dt \left( \overline{\overline{D}}{}^{-1} d_t\bm{X}_{t} \right)\cdot \left[ \overline{\overline{D}}(\bm{X}) \,\overline{\overline{M}} \bm{\nabla} V(\bm{X}) \right] \nonumber \\
    &\hphantom{aa} - (1-2\alpha + c(\alpha - 1)) \int\limits_{-T}^T dt \overline{\overline{D}}{}^{-1}\left(\overline{\overline{\Pi}}\nabla \overline{\overline{\Pi}}\right)\cdot  \left[ \overline{\overline{D}}(\bm{X}) \,\overline{\overline{M}} \bm{\nabla} V(\bm{X}) \right]\nonumber \\
    &\hphantom{aa} + \alpha \int\limits_{-T}^T dt \left( \bm{\nabla} \left(\overline{\overline{D}}\, \overline{\overline{M}}\bm{\nabla}  V\right)  + \frac{1}{2}(1-\alpha)(1-c) \nabla\cdot  \overline{\overline{\Pi}}{}^2+ \frac{\alpha c}{2}  (\nabla \overline{\overline{\Pi}})^2 \right) \nonumber \\
    &\hphantom{aa} -(1-\alpha) \int\limits_{0}^T dt \left( \bm{\nabla} \left(\overline{\overline{D}}\, \overline{\overline{M}}\bm{\nabla}  V\right)  + \frac{1}{2}(\alpha-c (\alpha - 1)) \nabla\cdot  \overline{\overline{\Pi}}{}^2 - c \frac{(1-\alpha)^2}{2\alpha}  (\nabla \overline{\overline{\Pi}})^2 \right).
\end{align}
Grouping up terms that depend on $c$, this can be rewritten as
\begin{align}
    \mathcal{T}\mathcal{A}_{\text{det}+\text{jac}} 
    &= \mathcal{A}_{\text{det}+\text{jac}}  + \ln p_0 (\bm{X}_{T}) - \ln p_0(\bm{X}_{-T}) -  \int\limits_{-T}^T dt \left( \overline{\overline{D}}{}^{-1} d_t\bm{X}_{t} \right)\cdot \left[ \overline{\overline{D}}(\bm{X}) \,\overline{\overline{M}} \bm{\nabla} V(\bm{X}) \right] \nonumber \\
    &\hphantom{aa} - (1-2\alpha) \int\limits_{-T}^T dt \overline{\overline{D}}{}^{-1}\left(\overline{\overline{\Pi}}\nabla \overline{\overline{\Pi}}\right)\cdot  \left[ \overline{\overline{D}}(\bm{X}) \,\overline{\overline{M}} \bm{\nabla} V(\bm{X}) \right] - (1-2\alpha) \int\limits_{-T}^T dt\bm{\nabla} \left(\overline{\overline{D}}\, \overline{\overline{M}}\bm{\nabla}  V\right)\nonumber \\
    &\hphantom{aa} + \alpha c \int\limits_{-T}^T dt \left(  \frac{1}{2}(\alpha-1) \nabla\cdot  \overline{\overline{\Pi}}{}^2+ \frac{\alpha }{2}  (\nabla \overline{\overline{\Pi}})^2 \right) \nonumber \\
    &\hphantom{aa} - c (1-\alpha) \int\limits_{0}^T dt \left( \frac{1}{2}(1 - \alpha) \nabla\cdot  \overline{\overline{\Pi}}{}^2 - \frac{(1-\alpha)^2}{2\alpha}  (\nabla \overline{\overline{\Pi}})^2 \right) \nonumber \\
    &\hphantom{aa} - c(\alpha - 1) \int\limits_{-T}^T dt \overline{\overline{D}}{}^{-1}\left(\overline{\overline{\Pi}}\nabla \overline{\overline{\Pi}}\right)\cdot  \left[ \overline{\overline{D}}(\bm{X}) \,\overline{\overline{M}} \bm{\nabla} V(\bm{X}) \right]
\end{align}
and, simplifying as many terms as possible,
\begin{align}
    \mathcal{T}\mathcal{A}_{\text{det}+\text{jac}} &= \mathcal{A}_{\text{det}+\text{jac}}  + \ln p_0 (\bm{X}_{T}) - \ln p_0(\bm{X}_{-T}) -  \int\limits_{-T}^T dt \left( \overline{\overline{D}}{}^{-1} d_t\bm{X}_{t} \right)\cdot \left[ \overline{\overline{D}}(\bm{X}) \,\overline{\overline{M}} \bm{\nabla} V(\bm{X}) \right] \nonumber \\
    &\hphantom{aa} + (1-2\alpha) \int\limits_{-T}^T dt \overline{\overline{D}}{}^{-1}\left(\overline{\overline{\Pi}}\nabla \overline{\overline{\Pi}}\right)\cdot  \left[ \overline{\overline{D}}(\bm{X}) \,\overline{\overline{M}} \bm{\nabla} V(\bm{X}) \right] - (1-2\alpha) \int\limits_{-T}^T dt\bm{\nabla} \left(\overline{\overline{D}}\, \overline{\overline{M}}\bm{\nabla}  V\right)\nonumber \\
    &\hphantom{aa} +  \frac{c}{2} (\alpha - 1) \int\limits_{-T}^T dt \nabla\cdot  \overline{\overline{\Pi}}{}^2 + c \frac{\alpha^3 - (1 - \alpha)^3}{2\alpha} \int\limits_{0}^T dt  (\nabla \overline{\overline{\Pi}})^2  - c(\alpha - 1) \int\limits_{-T}^T dt \overline{\overline{D}}{}^{-1}\left(\overline{\overline{\Pi}}\nabla \overline{\overline{\Pi}}\right)\cdot  \left[ \overline{\overline{D}}(\bm{X}) \,\overline{\overline{M}} \bm{\nabla} V(\bm{X}) \right].
\end{align}
Using the expression of the gradient of $D$ and the chain rule, this simplifies to yield
\begin{align}
    \mathcal{T}\mathcal{A}_{\text{det}+\text{jac}} &= \mathcal{A}_{\text{det}+\text{jac}}  +  \frac{c}{2} (\alpha - 1) \int\limits_{-T}^T dt \nabla\cdot  \overline{\overline{\Pi}}{}^2 + c \frac{\alpha^3 - (1 - \alpha)^3}{2\alpha} \int\limits_{0}^T dt  (\nabla \overline{\overline{\Pi}})^2  - c(\alpha - 1) \int\limits_{-T}^T dt \overline{\overline{D}}{}^{-1}\left(\overline{\overline{\Pi}}\nabla \overline{\overline{\Pi}}\right)\cdot  \left[ \overline{\overline{D}}(\bm{X}) \,\overline{\overline{M}} \bm{\nabla} V(\bm{X}) \right].
\end{align}

This expression has no obvious reason to cancel, even for a specific choice of $\alpha$, except if $c = 0$.

\subsubsection{Keeping $c$ fixed and changing $\alpha$ does not work}

\begin{figure}
    \centering
    \includegraphics[width=0.5\linewidth]{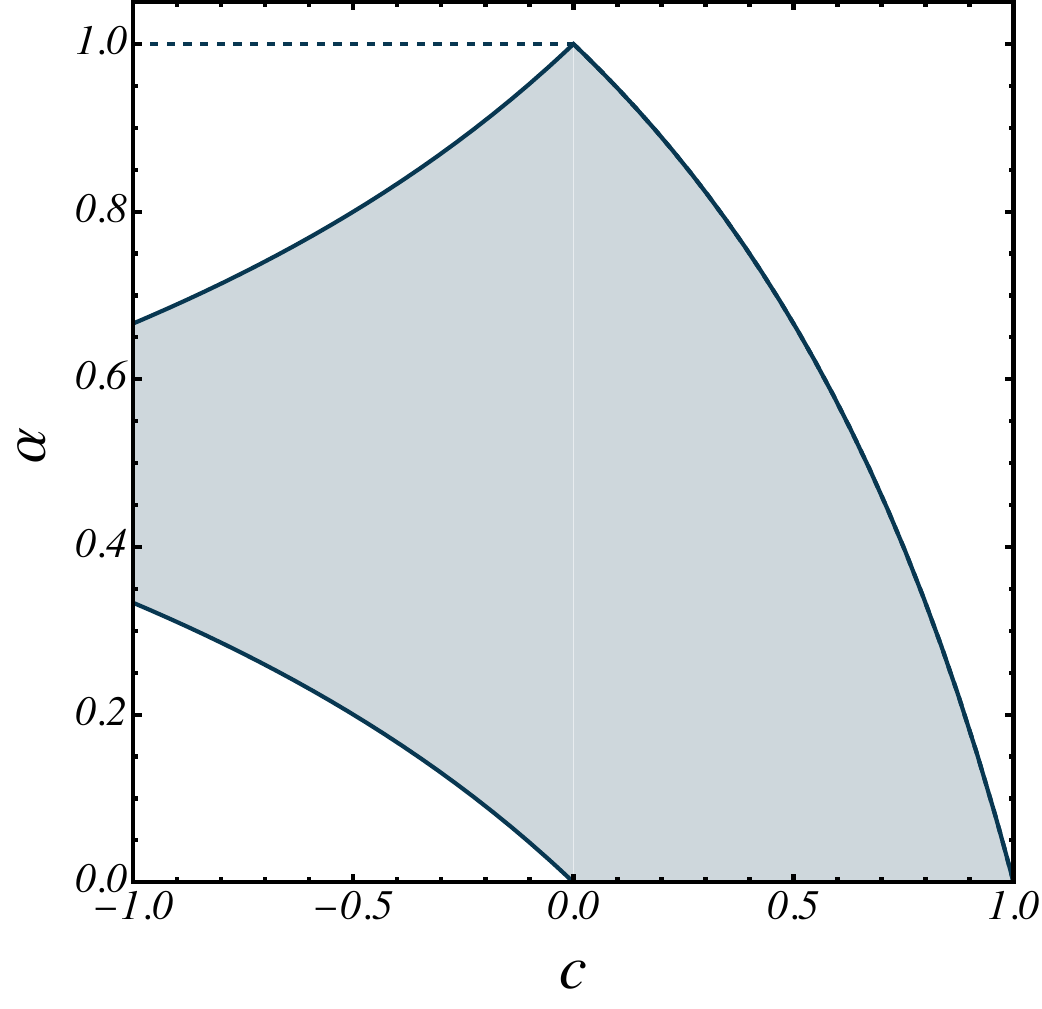}
    \caption{\textbf{Allowed $\alpha$ choices at a given $c$.}
    The safe choices are shown as a shaded region.
    For $c > 0$, the solid lines show the solutions of $\text{max}(0,\overline{\alpha}(\alpha_{\text{min}}(c),c) = 1)$ and $\text{min}(0,\overline{\alpha}(\alpha_{\text{max}}(c),c) = 0)$.
    For $c<0$, the second line is an overestimate of the safe region for $\alpha$, as applying the TR transformation takes $\alpha$ to an unsafe spot and TR thus cannot be applied a second time.
    For this reason, the raw solution $\alpha_{\text{max}}(c)$ is shown as a dashed line for $c < 0$, and the solid line is given by $\text{min}(\alpha_{\text{max}}(c),1-\alpha_{\text{min}}(c))$, which yields an interval for which $\alpha$ remains safe under two successive applications of the TR transformation. 
    }
    \label{fig:safe_alphas}
\end{figure}

Another perspective is to impose that $c$ remains unchanged by time reversal.
In that case, one gets the expression
\begin{align}
    \overline{c} &= c, \\
    \overline{\alpha} &= 1 - \alpha + \frac{c}{c-2}.
\end{align}
Once again, this expression may be problematic as the $\overline{\alpha}$ is not guaranteed to remain in $[0;1]$ for a generic $c$.
For instance, take $c = -1$, which yields $c/ (c-2) = 1/3$: this case is only ``safe'' if $\alpha \in [1/3;2/3]$.
Worse, for $c = 1$, $c /(c-2) = -1$ so that only $\alpha = 0$ works (and is stable by TRS).
The domain of ``safe'' $\alpha$ choices as a function of $c$ is shown as a shaded region in Fig.~\ref{fig:safe_alphas}.

Bravely accepting this non-standard transformation, which by definition leaves the dissipative part of the action unchanged, one needs to check its effect on the remaining terms of the action.
\begin{align}
    \mathcal{T}\mathcal{A}_{\text{det}+\text{jac}} &= \ln p_0 (\bm{X}_{T}) +  \int\limits_{-T}^T dt (i\bm{Q}_{-t} + \overline{\overline{D}}{}^{-1} d_t^{(\alpha, c)}\bm{X}_{-t})\cdot \left[ \overline{\overline{D}} \,\overline{\overline{M}} \bm{\nabla} V \right] \nonumber \\
    &\hphantom{aa} -(1 - \alpha +\frac{c}{c-2}) \int\limits_{-T}^T dt \left( \bm{\nabla} \left(\overline{\overline{D}}\, \overline{\overline{M}}\bm{\nabla}  V\right)  + \frac{1}{2}(\alpha - \frac{c}{c-2})(1-c) \nabla\cdot  \overline{\overline{\Pi}}{}^2+ (1 - \alpha+ \frac{c}{c-2}) \frac{c}{2}  (\nabla \overline{\overline{\Pi}})^2 \right)
\end{align}
It is quite clear that the term that depends on $V$ on the last time will, once again, not trivially cancel out.

As a conclusion: for $c \neq 0$, there is no generalized time-reversal operation that leaves the action invariant, as we explored all possibilities of $\alpha$ and $c$ updates that could have left the action unchanged.
The system is thus truly non-equilibrium and does not have a generalized fluctuation-dissipation theorem.

\section{Chain rule}

When considering the effect of time-reversal operations in spatial noise, Sec.~\ref{sec:TRS_Spatial}, it is useful to derive the generalized chain rule for this specific multiplicative noise.
For simplicity, we only consider the RO case ($\Lambda_{ij} = \Lambda_{ji}$ with $P_{ab} = \delta_{ab}$).
Consider a function $V(\bm{X})$.
Over a single step of the discretized dynamics, one may define the midpoint used for a specific choice of $\alpha$ as
\begin{align}
    \overline{\bm{X}_{n}} &= \bm{X}_n + \alpha (\bm{X}_{n+1} - \bm{X}_n), \\
    &= \bm{X}_{n+1} - (1 - \alpha) (\bm{X}_{n+1} - \bm{X}_n),
\end{align}
or equivalently
\begin{align}
    \bm{X}_n &= \overline{\bm{X}_{n}} - \alpha (\bm{X}_{n+1} - \bm{X}_n), \\
    \bm{X}_{n+1} &= \overline{\bm{X}_{n}} + (1 - \alpha) (\bm{X}_{n+1} - \bm{X}_n).
\end{align}
Using these expressions, 
\begin{align}
    V(\bm{X}_{n+1}) - V(\bm{X}_n) &= V\left(\overline{\bm{X}_{n}} +(1-\alpha) (\bm{X}_{n+1} - \bm{X}_n) \right) - V\left(\overline{\bm{X}_{n}} -\alpha (\bm{X}_{n+1} - \bm{X}_n) \right)\\
    &= (\bm{X}_{n+1} - \bm{X}_n)\cdot \nabla_{\bm{X}} V(\overline{\bm{X}_{n}}) \nonumber \\
    &\hphantom{aa} + \frac{1}{2}(1-2 \alpha) \sum\limits_{i,j = 1}^N \sum\limits_{a,b = 1}^d (x_{i,a}(t_{n+1}) - x_{i,a}(t_{n})) (x_{j,b}(t_{n+1}) - x_{j,b}(t_{n}))\partial_{x_{i,a}}\partial_{x_{j,b}} V(\overline{\bm{X}_{n}}) \nonumber \\
    &\hphantom{aa} + o(dt).
\end{align}
In particular, it will be useful to consider the average of that expression over the noise during the step (conditional on $\bm{X}_n$), 
\begin{align}
    \left\langle V(\bm{X}_{n+1}) - V(\bm{X}_n)\right\rangle 
    &= \sum\limits_{i = 1}^N \sum\limits_{a=1}^d \left\langle\delta x_{i,a} \right\rangle \partial_{x_{i,a}} V(\overline{\bm{X}_n})+ \frac{1}{2}(1-2 \alpha) \sum\limits_{i,j = 1}^N \sum\limits_{a,b = 1}^d \left\langle \delta x_{i,a} \delta x_{j,b}\right\rangle \partial_{x_{i,a}}\partial_{x_{j,b}} V(\overline{\bm{X}_{n}})  + o(dt) \\
    &=\sum\limits_{i=1}^N \sum\limits_{a=1}^d \left( F_{i,a}(\bm{X}) dt + \alpha \sum\limits_{j \neq i} \left(   \left[ \partial_{x_{i,a}} \sqrt{\Lambda_{ij} (\bm{X})} \right] \sqrt{\Lambda_{ij} (\bm{X})} + c   \left[ \partial_{x_{j,a}} \sqrt{\Lambda_{ij} (\bm{X})} \right] \sqrt{\Lambda_{ji} (\bm{X})}\right) dt\right) \partial_{x_{i,a}} V(\overline{\bm{X}_n}) \nonumber \\
    &\hphantom{aa} + (1-2 \alpha) \sum\limits_{i,j = 1}^N \sum\limits_{a,b = 1}^d D_{ijab}\mathcal{H}_{ijab}(\overline{\bm{X}_{n}}) dt + o(dt). 
\end{align}
This expression is often rewritten as a chain rule for a total derivative of $V$ by dividing on both sides by $dt$,
\begin{align}
    d_t \left\langle V (\bm{X}) \right\rangle &= \sum\limits_{i=1}^N \sum\limits_{a=1}^d \left( F_{i,a}(\bm{X}) dt + \alpha \sum\limits_{j \neq i} \left(   \left[ \partial_{x_{i,a}} \sqrt{\Lambda_{ij} (\bm{X})} \right] \sqrt{\Lambda_{ij} (\bm{X})} + c   \left[ \partial_{x_{j,a}} \sqrt{\Lambda_{ij} (\bm{X})} \right] \sqrt{\Lambda_{ji} (\bm{X})}\right) \right) \partial_{x_{i,a}} V(\overline{\bm{X}_n}) \nonumber \\
    &\hphantom{aa} + (1-2 \alpha) \sum\limits_{i,j = 1}^N \sum\limits_{a,b = 1}^d D_{ijab}\mathcal{H}_{ijab}(\overline{\bm{X}_{n}}) \\
    &= \left(\bm{F} (\bm{X}) + \alpha (\overline{\overline{\Pi}}\nabla \overline{\overline{\Pi}} -c {}^t\overline{\overline{\Pi}}\nabla\overline{\overline{\Pi}})\right)\cdot\nabla_{\bm{X}}V(\bm{X}) + (1-2\alpha) \overline{\overline{D}}\nabla_{\bm{X}}^2 V(\bm{X}) \label{eq:ChainRule}
\end{align}
where the last line introduces the short-hand notations used elsewhere in these notes (notice the sign flip in front of the term proportional to $c$ to keep the derivative over the first index).
In particular, for a symmetric $\overline{\overline{\Pi}}$,
\begin{align}
    d_t \left\langle V (\bm{X}) \right\rangle &= \left(\bm{F} (\bm{X}) + \alpha (1-c)\overline{\overline{\Pi}}\nabla \overline{\overline{\Pi}}\right)\cdot\nabla_{\bm{X}}V(\bm{X}) + (1-2\alpha) \overline{\overline{D}}\nabla_{\bm{X}}^2 V(\bm{X}) \label{eq:ChainRule_Symm}
\end{align}
It is also convenient to write this expression as
\begin{align}
        d_t \left\langle V (\bm{X}) \right\rangle &= \left\langle d_t \bm{X} \right\rangle\cdot\nabla_{\bm{X}}V(\bm{X}) + (1-2\alpha) \overline{\overline{D}}\nabla_{\bm{X}}^2 V(\bm{X}) 
\end{align}

\section{Aside: Multiplicative noise and SDE discretization\label{sec:Multiplicative_Aside}}

The most general discretization scheme for the SDE consists in choosing an $\alpha \in [0;1]$ such that discretized dynamics read
\begin{align}
    \bm{r}_n(t+dt) = \bm{r}_n (t) + \bm{U_n}( t + \alpha dt  ) dt,
\end{align}
where usual choices are $\alpha = 0$ (It\={o} convention), $\alpha = 1/2$ (Stratonovich convention) and $\alpha = 1$ (anti-It\={o}, kinetic, thermal, or Hänggi-Klimontovich convention).
This choice affects the expression of multiplicative parts of the stochastic processes in the velocity.

To illustrate this, consider the minimal case of a $1d$ process with multiplicative noise,
\begin{align}
    \dot{x} = f(x) + g(x) \zeta(t)
\end{align}
with $\zeta$ a zero-mean, unit-variance Gaussian white noise, $\langle \zeta\rangle = 0$ and $\langle \zeta(t) \zeta(t')\rangle = \delta(t - t')$.
In this case, one may write the Fokker-Planck equation associated to the stochastic variable $x$ as~\cite{Lau2007,GonzalezArenas2012}
\begin{align}
    \frac{\partial P}{\partial t}(x,t) = -\frac{\partial}{\partial x}\left[ \left(\vphantom{\int}f(x) + \alpha g(x) g'(x)\right) P(x,t)\right] + \frac{1}{2} \frac{\partial^2}{\partial x^2} \left[ g(x)^2 P(x,t) \right]. \label{eq:forward_FP}
\end{align}
It is immediately clear that the Fokker-Planck equation becomes independent of $\alpha$ when $g'(x) = 0$, \textit{i.e.} in the additive case.
The Fokker-Planck probability current $J$ for this minimal example is deduced from the mass conservation equation,
\begin{align}
    \frac{\partial P}{\partial t}(x,t) + \frac{\partial J}{\partial x}(x,t) = 0,
\end{align}
leading to 
\begin{align}
    J(x,t) &= \left(\vphantom{A^A}f(x) + \alpha g(x) g'(x)\right) P(x,t) - \frac{1}{2} \frac{\partial}{\partial x} \left[ g(x)^2 P(x,t) \right] \\
    &= \left(\vphantom{A^A}f(x) - (1 -\alpha) g(x) g'(x)\right) P(x,t) - \frac{1}{2} g(x)^2 \frac{\partial P}{\partial x} (x,t). \label{eq:FP_current}
\end{align}
Let us assume that the probability $P$ admits a steady-state $P_s(x)$ as $t \to \infty$.
The associated steady-state current $J_s$, according to mass conservation, should verify
\begin{align}
    \frac{d J_s}{d x}(x) = 0
\end{align}
so that $J_s(x) = J_s$ is constant in stationary states.
According to Eq.~\ref{eq:FP_current}, the expression of the steady-state current is
\begin{align}
    J_s &= \left(\vphantom{A^A}f(x) - (1 -\alpha) g(x) g'(x)\right) P_s(x) - \frac{1}{2} g(x)^2 \frac{d P_s}{d x} (x). 
\end{align}
In particular, it is convenient (and physically relevant) to write the stationary probability distribution as an exponential,
\begin{align}
    P_s(x) = \frac{1}{Z_s} e^{-S(x)}
\end{align}
with
\begin{align}
    Z_s \equiv \int dx e^{-S(x)}. 
\end{align}
With this rewriting, 
\begin{align}
    J_s &= \left[ \left(\vphantom{A^A}f(x) - (1 -\alpha) g(x) g'(x)\right) + \frac{1}{2} g(x)^2 S'(x)\right] P_s(x). \label{eq:steady_current}
\end{align}
This last equation can be solved for a given value of $J_s$ to find an expression of $S(x)$ and thus of $P_s(x)$.

In particular, an \textit{equilibrium} steady-state is characterized by $J_s = J_{\text{eq}} = 0$.
The solution in that case verifies, from Eq.~\ref{eq:steady_current},
\begin{align}
    \left(\vphantom{A^A}f(x) - (1 -\alpha) g(x) g'(x)\right) + \frac{1}{2} g(x)^2 S'(x) &= 0.
\end{align}
Assuming that $g \neq \overline{0}$, this expression yields
\begin{align}
    S_{\text{eq}}(x) = - 2 \int^x du \frac{f(u)}{g(u)^2} + (1 -\alpha) \ln\left[ g(x)^2\right]
\end{align}
where the lower bound of the integral can be absorbed into $Z_{\text{eq}}$.
Note in particular that in the case of conservative forces, 
\begin{align}
    f(x) = -\frac{1}{2} g^2(x) V'(x),
\end{align}
(where we absorb the temperature into the definition of the potential) leading to
\begin{align}
    S_{\text{eq}}(x) = V(x) + (1 -\alpha) \ln\left[ g(x)^2\right] \label{eq:eq_potential}
\end{align}
In other words, in cases in which the deterministic component is conservative, the combination of the choice of discretization and of multiplicative noise lead to an effective potential.
In order to recover the equilibrium distribution $P_{\text{eq}}(x) = e^{-V(x)} /Z_s$, the choice $\alpha = 1$ can be used.
However, calculations tend to be heavier with this choice.

One thus often chooses the Stratonovich convention $\alpha = 1/2$ (for reasons that will become clearer in the context of path-integral formalisms later).
In that case, it is necessary for the convergence of the distribution to the right steady state to compensate the spurious evolution due to $\alpha$ and $g$ by introducing a modified potential $V_\alpha(x)$ and using the replacement
\begin{align}
    V(x) \mapsto V_\alpha(x) \equiv V(x) -  (1 -\alpha) \ln\left[ g(x)^2\right]. \label{eq:V_replacement_nospurious} 
\end{align}
Applying this replacement, one recovers, in Eq.~\ref{eq:eq_potential}, $S_{\text{eq}}(x) = V(x)$ as expected.
Instead, one may choose to keep $V$ and $\alpha$ as well as the resulting non-equilibrium steady-state~\cite{GonzalezArenas2012}.

In the context of path-integral formulations, an important consideration is the effect of $g$ and $\alpha$ on the time-reversed process.
To study it, one may write~\cite{GonzalezArenas2012} the backwards Fokker-Planck equation for the reversed process, described by $\widetilde{P}(x,t)$ as
\begin{align}
    \frac{\partial \widetilde{P}}{\partial t}(x,t) = \frac{\partial}{\partial x}\left[ \left(\vphantom{\int}\widetilde{f}(x) + (1-\alpha) g(x) g'(x)\right) \widetilde{P}(x,t)\right] - \frac{1}{2} \frac{\partial^2}{\partial x^2} \left[ g(x)^2 \widetilde{P}(x,t) \right]. \label{eq:backward_FP}
\end{align}
as the time-reversal operation is defined via $x(t) \mapsto x(-t)$, $\alpha \mapsto 1-\alpha$ and  $f \mapsto \widetilde{f}$ to be determined.
In the backward evolution, assuming that we are interested in an equilibrium steady state ($\widetilde{J}_s = 0$), the equivalent of Eq.~\ref{eq:eq_potential} for backward dynamics is
\begin{align}
    \widetilde{S}_{eq} = \widetilde{V}(x) + \alpha \ln\left[g(x)^2\right].
\end{align}
Thus, in order for the dynamics to converge to the same equilibrium distribution in both forward and backward dynamics, one needs
\begin{align}
    \widetilde{f}(x) = f(x) - (1-2\alpha)g(x) g'(x). \label{eq:TRf_replacement_nospurious} 
\end{align}
For the choice $\alpha = 1/2$ (Stratonovich convention), $\widetilde{f}(x) = {f}(x)$, meaning that the deterministic dynamics are naturally time-reversal symmetric.

\bibliography{PostDoc-StefanoMartiniani,ref}